\def\ef{{$E_F$}}
\def\a {{\AA$^{-1}$}}
\def\bt {{Bi$_2$Te$_3$}}

\def\sn {{Sn 3$d_{5/2}$}}
\def\te {{Te 3$d_{5/2}$}}
\def\bi {{Bi 4$f_{7/2}$}}
\def\sb {{Sb 3$d_{5/2}$}}
\def\t {{$t_d$}}
\documentclass[preprint,onecolumn,amsmath,amssymb,aps,floatfix,prb,longbibliography]{revtex4-2}
\usepackage{graphicx}% Include figure files
\usepackage{dcolumn}% Align table columns on decimal point
\usepackage{bm}% bold math
\usepackage{appendix}
\usepackage{comment}
\usepackage{xr-hyper}
\usepackage{acronym}
\usepackage{xcolor}
\usepackage{txfonts}
\usepackage{multirow}
\pdfoutput=1 %% this is very important command to upload pdf figures in condmat
\newcommand{\srb} {\textcolor{black}}
\newcommand{\srbg} {\textcolor{black}}
\newcommand{\srbn} {\textcolor{black}}

\begin{document}
%	\begin{potrait}
%	\title{Bilayer stanene  on a magnetic topological insulator}% 
%	\title{Bilayer stanene  on a magnetic topological insulator with a sandwiched buffer layer}% 
%	\title{Bilayer stanene  on a magnetic topological insulator facilitated by buffer layer }% with a  buffer layer}% in between}% 
\title{Growth of bilayer stanene on a magnetic topological insulator aided by a buffer layer}
	% Bilayer stanene on a magnetic topological insulator facilitated by a buffer layer formation
%	Evidence of Stanene formation on Sb-doped MnBi$_2$Te$_4$ facilitated/aided/enabled by an interface layer}% Force line breaks with \\
%	\title{An interface layer enabled Stanene formation on Sb-doped MnBi$_2$Te$_4$ OR Evidence of Stanene formation on Sb-doped MnBi$_2$Te$_4$ facilitated/aided/enabled by an interface layer}% Force line breaks with \\
%	\title{Evidence of Stanene on Sb-doped MnBi$_2$Te$_4$}% Force line breaks with \\
%	\title{Electronic band structure of stanene and its development in the proximity of Sb-doped intrinsic magnetic topological insulator MnBi$_2$Te$_4$ surfaces}% Force line breaks with \\
	%\title{Stanene on Sb-doped Intrinsic Magnetic Topological Insulator MnBi$_2$Te$_4$}% Force line breaks with \\
	\author{Sajal Barman$^{1,\dagger}$, Pramod Bhakuni$^{1,\dagger}$,  Shuvam Sarkar$^{1}$, Joydipto Bhattacharya$^{2,3}$,  Mohammad Balal$^{1}$, Mrinal Manna$^{1}$, Soumen Giri$^{1}$,  Arnab Kumar Pariari$^{4}$, Tom\'a\v{s}  Sk\'ala$^5$,  Markus H\"ucker$^{4}$,  Rajib Batabyal$^{1}$, Aparna Chakrabarti$^{2,3}$, and Sudipta Roy Barman$^{1,*}$}
	\affiliation{$^1$UGC-DAE Consortium for Scientific Research, Khandwa Road, Indore, 452001, India}
	\affiliation{$^2$Theory and Simulations Laboratory,  Raja Ramanna Centre for Advanced Technology, Indore 452013, Madhya Pradesh, India}
	\affiliation{$^3$Homi Bhabha National Institute, Training School Complex, Anushakti Nagar, Mumbai  400094, Maharashtra, India}
	\affiliation{$^4$Department of Condensed Matter Physics, Weizmann Institute of Science, Rehovot, Israel}
		\affiliation{$^5$Charles University, Faculty of Mathematics and Physics, V Hole\v{s}ovi\v{c}k\'ach 2, CZ-18000 Prague 8, Czech Republic.}
		\affiliation{$^{\dagger}$Both authors contributed equally.}

\begin{abstract}
	Stanene, a two-dimensional counterpart to graphene, has the potential to exhibit novel quantum phenomena when grown on a magnetic topological insulator (MTI). This work demonstrates the formation of  up to bilayer stanene   on  
\srb{30\% Sb-doped MnBi$_{2}$Te$_4$ (MBST), a well known MTI, albeit with a buffer layer (BL) in between.
 ~Angle-resolved photoemission spectroscopy (ARPES), when combined with density functional theory (DFT), reveals stanene related bands such as two hole-like bands and an inverted parabolic band around the $\overline{\Gamma}$ point.  An outer hole-like band traverses the Fermi level  (\ef) and gives rise to  a  hexagonal Fermi surface, showing that stanene on MBST is metallic. In contrast,  a  bandgap of 0.8 eV is observed at the $\overline{K}$ point. We find that DFT  shows good agreement with ARPES only when the BL and hydrogen passivation of the top Sn layer are considered in the calculation.}	Scanning tunneling microscopy (STM) establishes  the honeycomb buckled structure of stanene. A stanene-related component is  also detected in the Sn $d$ core level spectra, in addition to a BL-related component. The BL, which forms because of the chemical bonding between Sn and the top two layers of MBST, has an ordered crystal lattice with random anti-site defects.    The composition of the BL  is estimated to be Sn:Te:Bi/Sb $\approx$ 2:1:1 from x-ray photoelectron spectroscopy. 		\srb{ Low energy electron diffraction  shows that the lattice constant of stanene is marginally larger than that of MBST, and the STM result aligns with this.} The BL bridges this disparity and provides a platform for stanene growth. 
	% with a step height of 0.35$\pm$0.03 nm and buckling height of 0.1$\pm$0.01 nm. % a lattice constant of 0.46$\pm$0.01 nm.  
	\end{abstract}
	\setcounter{figure}{0}
	\renewcommand{\figurename}{Figure}
	\renewcommand{\thefigure}{\arabic{figure}}
	
\maketitle
	
\section{\label{sec:intro}Introduction}
%\vskip 1.45cm
Stanene is  a two dimensional (2D) topological insulator (TI) or quantum spin Hall (QSH) phase %with a large gap of 0.3 eV, 
~characterized by the topologically protected helical boundary states that are immune to backscattering~\cite{cc_liu_prb2011,XuBinghai2013,Matthes2013}. It exhibits a honeycomb structure, similar to graphene~\cite{Novoselov2004,Geim2007}.   The first experimental realization of  monolayer stanene  was reported on  a TI,  namely, bismuth telluride  (\bt)~\cite{Zhu2015}. A later study  on Cu(111) reported the growth of ultra-flat stanene with a band inversion~\cite{Deng2018}.  The recent discovery of superconductivity in few-layer stanene grown on  Bi~\cite{Zhao_Sn_Bi_2022} and PbTe/\bt~\cite{Liao2018,Zang2018}  has revitalized this area of research. Superconductivity along with the QSH state might give rise to an interesting possibility of  realizing proximity-induced topological superconductivity~\cite{LiangFu2008,Ando_Shoman2015, Ando2020, Charpentier2017}. Liao \textit{et al.}~\cite{Liao2018} reported  that the superconducting transition temperature ($T_C$)  could be increased from 0.5~K to 1.2~K by enhancing the thickness of the PbTe buffer layer (BL).   \srb{Buffer layers such as Pd$_2$Sn and Ag$_2$Sn  were reported for  stanene that was grown on  Pd(111)~\cite{Yuhara2021} and Ag(111)~\cite{Yuhara2018}, respectively. BL formation} has also played a constructive role in  2D systems, such as for achieving quasi-free-standing bilayer graphene on SiC~\cite{Oliveira2015,Mesple2023} and silicene on Ag~\cite{Du2016}.  %, as well as a high capacitance dielectric layer for the fabrication of field effect transistors~\cite{Xu2023} and growth of nano-crystals~\cite{Huang1998}   
 %\cite{Yi2022}
 
Tellurides have been at the forefront of  materials research, such as  the discovery of  the TI phase  in \bt~\cite{Chen2009, Zhang2009}; high thermoelectric performance in BiSbTe~\cite{Poudel2008};  a new topological phase Kramers nodal line  in rare earth chalcogenide  LaTe$_3$~\cite{Sarkar2023}; large magnetoresistance,   vortices in electron flow,  superconductivity, and higher order topology  in transition metal dichalcogenides~\cite{Ali2014,  Steinberg2022, Kang2015, Wang2019}; and the intrinsic magnetic topological insulator (MTI) phase in MnBi$_2$Te$_4$ (MBT)~\cite{Otrokov2019, JiangLi2019,Padmanabhan2022,Gao2023}, to mention a few.   MBT is a layered material  with  the  (0001) surface being Te terminated,  Te forms the top layer of a Te-Bi-Te-Mn-Te-Bi-Te ``septuple", where multiple septuple layers  are coupled by van der Waals  interaction. MBT  hosts the topological surface state located within the energy gap between the bulk valence and the conduction band  stemming from the topologically nontrivial bulk bands~\cite{Otrokov2019,chenMBT2019}.   However, because of  n-type doping intrinsically present in MBT, these bands  are shifted to higher binding energy such that the Dirac point appears $\sim$0.28 eV below the Fermi level (\ef)~\cite{Otrokov2019, chenMBT2019, HangLi2019}. It was shown that  substitution of  Bi by Sb in MBT %forming Mn(Bi$_{0.74}$Sb$_{0.26}$)$_2$Te$_4$  
 ~shifts back the bands towards \ef\, such that the topological surface states  cross  \ef~\cite{ChenSbMBT2019, Ma2021, Glazkova2022}.   
	
Stanene is a member of the X-ene family~\cite{cc_liu_prb2011,cc_liu2011,XuBinghai2013,Kamal2015} and has great promise for application in  spintronics~\cite{Pesin2012}, nanoelectronics~\cite{Molle2016}, catalysis~\cite{Wang2023}, and  biomedicine~\cite{Tao2019}.  Deposition of Sn on chalcogenides  and related substrates has produced stanene~\cite{Zhu2015,Xu2017,Deng2018,Liao2018,Zang2018,Yuhara2018, Yuhara2021,Li2020,Zheng2019,Gou2017,Zhao_Sn_Bi_2022} as well as  surface alloys of Sn~\cite{Sadhukhan_appSS2020,Shah2021,Sadhukhan2019}  with  interesting properties.  \srb{Therefore, although stanene has previously been grown on non-magnetic topological and non-topological substrates, the impact of a magnetic topological substrate like MBT or Sb-doped MBT on its properties may be  intriguing, given that the quantum anomalous Hall effect can result from magnetic interaction with the QSH state~\cite{Deng_qah2020, Haldane1988, Yu2010,XuBinghai2013,  ZhangScience2013}. Furthermore, proximity-induced superconductivity might appear in an MTI  due to a two-dimensional superconducting stanene grown on it, which might facilitate the existence of Majorana fermions and find applications in the field of quantum computing. }  While no such work exists in the literature to date,  a study of stanene on \bt~\cite{Zhu2015} --  structurally similar to MBT with a Te-terminated ``quintuple" layer (Te-Bi-Te-Bi-Te) --  suggested  weak-coupling  between stanene and the substrate with no  chemical bonding. In contrast, a relatively recent research~\cite{Li2020} has provided evidence of  a BL formed by an interfacial chemical reaction of Sn with \bt.  

 \srb{Based on our initial investigations, it appears that both Sb-doped MBT and MBT, which have similar structure, termination, and lattice constants, have the ability to promote the growth of stanene. However, for this study, we have chosen to focus on conducting a thorough analysis of the former to examine the effect of Sn deposition on the bulk bands that disperse up to the \ef.}
 	%Although our preliminary investigations indicate that both MBT and Sb-doped MBT, both having similar structure, termination, and close lattice constants, have the potential to facilitate stanene growth, in the present work, we opt to conduct a comprehensive study on the latter in order to analyze the impact of  Sn deposition on the topological bulk bands that disperse up to the \ef.
~Combining multiple techniques such as angle resolved photoemission spectroscopy (ARPES), scanning tunneling microscopy (STM), density functional theory (DFT),   x-ray photoelectron spectroscopy (XPS),   and low energy electron diffraction (LEED), we demonstrate  the formation of  monolayer and bilayer stanene on $\sim$30\% Sb-doped MnBi$_2$Te$_4$ (henceforth referred to as MBST)  at room temperature.  This is the first report of stanene growth, or any X-ene growth, on an MTI.  	 Height profile analysis of atomic resolution STM  topography images further establishes stanene formation, which is preceded by the growth of  a BL. The XPS analysis reveals that the BL forms through chemical bonding between Sn and the two uppermost  layers of MBST. It has a thickness of 0.9 nm and a composition ratio of  $\approx$ 2:1:1 (Sn:Te:Bi/Sb). \srbn{ARPES shows  two hole-like and an inverted parabolic stanene bands around the $\overline{\Gamma}$ point.  An outer hole-like band crosses the Fermi level (\ef) creating a hexagonal Fermi surface, but a bandgap  is seen at the $\overline{K}$ point.  DFT calculation that includes the BL and hydrogen passivation of the top stanene layer exhibits good agreement with ARPES.} 
%1st: monolayer stanene
%2nd: bilayer stanene

\section{\label{sec:methods}Methods}
%\section{Methods}
\noindent\underline{Photoemission spectroscopy:} The ARPES and XPS measurements were performed  using a helium discharge lamp equipped with a monochromator and a monochromatized Al K${\alpha}$ source, respectively that are mounted on an analysis chamber with a base pressure of 1$\times$10$^{-10}$ mbar.  The  photoelectrons were detected   using the R4000 hemispherical electron energy analyzer from Scienta Omicron GmbH   with the  energy resolution set at better than 15 meV,  while the angular resolution was 1$^{\circ}$. The analyzer is equipped with a wide angle lens that has an acceptance angle of $\pm15^\circ$.   The XPS measurements were performed using the transmission lens mode of the analyzer with an energy resolution  of  0.33 eV.  The LEED equipment comprised of a four-grid rear-view optics system. 
\srb{Core level spectroscopy (CLS) using low energy photons (90 eV) was performed using synchrotron radiation at the Materials Science Beamline at Elettra in Trieste, Italy. Hard x-ray photoelectron spectroscopy (HAXPES) using 6 keV photon energy was performed at the P22 beamline in  PETRA III of  Deutsches Elektronen-Synchrotron in Hamburg, Germany.} % with 90 eV and He II radiation from the helium discharge lamp. }

The analysis of the photoemission data was performed using Igor Pro, ver 9 from Wavemetrics Inc. The core level spectra were fitted with the least square error minimization method as in our earlier works~\cite{Biswas2004} with the instrumental resolution represented by a Gaussian function of fixed width.  The energy position, intensity, life-time broadening,  and the Tougaard background~\cite{Tougaard1989} parameter were varied  during the fitting. \srb{ The position of the bands  has been determined from the raw ARPES intensity plot by taking cuts  along the momentum  axis at a fixed energy  to obtain the momentum distribution curves (MDCs). The energy distribution curves (EDCs) are obtained  from cuts  along the energy  axis at a fixed momentum.  The band positions are identified  by least-squares curve fitting of the  peaks of the cuts with Lorentzian functions and  a linear background.   The second derivative analysis of the raw ARPES intensity plots has been performed  with respect to the energy axis to show  the  bands with better contrast, as in our earlier work~\cite{Sadhukhan2019}.} 
\vskip 5mm
\noindent\underline{STM:} The STM measurements were performed using a variable temperature STM from Scienta Omicron GmbH  at a base pressure of 4$\times$10$^{-11}$ mbar.  Electrochemically etched polycrystalline W tips as well as  mechanically grinded Pt-Ir tips  were used. These tips were cleaned \textit{in-situ} using Ar$^{+}$ ion sputtering and applying voltage pulses. The tip was biased and the sample was kept at the ground potential. The average height of each layer is calculated by averaging over more than 50 height profiles and also by height histogram. %, as in our earlier work~\cite{Singh2023}.  %In a height profile, the difference of the average $z$ corrugation on the  adlayer and the substrate provides the height of the former.
%~ All the measurements were performed at RT unless otherwise mentioned. %srbPtoPB: **other analysis method?**
~The STM images were analyzed using the SPIP and Gwyddion software~\cite{Neas2011}. Atomic resolution STM images are enhanced using Fourier filtering. % in which we take the fast Fourier transform of the image, and then select all the characterisitic Bragg peaks and the background centered around k= 0 which reduces the experimentally induced diffuse features due to noise. 
%~The height profiles extracted from STM images were  fitted with Gaussian function using the multipeak fit package of Igor Pro. 
%~To calculate the mean square roughness ($S_q$) for the BL, we have used a masking procedure %given in Gwyddion, in which we 
%to select only the islands related to the BL. % and leave out the uncovered MBST region to calculate the roughness of the BL islands.

\vskip 5mm
\noindent\underline{Density functional theory:} 
The DFT based first-principles  calculations were performed  using the projector-augmented-wave (PAW)~\cite{Kresse1999} potential and the exchange correlation functional  Perdew–Burke–Ernzerhof (PBE) generalized gradient approximation (GGA)~\cite{Perdew1996}, as implemented in the Vienna ab initio simulation package (VASP)~\cite{Kresse1996}. Periodic  slab calculations have been performed with  a vacuum of 1.5~nm   along the $z$ direction to overcome the interactions between the  periodic images. The slab calculations were performed with  a plane wave basis with an energy cutoff of 400 eV, a 13$\times$13$\times$1 Monkhorst-Pack $k$ grid~\cite{Monkhorst1976} was used for the $k$-space sampling. A force convergence criterion of 0.01 eV \a\,  was selected for structural optimization. The spin–orbit coupling (SOC) interaction was included in  the electronic structure calculation.  The band structure calculations were performed along the high-symmetry path  $\overline{M}$ (0.5, 0.0, 0.0) – $\overline{\Gamma}$ (0.0, 0.0, 0.0) – $\overline{K}$ (1/3, 1/3, 0.0). The VESTA program~\cite{Momma2011} has been employed for the representation and visualization of the crystal structures. The orbital character of the bands is obtained by projecting the Bloch wave function on the orbitals. 

 %srb: taken to DFT discussion: Monolayer stanene  was considered in which the lower (upper) Sn sublattice is located at the hollow (face centered cubic) site of the substrate. The bilayer stanene was considered following the $\alpha$-Sn structure in the (111) direction.  The MBST(0001)  surface was modeled by a periodic slab including three septuple layers, where 2 of the 6 Bi atoms were replaced by Sb so that the doping level is 33\%, which is close to the experiment. The lattice constant is taken to be 0.43 nm based on  x-ray crystallography data~\cite{Yan2019}. The input structure includes  hydrogen at the on top positions for hydrogen passivation calculation. The atom positions and lattice constant??  were relaxed for stanene and the top  septuple of MBST, whereas the two bottom septuples were kept fixed.
\vskip 5mm
\noindent\underline{Sample preparation and characterization:}
Stanene was grown by deposition of Sn ($99.99\%$) using a water-cooled  Knudsen cell~\cite{Shukla2004}  equipped with a mechanical shutter at an operating  temperature of 1073 K  and chamber pressure of $2\times10^{-10}$ mbar. The substrate temperature was  303-313~K and its distance from the Knudsen cell was $\sim$70 mm and the angle of deposition was $\sim$45$^\circ$.  %Sequential Sn deposition was carried out for a total cumulative time of 432 sec where initially the duration of the growth in each step was 18 sec and latter this was increased to the duration of 72 sec after four round of the growth. 

\begin{figure*}[t] 
	\includegraphics[width=\linewidth,keepaspectratio,trim={0.3cm 1.15cm 0cm 0cm },clip]{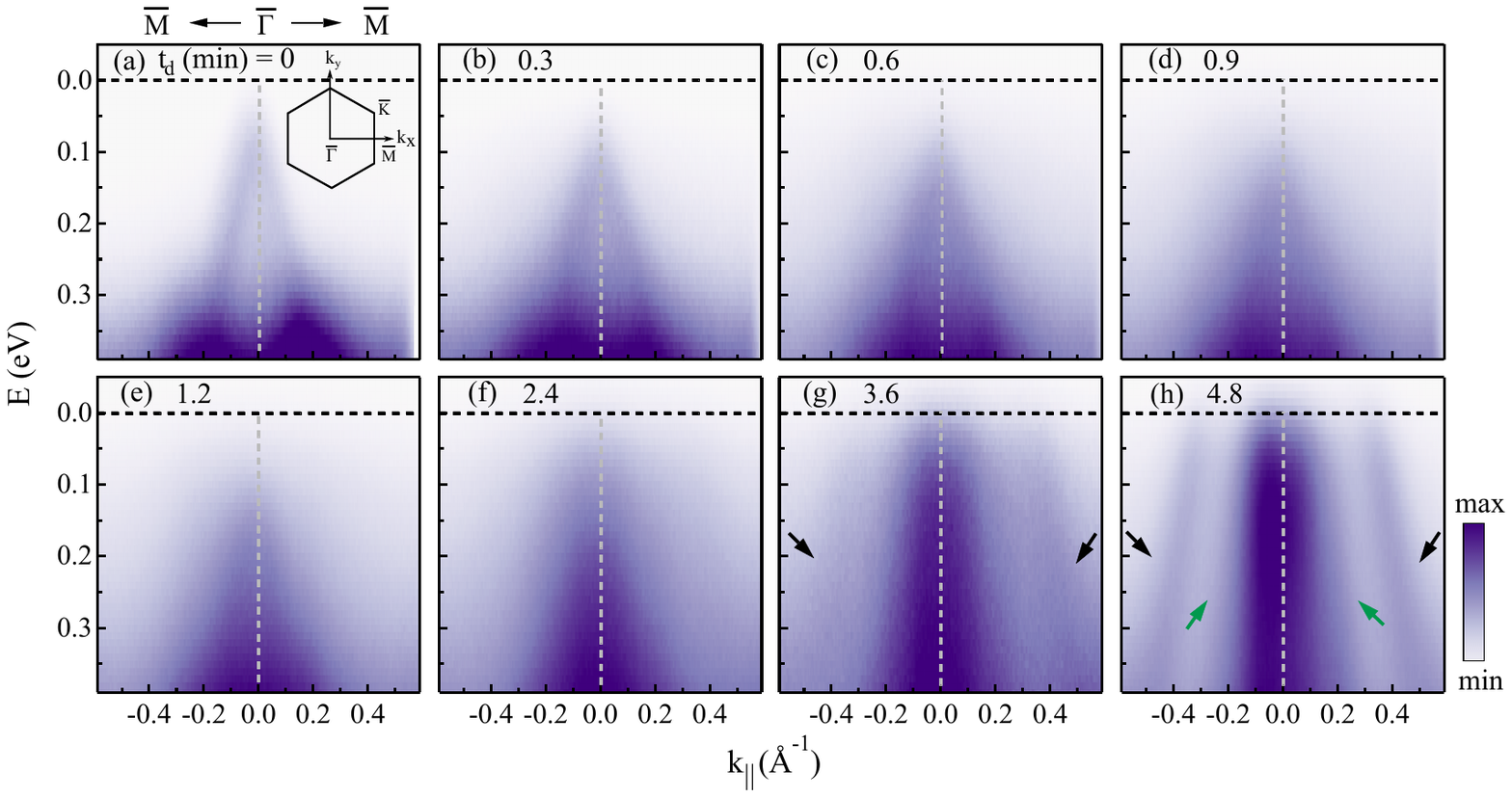}%{ARPES_All_2nd_deri_v6}
	\caption{ %\textbf{ARPES as a function of   Sn deposition on MBST.}  
		 (a-h)  ARPES intensity plots as a function of   Sn deposition on MBST,  \srbg{where  the binding energy ($E$)} \srb{in the vertical axis is plotted against the momentum parallel to the surface ($k_{||}$) in the horizontal axis}  from $\overline{\Gamma}$ towards $\overline{M}$ with increasing deposition time ($t_d$), as mentioned in unit of min at the top of each panel. The black (green) arrows show the outer (inner) band related to stanene. The Brillouin zone (BZ) is shown in the inset of panel \textbf{a}.  The zero of  $E$ corresponds to the \ef.}
	\label{arpes_All}
\end{figure*}

\textit{In-situ} peeling with a scotch tape  exposed an atomically clean (0001) surface of  the substrate MBST crystal. The peeling was performed at room temperature in the preparation chamber at a pressure of  $1\times10^{-10}$ mbar. % in the preparation chamber.  
~The  MBST single crystal was  grown by the flux method following a similar temperature profile described in previous reports~\cite{Lee_prx2021, Ma2021}. In order to grow $n$\% Sb-doped MBST crystal, a mixture of  high purity ($\geq$99.999\%) Mn powder, %(Alfa Aesar, 99.95\%), 
~Bi shots, %(Furuchi Chemical Corporation, 99.999\%), 
~Te chunks, %(Furuchi Chemical Corporation, $99.9999\%$), 
~and Sb shots %(Alfa Aesar, 99.999\%) 
~in a molar ratio Mn: Sb: Bi: Te = 1:\,10$\times$$n$:\,10$\times$$(1-n)$:\,16 were placed in an alumina crucible of a Canfield crucible set. %A thin quartz wool pad was also placed between the bottom crucible containing the above mentioned chemicals and the frit-disk to minimize cracks on crystals during centrifuge. %All the steps described above were done in an Ar glove box (MBRAUN). 
~Then, the entire crucible set was sealed in an evacuated silica ampule at $<10^{-5}$ mbar pressure. The sealed ampule was heated to 1173 K % 900$^\circ$C 
~for two days and kept at that temperature for 12 hours followed by a slow cooling 10 K/hr %10$^\circ$C/h 
down to 878~K. %605$^\circ$C. 
~At the final stage of the temperature profile, the ampule was cooled from 878~K%605$^\circ$C 
~to 863~K%590$^\circ$C 
~in 336 hrs and kept at that temperature for 72 hrs followed by centrifuge at 3000 rpm to remove the excess flux from crystals. Plate-like shiny crystals %with dimensions \srb{of  4-5 mm,} shown in  Fig.~S1(b) of the SM~\cite{supple}, 
~are obtained from the bottom crucible. %The transfer of the silica ampule at 590$^\circ$C from the furnace to the centrifuge setup in less than five seconds to avoid the formation of the Bi$_2$Te$_3$ or Sb$_2$Te$_3$ phase. After this procedure, successfully extracted thick, irregular block-shaped, well separated pieces of MBST single crystals with planner dimensions of the order of few mm were obtained from the bottom alumina crucible. 

\srb{The amount of Sb doping in our MBST crystal has been experimentally determined from the  Sb 4$d$ and Bi 5$d$ peaks measured by HAXPES. It is an appropriate method for obtaining information about the bulk electronic states~\cite{Nayak2012}. This is because of the large inelastic mean free path (IMFP) of the photoelectrons, which in the case of MBST  is estimated to be about 9~nm~\cite{Tanuma2011}. From the intensities of the Sb 4$d$ and Bi 5$d$ peaks shown in Fig.~S1 of the supplementary material (SM)~\cite{supple} obtained from multi parameter core level fitting  and considering the respective photoemission cross sections~\cite{Trzhas2018}, we find the Sb doping to be $\sim$30\%. Also, earlier research~\cite{ChenSbMBT2019,Glazkova2022} has shown that the Sb doping level 
influences 	 the position of the valence and conduction bands of MBST. It is clear that our crystal has  30\% Sb level because  our ARPES intensity plot in Fig.~\ref{arpes_All}(a) is similar to  Fig.~2(e) of Ref.~\onlinecite{ChenSbMBT2019} and Fig.~2(d) of Ref.~\onlinecite{Glazkova2022}, both of which also have a 30\% Sb doping.}% Therefore, henceforth in the text, the abbreviation MBST will represent MBT  with 30\% Sb doping. } % i.e., the composition being Mn(Bi$_{0.7}$Sb$_{0.3}$)$_2$Te$_4$ (MBST). 
	%srb: no need of XPS then: XPS with smaller IMFP of 2.9 nm, also  give similar estimate of Sb doping.
	 ~Futhermore, the magnetization ($M$) versus temperature  at a magnetic field ($H$) of 0.1 Tesla  shows the Neel temperature to be 24~K, whereas  $M(H)$ measurement at 2 K  shows the spin flop transition from antiferromagnetic to canted antiferromagnetic at 2.3 Tesla (Fig.~S2 of the SM~\cite{supple}), in  agreement with  literature~\cite{ChenSbMBT2019,Lee_prx2021}.  
	 The unit cell of MBST is shown in  Fig.~S3(a) of the SM~\cite{supple}. The MBST crystals with dimensions \srb{of  4-5 mm} that have been used in this work are  shown in  Fig.~S3(b) of the SM~\cite{supple}.

\section{\label{sec:results}Results and Discussion}
\subsection{ARPES study of sequential Sn depositions on MBST}
Figure~\ref{arpes_All}(a) shows an  inverted ``V"-like valence band  of MBST    with its maximum  close to \ef\, at the $\overline{\Gamma}$  point, as has been reported in literature~\cite{ChenSbMBT2019,Ma2021,  Glazkova2022,Volckaert2023}.  It disperses   from the maximum down to   about 0.3 eV \srbg {binding energy ($E$)} at  0.2~\a.  The substitution of Bi by Sb induces p-type doping in MBST~\cite{ChenSbMBT2019} causing the band maximum to appear close to \ef\ in contrast to  MBT, where it is observed  $\sim$0.4 eV below \ef\ due to the inherent n-type doping~\cite{chenMBT2019}.

An intriguing observation  with  Sn deposition   is a gradual, rigid band-like shift of the MBST  bulk valence band  maximum towards  larger $E$, as shown in \srb{Figs.~\ref{arpes_All}(b) to \ref{arpes_All}(e) measured towards the $\overline{\Gamma}$-$\overline{M}$ direction (see Fig.~S4 of SM~\cite{supple} for the second derivative intensity plots). }   A quantitative estimate of this shift  is  about 65 meV for a deposition time (\t) of  1.2 min, which is obtained by fitting the energy distribution curves (EDCs) drawn at $k_{\parallel}$= 0 with Lorentzian functions (Fig.~S5 of the SM~\cite{supple}). The ARPES intensity plots towards the $\overline{\Gamma}$-$\overline{K}$ direction also show a similar  shift of the band maximum; \srb{see  Fig.~S6 and S7  of the SM~\cite{supple} for the raw and second derivative images, respectively}.    
~The observed shift   contrasts with  the effect of p-type doping by Sb and  indicates that Sn deposition results in n-type doping. This  implies  the formation of a chemical bond between Sn and the substrate due to charge transfer. %srbT: too premature to say this here now: , which is an indication of formation of a buffer layer (BL) at the outset. 
~Note  that Figs.~\ref{arpes_All}(b) to \ref{arpes_All}(e) show  that the   shift is  gradual with \t, %   (figure~\ref{supp-SM_EDC_BE_Shift} -\ref{supp-GK_arpes_td} of SM~\cite{Misc}), 
~while, on the other hand, a discontinuous shift is  expected across  an intrinsic bulk band gap  (0.2 eV for MBST).  So, the observed gradual shift suggests the presence of the in-gap topological surface states~\cite{ChenSbMBT2019,Ma2021},  %up to \t= 1.2 min ~in spite of  possible formation of BL, 
~but these states are not observed directly in our ARPES data  recorded with 21.2 eV He~I radiation due to the photoemission matrix element  effect~\cite{chenMBT2019}. 

For  the largest deposition of  \t= 4.8 min, the ARPES intensity plot shows emergence of  new bands [Fig.~\ref{arpes_All}(h)]. An outer hole-like band indicated by black arrows is observed to disperse from   close to \ef\, at about 0.36~\a\,   to  0.4 eV at 0.5~\a.  In addition, a weak inner hole-like band -- highlighted by green arrows --  is also visible. At a relatively smaller Sn deposition of \t= 3.6 min in Fig.~\ref{arpes_All}(g), the outer  band is also visible (black arrows). %At  \t= 2.4 min, this band is faint [Fig.~\ref{arpes_All}(f)], but is  identified (white arrows) in the second derivative plot in  Fig.~S4(f)  of the SM~\cite{supple}.   
~\srb{The ARPES bands and the Fermi surface for  monolayer stanene are discussed in  section III.E. Results of our DFT calculation are presented in section III.F and compared with ARPES. Prior to that in the next sections, we present the evidence of  stanene and the BL formation from XPS, STM and LEED in sections III.B, III.C, and III.D, respectively.}

\subsection{Signature of stanene and buffer layer from core level spectroscopy}
	\begin{figure*}[t] 
	\includegraphics[width=\linewidth,keepaspectratio,trim={0cm 0cm 0cm 0cm},clip]{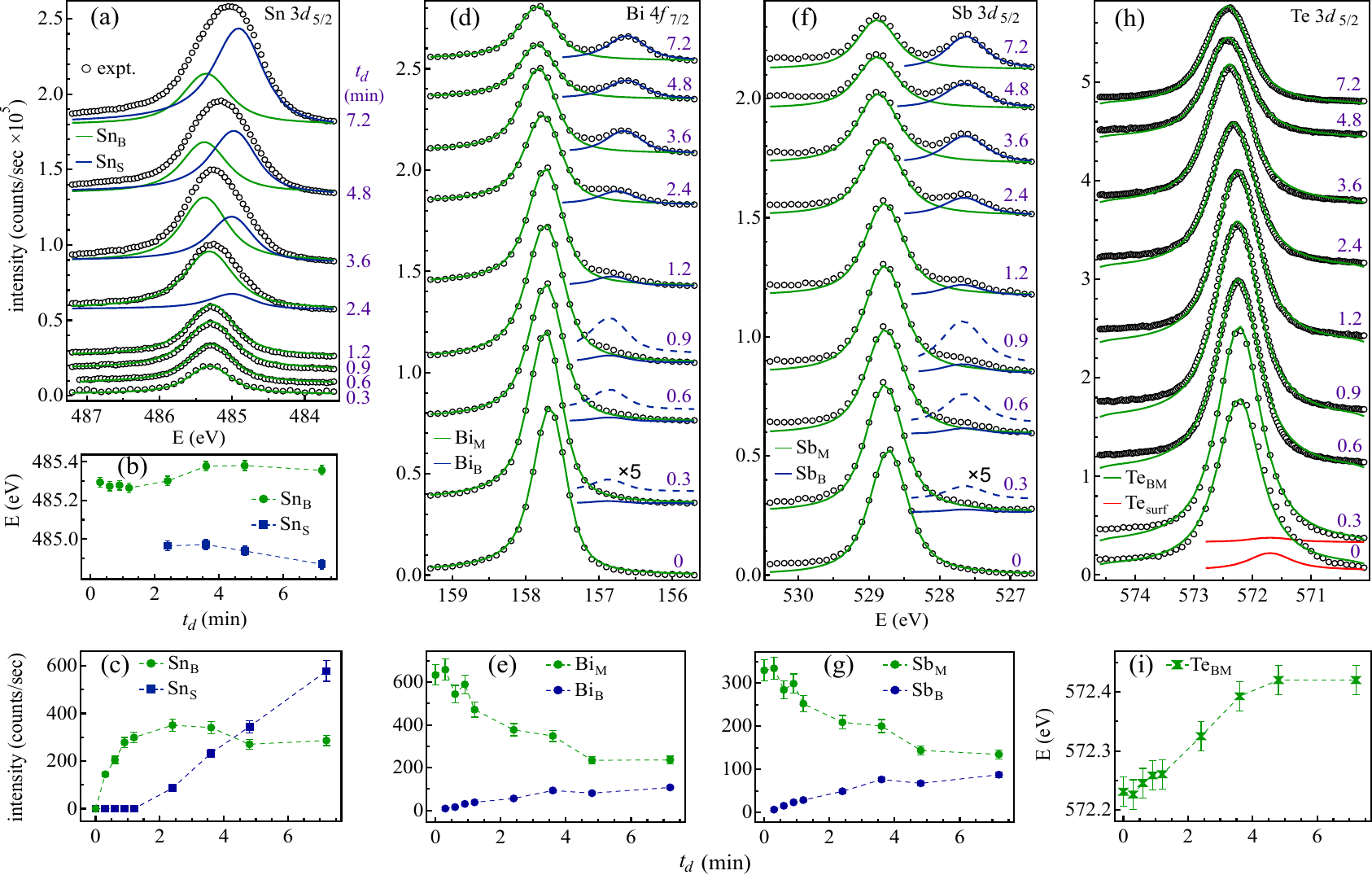}
	\caption{\label{xps} %\textbf{ XPS   as a function of Sn deposition on MBST.}   
		(a) Sn 3$d_{5/2}$ core level spectra  with two fitted components Sn$_{\rm B}$  and Sn$_{\rm S}$  for different \t, as labeled in purple color on the right.   (b) Binding energy ($E$) and (c) intensity of  the components as a function of \t. 
				(d)	Bi 4$f_{7/2}$ and (f) Sb 3$d_{5/2}$ core level spectra with two fitted components (blue and green curves).   Intensity as a function of \t\, for (e) Bi$_{\rm M}$ and Bi$_{\rm B}$  and (g) Sb$_{\rm M}$ and Sb$_{\rm B}$ components.  	(h) \te\ core level spectra with the fitted components Te$_{\rm BM}$ and Te$_{\rm surf}$, the latter becomes zero for \t$\geq$0.6 min;  (i) $E$(\t) for Te$_{\rm {BM}}$.  The spectra in panels \textbf{a, d, f, h} are staggered  vertically. }
\end{figure*}

\srb{The growth mechanism of Sn on MBST has been determined by studying the core level spectra of Sn and the components of the substrate as a function of \t. The discussion begins below with the \sn\ spectra. }

\begin{figure*}[t!] 
	\includegraphics[width=\linewidth,keepaspectratio,trim={0.4cm 2.8cm 9.6cm 0cm},clip]{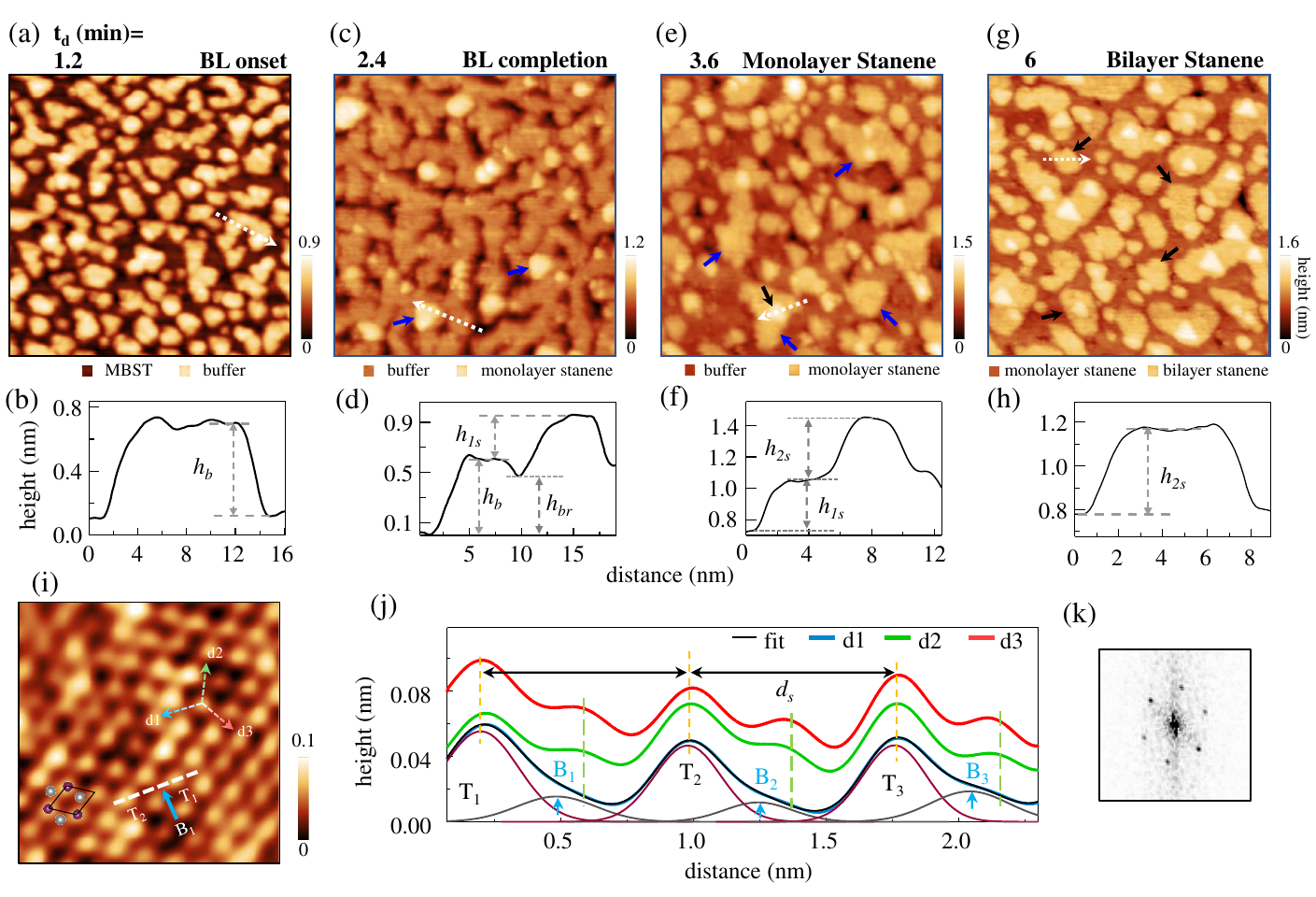} %{STM_growth_evolution_v4}
	\caption{ %\textbf{STM as a function of Sn deposition on MBST.} 
		STM topography images of size 70 nm $\times$ 70 nm for  \t= (a) 1.2 min with $I_T$=  0.4 nA and $U_T$= 1 V, (c) 2.4 min with  $I_T$= 0.14 nA and $U_T$= 0.07 V, (e) 3.6 min with  $I_T$= 0.9 nA and $U_T$= 0.4 V, and (g) 6 min with  $I_T$= 0.1 nA and $U_T$= 0.5 V. The legends at the bottom of each image indicate the color of the different regions. \srb{The blue [black] arrows in panels  \textbf{c, e} [\textbf{e, g}] highlight the monolayer (bilayer) stanene islands. 	  (b, d, f, h) Height profile along the white dashed arrow  for  panels \textbf{a}, \textbf{c}, \textbf{e}, and \textbf{g}, respectively. The heights of the BL, monolayer and bilayer stanene are represented by $h_b$,  $h_{1s}$ and $h_{2s}$, respectively.}  (i) An atomically resolved image of   bilayer stanene  (4 nm $\times$ 4 nm, $I_T$= 1.5 nA, $U_T$= -0.5 V, \t= 6 min).   
		 (j) Average height profiles of panel \textbf{i}  along  d1,  d2 and d3 directions.  The  curve along d1 is  fitted with  Gaussian functions. White dashed line in panel \textbf{i} defines T$_{\rm 1}$, B$_{\rm 1}$ and T$_{\rm 2}$. Peaks marked by  T$_{\rm n}$ ($n$ = 1-3) (magenta curves, maxima shown by  yellow dashed lines )  represent the upper sublattice atoms of stanene. B$_{\rm n}$ (gray curves), whose peak positions are  highlighted  by blue arrows, represent the lower sublattice atoms that are shifted  from $d_s$/2 (green dashed lines). (k) The Fourier transform of panel \textbf{i}. }% 
			   \label{stmseries}
\end{figure*}

\vskip 5mm
\noindent \textbf{\underline {BL formation from \sn}}:~~Figure~\ref{xps}(a) shows that for \t$\leq$ 2.4 min, the spectra are well fitted with only one component, Sn$_{\rm B}$ (green curve).  Sn$_{\rm B}$  appears at   485.3%$\pm$0.05 
~eV  [Fig.~\ref{xps}(b)]  that is significantly shifted to a larger $E$ from bulk Sn metal (484.7 eV). Rather, $E$ of   Sn$_{\rm B}$  is  close to SnTe (485.4 eV~\cite{Li2020,Shalvoy1977}) indicating that the Sn atoms react  with the top Te  layer of  the (0001)  surface of MBST (Fig.~S3(a) of the SM~\cite{supple}) forming Sn-Te bonds by transferring electrons, which is also supported by ARPES [Figs.~\ref{arpes_All}(b) to \ref{arpes_All}(e)]. %, which in turn, explains the shift of the MBST band maximum  observed in Fig.~\ref{arpes_All}(b-e). 
~This  is also consistent with  the smaller  tendency  of Sn to gain electrons  because the electronegativity of Sn is 1.96 in the Pauling scale compared to Te with electronegativity of 2.1~\cite{Mann2000}.  The reaction between Sn and Te contributes to the formation of a BL. \srb{Interaction of Sn with Bi/Sb also contributes to the BL formation, as shown by the \bi\ and \sb\ core level spectra discussed below.}  The BL related component, Sn$_{\rm B}$, %(``B" stands for ``buffer"), 
~is visible right from the beginning at \t= 0.3 min and increases  rapidly in intensity up to \t= 2.4 min as the BL grows, but stagnates thereafter [Fig.~\ref{xps}(c)]. As discussed later in section III.C, a STM topography image  [Fig.~\ref{stmseries}(c)]  shows that the BL is almost fully formed and covers more than  90\% of the surface at  \t= 2.4 min. 
\vskip 5mm
\noindent \textbf{\underline {Evidence of Stanene  from \sn}}:
For \t$\geq$ 2.4 min, \sn\ in Fig.~\ref{xps}(a) starts to show an extra component, Sn$_{\rm S}$  (blue curve), which is identified from our least square fitting.  %As discussed later in this section, this is assigned to the stanene component. These components  are obtained by  least squares  fitting, 
~The details of the fitting are shown in  Fig.~S8 of the SM~\cite{supple}, where %each spectrum is normalized to the same height. 
~the components, background, and residual of the fitting are shown.  Table~S\,I  enlists the values of the converged fitting parameters. %with their errors.

The Sn$_{\rm S}$ component of \sn\, is related to stanene  because it is  observed only for \t$\geq$\, 2.4 min, i.e., after BL formation is completed [Fig.~\ref{stmseries}(c)] and stanene related bands are  visible in ARPES (Fig.~\ref{arpes_All}, discussed further in sections III.E and III.F). It has zero intensity for \t$\leq$~1.2 min [Fig.~\ref{xps}(c)], where neither ARPES  nor STM [Fig.~\ref{stmseries}(a)] detect the stanene signature.    Sn$_{\rm S}$ appears at $E$= 484.95 eV % this is not in agreement with the reported position of the stanene on \bt\, 485.3 that rather agrees with our BL position, this shows that bt also possibly forms the BL, but let's not mention this here
~and becomes more intense than Sn$_{\rm B}$ at \t= 4.8 min, where the stanene bands are unambiguously identified from ARPES [Fig.~\ref{arpes_All}(h)].   
\vskip 5mm
\noindent \textbf{\underline {Evidence of BL from \bi\, and \sb}}:  The analysis of the  \bi\, and \sb\, core level spectra in Figs.~\ref{xps}(d) and ~\ref{xps}(f), respectively, provide additional insight  into the  BL, where  both Bi and Sb (referred to together as ``Bi/Sb")  show a component (Bi/Sb)$_{\rm B}$ (blue curve, BL component) in addition to the main peak, (Bi/Sb)$_{\rm M}$ (green curve, bulk component from deeper layers). See also Figs.~S9 and S10 and Tables~S\,II and S\,III of the SM~\cite{supple} for the details of the fitting.  Bi$_{\rm B}$ and Sb$_{\rm B}$ appear at 156.85 eV and 527.65 eV, which are shifted by 0.85 %"-1.2": this was a typo that creeped in after the R version before submission to Nat Mater!! 
~and 1.15~eV, respectively,  towards  lower $E$  compared to their  respective main peaks.  The significant shift in (Bi/Sb)$_{\rm B}$ suggests  chemical bonding between Sn and Bi/Sb from the beginning, i.e., \t= 0.3 min. $E$'s of  (Bi/Sb)$_{\rm B}$ are smaller compared even to their corresponding elemental bulk values (156.9 and 528.2 for Bi and Sb, respectively~\cite{Shalvoy1977}). This points to a bonding with  Sn that would transfer electrons to Bi/Sb due to the larger electronegativity of Bi/Sb (2.02/2.05~\cite{Mann2000}) compared to Sn (1.96). 
The (Bi/Sb)$_{\rm B}$  component has finite intensity from the outset  (\t= 0.3 min)   [see the five times magnified blue dashed curves  in Figs.~\ref{xps}(d) and \ref{xps}(f) and the intensity variation in Figs.~\ref{xps}(e) and \ref{xps}(g)]    indicating the formation of Sn-Bi/Sb bonds, although Bi/Sb is below the Te layer (Fig.~S3(a) of the SM~\cite{supple}). 
\vskip 5mm
\noindent \textbf{\underline {\te\ and Mn 2$p$ spectra }}:~~ In Figs.~\ref{xps}(h) and \ref{xps}(i), the effect of Sn-Te bonding on  the \te\,  core level spectra is manifested by a shift towards larger $E$ with respect to bare MBST. The spectra can be fitted with a single component denoted as Te$_{\rm BM}$ comprising of  (i)  Te$_{\rm B}$ i.e., the \te\ signal from the BL that reacts with Sn forming Sn-Te like bonds,  and (ii) Te$_{\rm M}$ i.e., the \te\ signal from  the deeper Te layers that do not react with Sn  (see Fig.~S11 and Table~S\,IV of the SM~\cite{supple} for details of fitting). Te$_{\rm M}$ and Te$_{\rm B}$ are not resolved as separate peaks within our resolution.  This is because the $E$ of  Te$_{\rm M}$ is 572.2 eV [determined from  the bottom spectrum i.e., that of  MBST in Fig.~\ref{xps}(h)], which  is close to that of  Te$_{\rm B}$ ($\sim$572.3 eV~\cite{Shalvoy1977}, which  is the $E$ of \te\ in SnTe).  

Note that a surface core level peak, Te$_{\rm surf}$,  is observed for   MBST (\t= 0) that is shifted by 0.5 eV %0.53 
~towards lower $E$ from the main peak  [red curves in Fig.~\ref{xps}(h)]. In  Fig.~S12 of the SM~\cite{supple}, it is confirmed by a difference spectrum between \t= 0 min and 0.3 min, where it is suppressed in the latter due to a change in the surface potential caused by Sn deposition.  %reveals that its position is same as that obtained by the curve  fitting. 
~ Such surface core level peak -- reported in other Te terminated surfaces~\cite{Leiro2006, Prince1988}-- are  generally related to a reduction in the coordination number of the surface Te atoms. % and a change in the surface potential with small Sn deposition (\t= 0.6 min) makes it disappear.    

The Mn 2$p$ spectrum of MBST~\cite{chenMBT2019} is hardly influenced by Sn deposition (Fig.~S13 of the SM~\cite{supple}). This observation indicates that the Mn layer, which is the fourth layer of the septuple, has minimal interaction with Sn and is unlikely to play a role in the formation of the BL. 

\vskip 5mm
\noindent \textbf{\underline {Composition of the BL from XPS}}:~The composition of the BL has been determined from the areas of  the  Sn$_{\rm B}$, Te$_{\rm B}$, Bi$_{\rm B}$, and Sb$_{\rm B}$ components %at \t~= 2.4 min (where the BL is almost fully formed and stanene formation has not started),
~after dividing by their respective photoemission cross sections~\cite{Trzhas2018} and the analyzer transmission function.  %After that we calulate the total area of the BL components (i.e., total area of (Sn$_{\rm B}$+Te$_{\rm B}$+Bi$_{\rm B}$ +Sb$_{\rm B}$)). then Sn$_{\rm B}$ , Te$_{\rm B}$, Bi$_{\rm B}$ , and Sb$_{\rm B}$ components area divided by the total area. we obtain composition of the BL Sn$_{55}$Te$_{21}$Bi$_{13}$Sb$_{11}$ i.e., ratio of Sn:Te:Bi/Sb is about 2:1:1 (total Bi and Sb contribution 13+11= 24).
~Since \te\ does not show a distinguishable BL component,  Te$_{\rm B}$ has been extracted from the total Te signal, Te$_{\rm BM}$,  by calculating the  contribution of the top or first Te layer that is involved in the BL  formation using the equation $I_i=I_1$e$^{-d_{i}/\lambda_{i}}$. %defining, $f(d_i)= e^{-d_{i}/\lambda_{i}}$ where, 
~Here, $I_i$= intensity of the $i^{\rm th}$ Te layer, $I_1$= intensity of the top layer, $d_{i}$= depth of the $i^{\rm th}$ Te layer from the top surface, and $\lambda_{i}$ is the   inelastic mean free path~\cite{Tanuma2011} calculated by weighted averaging for the atoms in the region through which the electrons emitted from the $i^{\rm th}$ Te layer passes to reach the surface.  The composition of the BL turns out  to be   54\% Sn, 23\% Te, 13\% Bi, and 10\% Sb after averaging for \t$\geq$ 2.4 min i.e., after it is fully formed. Therefore, the  ratio of Sn:Te:Bi/Sb atoms in the BL is $\approx$~2:1:1, which   shows that it comprises of Sn atoms that bond with a nearly equal number of  Te and  Bi/Sb atoms. This is facilitated by the presence of random anti-site defects in the BL, as shown by our LEED study (section III.D).
%~implying that two Sn atoms per unit cell are required ~to form the BL.

\vskip 5mm
\noindent \textbf{\underline {Sn 4$d$ and Bi 5$d$  core level spectra}}:~The Sn 4$d_{3/2}$ and 4$d_{5/2}$ peaks are observed at 25.5 and 24.40 eV as Sn deposition commences (Fig.~S14 of the SM~\cite{supple}). This is the BL related Sn$_{\rm B}$ component that is observed in \sn\ spectra discussed above. A black dashed line shows that the Sn 4$d_{3/2}$ intensity increases  with position nearly unchanged up to \t= 1.1-1.2 min. However, at \t= 2.4-2.6 min, there is a small shift of $\sim$0.09 eV to lower $E$  with a change in shape, indicating appearance of a second component.  This component becomes more intense compared to Sn$_{\rm B}$ for \t$>$ 4.5 min and appears at  $E$= 25.1 eV, which is  shifted with respect to the Sn$_{\rm B}$ component appearing at $E$= 25.5 eV.  This is the stanene related Sn$_{\rm S}$ , as shown by the red dashed lines. \srb{The significance of investigating the Sn 4$d$ shallow core level is that the Sn$_{\rm S}$ component is observed as a distinct peak in both the Sn 4$d$ spin orbit components (red  double sided horizontal arrows in  %the \t= 6.4 min spectrum in 
~Fig.~S14(a) of SM~\cite{supple}) in contrast to \sn.  This is because of  smaller life time broadening of the Sn 4$d$ core level as well as use of  synchrotron radiation that result in better energy resolution.} The Sn$_{\rm S}$  component is shifted  from the Sn$_{\rm B}$ component in Sn 4$d$ by 0.4 eV, which is nearly similar to that obtained for \sn\ from curve fitting. %Moreover,  suitable choice of photon energy (90 eV) results in enhanced  surface sensitivity (IMFP of ?? nm) and larger photoemission cross section for the Sn 4$d$ photoelectrons.  

The spectra in Fig.~S14 of the SM~\cite{supple} also show the Bi 5$d$ core level, which furthermore enables us to comment about the BL.  The  Bi 5$d_{5/2}$ and 5$d_{3/2}$ peaks for MBST (\t=0)  at $E$= 24.8 and 27.9 eV, respectively,  represent the main component (Bi$_{\rm M}$). As in case of \bi,  an additional component (Bi$_{\rm B}$) related to the BL appears for \t$>$0. This component at $E$= 26.75 eV is evidently  for  Bi 5$d_{3/2}$ (blue dashed line), since it does not coincide with  Sn 4$d$. It is also observed for Bi 5$d_{5/2}$ at similar separation, as shown by blue arrows. The separation between  Bi$_{\rm B}$  and Bi$_{\rm M}$ is  1.15~eV, which is similar to that observed for \bi\ [Fig.~\ref{xps}(d)]. Bi$_{\rm B}$ increases in intensity as the BL forms up to \t= 2.4-2.6 min and thereafter remains almost unchanged. The underlying Bi$_{\rm M}$ component is very weak for \t$\geq$2.6 min when the BL is fully formed, unlike in \bi, because enhanced surface sensitivity of the former with IMFP of 0.5 nm~\cite{Tanuma2011} compared to  relatively less surface sensitive \bi\  with IMFP of 2.9 nm.

%% srbtosrb: this sentence was plain and simple wrong, in both Sn 4d and 3d the Sn_S is more intense at t=4.8 min !!! Because of  surface sensitivity, here the Sn$_{\rm S}$ is more intense than the Sn$_{\rm B}$ from the underlying BL unlike in  XPS; compare Fig.~\ref{xps}(a) and  Fig.~S17 of the SM~\cite{supple}  for \t$\geq$ 4.8 min.  
%srbtosb: why no error bars in 2(b)?? Reply: yes it is there, small
%srbtosb: Sn_S bilayer vs monolayer position??  well there is a small shift of 0.05 in XPS between 3.6 and 4.8, but it is within error perhaps.  It could be said more accurately from recent core level data or He II, if fitting is done. If shift is small or no shift, it can be related to less bl interaction with monolayer stanene, related to referee's comment !! But, it is not a tight argument, so for now say nothing.
  %srbQ: leave this out! it could be slightly different depending on the substrate and analyzer EF etc: The $E$ of Sn$_{\rm S}$ here is $\sim$25.1 eV, which is in good agreement with that reported for stanene on Ag and Pd2Sn (~24.9)   

\subsection{\label{subsec:Stn_STM} STM study of  stanene and the buffer layer}

%\subsubsection{BL onset}
 \noindent \textbf{\underline {BL onset}}:
 At the initial stage of Sn deposition on MBST (\t= 1.2 min), where both ARPES and XPS indicate the formation of BL without any signature of stanene, Fig.~\ref{stmseries}(a) shows that about 60\% of the substrate is covered  by the BL  that are manifested as islands of height $h_b$=  0.5$\pm$0.05 nm %with respect to the MBST surface 
 ~measured from the uncovered Te top layer of MBST (dark brown region) (Fig.~\ref{stmseries}(b),  see also the  height histogram in Fig.~S15(a) of the SM~\cite{supple}). Thus, $h_b$ is  due to  Sn deposition and is larger than the typical thickness of  a  Sn monolayer (0.18-0.38 nm) grown on different substrates~\cite{Deng2018,Yuhara2018,Zhu2015,Zheng2019,Gou2017,Zhao_Sn_Bi_2022,Singh2020prr}.%,srb: any other where there is no alloying but forms triangular Sn??? where there is no surface alloying. 
 ~Rather, $h_b$ is close to twice  the covalent diameter of Sn (0.278 nm~\cite{Cordero2008}), indicating that two Sn atoms per unit cell would be required for BL formation, as has been shown earlier by its composition determined from XPS. Despite the  admixture of Sn with Te and Bi/Sb in the BL islands caused by chemical bonding and anti-site defect formation, the validity of the aforementioned argument holds due to the similarity in their sizes~\cite{Cordero2008}.
 %  *The validity of the above argument holds even though  the BL islands would have admixture of  Sn with  Te and Bi/Sb due to chemical bonding and anti-site defect formation,  because of the similarity of their sizes~\cite{Cordero2008}. %, as mentioned earlier. %However, this difference would not be large since the covalent diameters~\cite{??} of Sn (0.28 nm), Te (0.272 nm) and Bi (0.302 nm)/Sb (0.28 nm) are rather similar. % Thus, $h_b$ of 0.5 nm indicates that   two monolayers  of  Sn  form the BL, since the covalent diameter of an Sn atom is  0.28 nm. 
 ~The total thickness of the BL, including the thickness of the Te and Bi/Sb  layers,   is estimated  to be $\sim$0.9 nm, 
 where the thickness of the top two layers of MBST is taken to be  $\sim$0.4 nm from x-ray crystallography~\cite{Yan2019}.    % (0.27 and 0.30/0.28 nm, respectively from www.webelements.com) to be slightly larger than 1 nm.}
%[the covalent diameters of Sn, Te, Sb and Bi are   0.28, 0.272, 0.28, and 0.302 nm, respectively; from www.webelements.com] and outer shell configurations ($s^2$$p^{2-4}$). 
% from STM is also consistent with its composition from XPS.

%\subsubsection{BL completion}
 \vskip 5mm\noindent \textbf{\underline {BL completion}}:
In Fig.~\ref{stmseries}(c) for \t= 2.4 min deposition, the almost fully formed BL covers more than  90\% of the substrate. Its height is similar to that of previous deposition, but there are some less darker regions, where the  height  is however smaller than 0.5 nm. An example of the reduced  height of the BL  is shown by  $h_{br}$ in the height profile [Fig.~\ref{stmseries}(d)].  This is also indicated by the asymmetry in the height histogram  around 0.3 nm in Fig.~S15(b) of SM~\cite{supple}.  Thus,  the BL has a somewhat non-uniform thickness. This is reflected by its enhanced  mean square roughness ($S_q$) of 0.12 (0.14)$\pm$0.025 nm at \t= 1.2 (2.4) min estimated by masking the uncovered substrate.

\begin{figure*}[t!] 
	\includegraphics[width=\linewidth,keepaspectratio,trim={0cm -.5cm 0cm 0cm},clip]{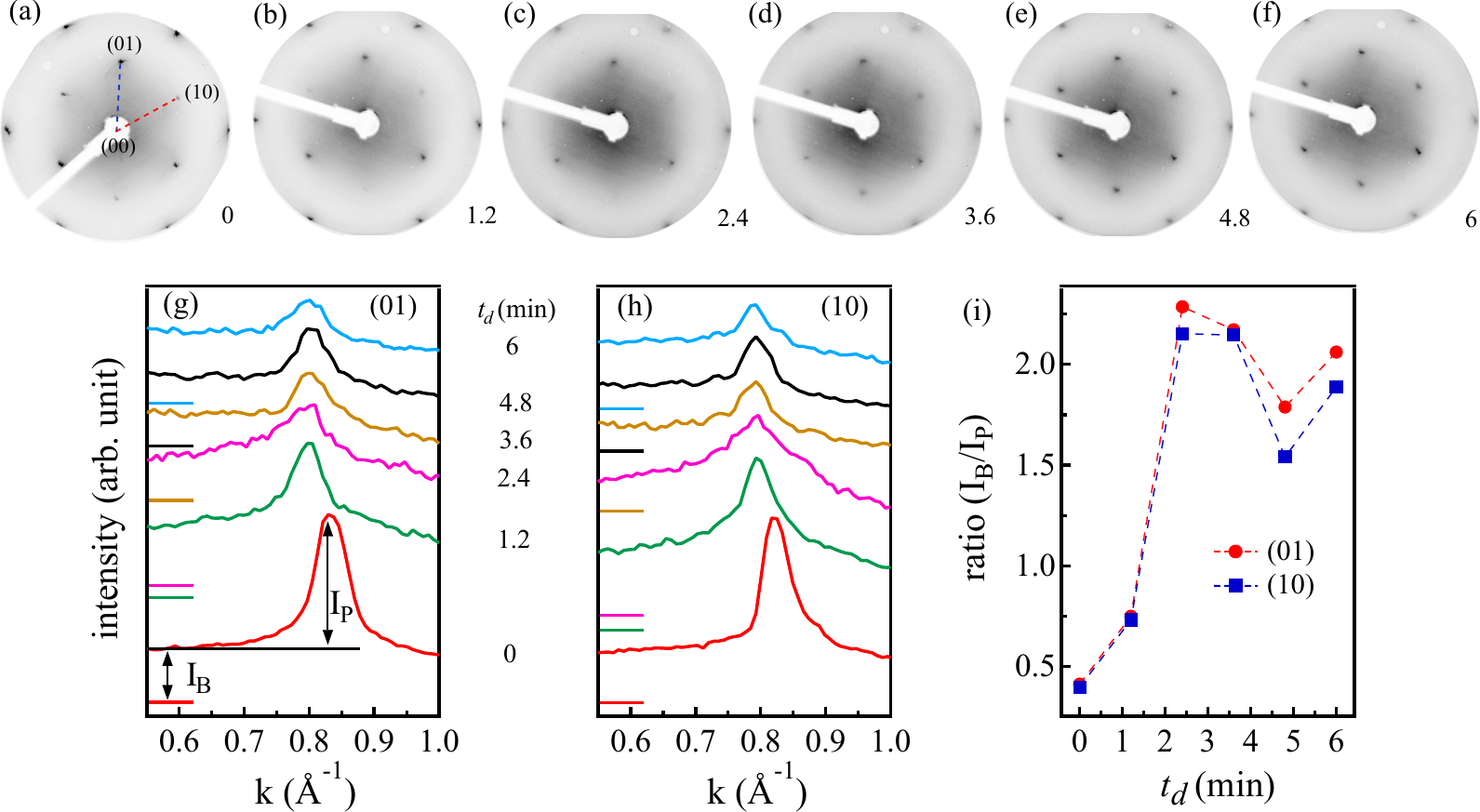} %{STM_growth_evolution_v4}
	\caption{(a-f) A set of LEED patterns in inverted gray scale with primary electron beam energy ($E_p$) of  55 eV,  \t\, in min is indicated at the lower right corner of each panel. The intensity profiles from the LEED patterns in panels \textbf{a-f}  along (g)  (00) to (01) spot [blue  dashed line in panel \textbf{a}] and  (h) (00) to (10) [red dashed line in panel \textbf{a}]. All the profiles are  staggered in the vertical direction and are aligned to (00) at the center of the pattern that is covered by the shadow of the electron source. The zero of each profile is shown by a short horizontal line of same color on the left vertical axis. (i) Ratio of  the background intensity ($I_B$) and intensities of the (01) and (10) spots ($I_P$), measured as shown in panel \textbf{g}.}
	\label{leed}
\end{figure*}

%\subsubsection{Stanene formation}
  \vskip 5mm\noindent \textbf{\underline {Monolayer stanene}}:
 \srb{In Fig.~\ref{stmseries}(c), blue arrows indicate the presence of a few bright islands on the BL that have a step height of $h_{1s}$ and cover $\sim$10\% of the area. Note that at this \t~ of 2.4 min, signature of stanene formation is also visible from XPS  [Fig.~\ref{xps}(a)].}
  At \t= 3.6 min, these  islands increase in size with similar $h_{1s}$   that  covers about 65\% of the BL [blue arrows in  Fig.~\ref{stmseries}(e)]. 
 \srb{Based on the height profile shown in Fig.~\ref{stmseries}(f) and the histogram in Fig.~S15(c) of the SM~\cite{supple}, it can be concluded that the value of $h_{1s}$ is 0.35$\pm$0.02 nm, suggesting that the observed islands are composed of a monolayer of stanene. This is corroborated by the height profile analysis of the atomic resolution image depicted in Fig.~S16 of the SM~\cite{supple}. Subsequently, this will be elaborated upon, particularly in relation to bilayer stanene in Figs.~\ref{stmseries}(i) and \ref{stmseries}(j).
 %From the height  profile in Fig.~\ref{stmseries}(f) and histogram in Fig.~S15(c) of SM~\cite{supple}, $h_{1s}$ turns out to be 0.35$\pm$0.02 nm indicating that these are monolayer stanene islands. This is confirmed by the height profile analysis of the atomic resolution image in Fig. S?? of SM\cite{supple}. This is discussed in details  later in conjunction with bilayer stanene. 
~In Fig.~\ref{stmseries}(e),  the regions that are  not covered by monolayer stanene represent the  BL. These BL regions are smoother  with $S_q$ = 0.08$\pm$0.01 nm compared to Fig.~\ref{stmseries}(c). In fact, the smoothness of the BL here is comparable to the substrate ($S_q$ = 0.07$\pm$0.01 nm).  An atomic resolution STM image of the BL region  in Fig.~S17 of the SM~\cite{supple}  shows an ordered structure, as also indicated by  LEED (Fig.~\ref{leed}). }% which  enables it to support the growth of stanene.  } %srbv1: At this point, the BL is able to support the growth of stanene. % on the BL. %This could result in enhanced roughness, which is evident from the mean square roughness ($S_q$) of 0.125$\pm$0.025 nm that is significantly larger than MBST (0.07$\pm$0.01 nm).   

\begin{comment}
	srb: leave this out, as stanene cannot be identified unless these are topological:
\srb{Another possible evidence of  the formation of stanene is existence of  the  edge states i.e. states localized at the edge of a stanene island. In Fig. ??, scanning tunneling spectra (STS) taken on the bilayer stanene island (marked by ``A" and ``B", as shown in Fig. ??) differs from the spectrum taken at the edge (marked by ``C"). The latter shows an increased LDOS around 0.2 V, which is signature of edge state. However, this enhancement in the LDOS at the edge can also be due to non-topological edge states, such as dangling bonds~\cite{zheng_2dmat20}. Additional experimental and theoretical investigations are needed, to comment about the topological nature of these edge states.} 
\end{comment}

%srbT: look for paper where both st height and buckling are given:  InSb (0.35 nm)~\cite{Zheng2019}, Sb (0.37 nm)~\cite{Gou2017} and Bi (0.38 nm)~\cite{Zhao_Sn_Bi_2022}, but is larger compared to ultra-flat stanene on Cu (0.18 nm)
%~\cite{Deng2018} or Ag (0.25 nm)~\cite{Yuhara2018}.   This comparison indicates buckled stanene formation on MBST.  The relaxed structure of stanene obtained from our DFT calculation also shows a buckling of  0.1 nm.

%\subsubsection{Bilayer stanene}
 \vskip 5mm\noindent \textbf{\underline {Bilayer stanene}}:
Figure~\ref{stmseries}(e) shows evidence of  small islands  on monolayer stanene, \srbn{one of these is highlighted by a black arrow.}  These islands cover about 3\% area, but  grow in size and number with further Sn deposition [\t= 6 min, Fig.~\ref{stmseries}(g), islands shown by black arrows] covering more than  $\sim$60\% of the total area. %*srbtopb: show 4-5 black arrows*   
~The height of these Sn islands ($h_{2s}$) shown by height profiles along the white dashed lines in Figs.~\ref{stmseries}(f) and \ref{stmseries}(h)  is 0.35$\pm$0.03 nm. The difference between the two peaks of the height histogram in Fig.~S15(d) of SM~\cite{supple} also agrees with this estimate. Note that $h_{2s}$ is similar to $h_{1s}$ indicating formation of bilayer stanene.  Its growth  commences  on the monolayer stanene islands prior to the  latter's complete coverage of  the BL, which is a characteristic of  Stranski-Krastanov (SK) growth.

\srb{The step height of 0.35 nm determined above is close to the step height (0.35-0.36 nm) of stanene on other substrates such as  \bt~\cite{Zhu2015} %0.35, 0.12
~and PbTe~\cite{Zang2018},%0.36, 0.09 from DFT
%InSb~\cite{Zheng2019},%0.35, buckled but no value
%Sb~\cite{Gou2017}%0.37, buckled but no value
%and Bi~\cite{Zhao_Sn_Bi_2022},%0.38, buckled but no value
~where buckling  of  0.09-0.12 nm was  reported.  Zhu \textit{et al.}~\cite{Zhu2015} determined buckling height from a step height between the upper and  lower stanene sublattice.  Following a similar approach, we find the  buckling height  of stanene on MBST to be 0.1$\pm$0.015 nm  (Fig.~S18 of SM~\cite{supple}). The different structural parameters, such as lattice constant, step height, and buckling height, determined in this work are compared to those on other substrates in the literature  in Table~I.  According to Table~I, the structural characteristics of stanene on MBST  are similar to those of other chalcogenide substrates with Te termination, such as \bt~\cite{Zhu2015} and PbTe~\cite{Zang2018}.} %, where BL formation has not been reported %srbtorb: OK. agreed. This suggests that the structure of stanene on MBST is not significantly influenced by the BL.}
%The closeness of the  structural parameters of stanene on MBST and those of  other chalcogenide substrates with Te termination such as \bt\ and PbTe, where BL formation is not formed~\cite{Zhu2015,Zang2018}, indicates that stanene's structure on MBST is largely unaffected by BL. }
\vskip 5mm
\noindent \textbf{\underline {Atomic resolution image of stanene}}:
An atomic resolution image of  a bilayer stanene island is shown in Fig.~\ref{stmseries}(i), where the unit cell (black line) joins four Sn atoms visible as bright spots (magenta circles)  of the upper sublattice (represented by T$_{\rm n}$).  The lower sublattice atoms are shown by gray circles and represented by B$_{\rm n}$. In Fig.~\ref{stmseries}(j), the average height profiles along d1, d2 and d3 directions 120$^{\circ}$ apart are shown staggered along the vertical axis. These directions are shown  in  Fig.~\ref{stmseries}(i). In addition  Fig.~S19 of SM~\cite{supple} shows all the parallel dashed lines in each direction over which averaging was performed. %to ensure that thermal drift does not affect our conclusions 
~These height profiles show distinct peaks related to T$_{\rm n}$ and shoulders related to B$_{\rm n}$, n= 1-3. In order to ascertain their  positions,  Gaussian functions with nearly  similar full width at half maxima have been used to fit the profile. The quality of the fitting is good, as shown by the  fitted curve in Fig.~\ref{stmseries}(j). 

  A confirmation that Fig.~\ref{stmseries}(i)  represents  stanene is based on the observation that  B$_{\rm n}$ related peaks (gray curves, peak position  highlighted  by blue arrows) are not at  $d_s$/2 [green dashed vertical line at the mid points of  T$_{\rm n}$ and T$_{\rm n+1}$ in Fig.~\ref{stmseries}(j)], but is  closer to T$_n$'s. This asymmetry is the signature of  stanene formation  that stems from the contribution of intensity to B$_{\rm n}$ by the Sn atoms in the lower sublattice. An atomic resolution image for monolayer stanene (\t= 3.6 min) analyzed in a similar way also shows this behavior (Fig.~S16 of SM~\cite{supple}). 
 
 The  average separation between   T$_{\rm n}$ and B$_{\rm n}$ related peaks  is 0.34$\pm$0.04~nm. This is comparable to the  ideal value of $d_s$/3 (= 0.27 nm)  obtained when a buckled hexagon is projected on the horizontal plane.  The average separation between T$_{\rm n}$ and T$_{\rm n+1}$ is $d_s$ = $0.8\pm$0.02~nm, which  is the longer diagonal of the primitive unit cell.  $d_s$  is related to   the lattice constant $a_s$ by   $d_s$/$\sqrt{3}$, which turns out to be 0.46$\pm$0.01~nm.   \srb{A similar value of $a_s$ is also obtained from the Fourier transform shown in Fig.~\ref{stmseries}(k).  }
   It is worth noting that $a_s$  is close to that of  freestanding stanene (0.467 nm)~\cite{Matthes2013,XuBinghai2013}. 
      The unequal heights of the upper (T$_{\rm n}$) and the lower sublattice (B$_{\rm n}$) from our height profile analysis is also an indication of  buckling in stanene.  %But difference of  T$_n$ and B$_n$ ($\sim$0.04 nm) from STM (Figs. ??)  largely underestimates  the buckling because of  undertermined contribution from the electron density arising from upper sublattice atoms as in case of the substrate discussed below. 
        
        \begin{table}[!ht]
        	\caption{The lattice constant, step height and buckling height   (in nm) of stanene on MBST  compared with stanene on other substrates ($^{\dagger}$lattice matched with substrate, $^{\ddag}$DFT value,$^{\dagger\dagger}$(2$\times$2) structure).}
        	\vspace*{-0.1cm}
        	\begin{center}
        		\renewcommand{\arraystretch}{1.9} % adds more space between rows
        		\begin{tabular}{|p{0.30\linewidth}|p{0.2\linewidth}|p{0.22\linewidth}|p{0.16\linewidth}|p{0.16\linewidth}|p{0.1\linewidth}|}
        			
        			\hline
        			\multirow{1.0}{\linewidth}{\centering Name of substrate [Reference]}  & %\multirow{1.5}{\linewidth}{\centering Lattice constant of substrate }  &
        			\multirow{1.0}{\linewidth}{\centering Lattice constant } &
        			\multirow{1.0}{\linewidth}{\centering Step height } &
        			\multirow{1.0}{\linewidth}{\centering Buckling height }  \\
        			%& & &   \\
        			\hline
        			%%%%%%%%% ($h_{1s}$, $h_{2s}$) ,  ($a_s$)
        			
        			\multirow{1}{\linewidth}{\centering free standing \cite{XuBinghai2013,Matthes2013}}  & %\multirow{1}{\linewidth}{\centering --}  &
        			\multirow{1}{\linewidth}{\centering 0.467$^{\ddag}$ } & \multirow{1}{\linewidth}{\centering --} & \multirow{1}{\linewidth}{\centering 0.085$^{\ddag}$}  \\
        			
        			%\hline
        			\multirow{1.0}{\linewidth}{\centering MBST~[our work]}  &
        			%\multirow{1}{\linewidth}{\centering 0.430 }  &
        			\multirow{1}{\linewidth}{\centering 0.46$\pm$0.01 } &
        			\multirow{1}{\linewidth}{\centering 0.35$\pm$0.03 } &
        			\multirow{1}{\linewidth}{\centering 0.1$\pm$0.015 }  \\
        			
        			%\hline
        			\multirow{1}{\linewidth}{\centering Bi$_2$Te$_3$~\cite{Zhu2015}}  & %\multirow{1}{\linewidth}{\centering 0.438}  &
        			\multirow{1}{\linewidth}{\centering 0.438$^{\dagger}$} & \multirow{1}{\linewidth}{\centering 0.35$\pm$0.02 } & \multirow{1}{\linewidth}{\centering 0.12$\pm$0.02 }  \\
        			
        			%\hline
        			
        			\multirow{1}{\linewidth}{\centering PbTe~\cite{Zang2018,Liao2018}}  & %\multirow{1}{\linewidth}{\centering 0.452}  &
        			\multirow{1}{\linewidth}{\centering 0.452$^{\dagger}$ } & \multirow{1}{\linewidth}{\centering 0.36} & \multirow{1}{\linewidth}{\centering 0.093$^{\ddag}$ }  \\
        			
        			%\hline
        			\multirow{1}{\linewidth}{\centering Bi \cite{Song2021,Zhao_Sn_Bi_2022}}  & %\multirow{1}{\linewidth}{\centering 0.454  }  &
        			\multirow{1}{\linewidth}{\centering 0.454$^{\dagger}$ } & \multirow{1}{\linewidth}{\centering 0.4 } & \multirow{1}{\linewidth}{\centering 0.13}  \\

        			%\hline
        			\multirow{1}{\linewidth}{\centering InSb \cite{Zheng2019}}  &
        			%\multirow{1}{\linewidth}{\centering 0.458 }  &
        			\multirow{1}{\linewidth}{\centering 0.462 } & \multirow{1}{\linewidth}{\centering 0.35 } & \multirow{1}{\linewidth}{\centering --}  \\
        			
        			%\hline
        			\multirow{1}{\linewidth}{\centering Sb \cite{Gou2017}}  &
        			%\multirow{1}{\linewidth}{\centering 0.43}  &
        			\multirow{1}{\linewidth}{\centering 0.43$^{\dagger}$ } & \multirow{1}{\linewidth}{\centering 0.51$\pm$0.01} & \multirow{1}{\linewidth}{\centering --}  \\

        			%\hline
        			\multirow{1}{\linewidth}{\centering Pd \cite{Yuhara2021}}  &
        			%\multirow{1}{\linewidth}{\centering 0.275??}  &
        			\multirow{1}{\linewidth}{\centering 0.48 } & \multirow{1}{\linewidth}{\centering -- %0.23 from DFT possibly
        				~} & \multirow{1}{\linewidth}{\centering 0.02}  \\
        			
        			%\hline
        			\multirow{1}{\linewidth}{\centering  Cu \cite{Deng2018}}  &
        			%\multirow{1}{\linewidth}{\centering 0.255 }  &
        			\multirow{1}{\linewidth}{\centering 0.51$^{\dagger\dagger}$ } & \multirow{1}{\linewidth}{\centering 0.18 } & \multirow{1}{\linewidth}{\centering 0}  \\

        			\hline
        		\end{tabular}
        		\label{All_lit2}
        		\vspace*{-0.5cm}
        	\end{center}
        \end{table}

 \noindent \textbf{\underline {MBST substrate}}:
 % The asymmetry in position of B$_{\rm n}$'s   distinguishes the honeycomb structure of stanene from the triangular lattice of   the MBST surface. 
Figures~S20(a,b) of the SM~\cite{supple} show a wide area topography image of MBST, where the step height of 1.4$\pm$0.03~nm  corresponds to the height of a  septuple layer. % [Fig.~S1(a) of SM~\cite{supple}].  
~A similar analysis, as discussed above for stanene, has been performed for an atomic resolution STM image of MBST  in Figs.~S20(c) and S20(d) of the SM~\cite{supple}. Curve fitting of the height profiles along d1-d3 shows that, unlike stanene,  the shoulder labeled by R is  at $d_s$/2 i.e., at the middle of  T$_{\rm n}$T$_{\rm n+1}$. R is attributed to the residual contribution of  the Te atoms connected by the smaller diagonal of the unit cell  shown in Fig.~S20(c) of SM~\cite{supple} by dashed blue lines.  \srb{The lattice constant of MBST is  determined to be 0.44$\pm$0.01 nm, which agrees well %a recent STM study~\cite{Lupke2023} and  
	with x-ray diffraction result (0.430 nm for  $\sim$30\% Sb-doped MBST)~\cite{Yan2019}. The uniformity of the Sb doping in MBST is demonstrated by the resemblance of scanning tunneling spectra, which depict the local density of states, at various spatial positions, as shown in Figs.~S20(e) and S20(f)  of the SM~\cite{supple}.  } %(0.43043(4) for $x$=  0.315 nm)
~Interesting to note is that the  lattice constant of stanene (0.46 nm)  turns out to be larger than MBST. This  lattice mismatch is bridged by the BL, as shown by our LEED study presented below. 

 \subsection{Low energy electron diffraction for different \t} 
 A set of LEED patterns in Figs.~\ref{leed}(a) to \ref{leed}(f)  demonstrate epitaxial growth  for all depositions, in spite of the BL formation.
 Sharp  (1$\times$1) spots are observed for  the  beam energy ($E_p$) range 41 eV $<$$E_p$$<$ 130 eV  without any splitting or streaking. %, as shown by the video files named  monostanene.mp4, BL.mp4 and MBST.mp4 at a step of 1 eV for  monolayer stanene (\t= 4.8 min), BL (\t= 2.4 min) and substrate MBST (\t= 0 min), respectively  in the SM~\cite{supple}. %(1$\times$1) spots are observed throughout with no  evidence of splitting or broadening. 
 ~The symmetry of the BL remains similar to the substrate, as shown by Figs.~\ref{leed}(a) to \ref{leed}(c). The intensity profiles along   (00) to (01) spot [Fig.~\ref{leed}(g)]  and  (00) to (10) spot  [Fig.~\ref{leed}(h)] i.e., along the blue and red dashed lines, respectively, do not show any splitting. % [the dashed lines are shown in  Fig.~\ref{leed}(a)]. 
 \srb{There is however a shift in the positions of both the (01) and (10) spots: the peaks in the corresponding intensity profiles  for stanene e.g., for \t$\geq$ 3.6 min are shifted towards lower $k$ by similar amount compared to that of  MBST (\t= 0 min). This  shows a decrease in their separation with the   (00) spot  ($k$= 0). This indicates  that the lattice constant of  stanene is larger compared to MBST. This  increase is estimated to be 3.1\%$\pm$0.5\% considering  more than 80 intensity profiles of  different $E_p$'s for \t= 3.6-6 min (the lattice constants of monolayer and bilayer stanene are similar). %36 profiles were considered for the substrate
 	 ~The lattice constants of stanene and MBST determined by STM  are consistent with the LEED results. }%This indicates  that the lattice constant of  stanene is larger compared to MBST; the  %(for about 90 profiles from ?? patterns at different beam energies) 
 %	~increase   is estimated to be 3.2\%$\pm$0.8\% by averaging over different intensity profiles. The lattice constants from STM  discussed earlier aligns with this result. 
 %SRBtoSB: can be included after reevaluation:  ~From the similar position of the (01) and (10) peaks, it can also be concluded that the lattice constant of monolayer stanene (\t=3.6 min and 4.8 min) is similar to bilayer stanene (\t= 6 min).} %The change in the lattice constant STM discussed in the previous section. 
  ~Figures~\ref{leed}(g,h) further show that when the BL formation is completed (\t= 2.4 min), both the (01) and (10) peaks are nearly aligned to that of stanene i.e., are shifted towards lower $k$.  This indicates  that the lattice constant of the BL is larger than MBST by 3\%$\pm$0.5\%. Thus, the lattice constant of the BL  is similar  to that of stanene within the error. This shows that the lattice mismatch between stanene and MBST is bridged by the  BL.  \srb{A similar role of the  BL  has been reported in the case of  Bi$_2$Se$_3$ grown on   2H-NbSe$_2$ substrate~\cite{Wang_2014,Yi2022}.  }
 
 An additional important observation from Figs.~\ref{leed}(a) to \ref{leed}(f) is that the  background  to the spot intensity ratio increases quite substantially  for the BL  compared to MBST, as quantified in Fig.~\ref{leed}(i) from the ratio of  the background ($I_B$) and the spot ($I_P$) intensities. This ratio is maximum at \t= 2.4 min where the BL is fully formed. Such an increase in the background intensity in the LEED pattern has been related to the existence of random anti-site defects in previous literature~\cite{Janzen2001, Zielasek2000}.    (Bi/Sb)$_{\rm Te}$ anti-site defects, where Bi/Sb atoms of the second layer are exchanged with the Te atoms of the top layer, are already present in MBST.  In agreement with previous literature~\cite{Garnica2022, Lupke2023}, signature of (Bi/Sb)$_{\rm Te}$ anti-site defect has been observed as bright circular spot %and dark triangles 
 in the STM image  (red dashed circles in Fig.~S20(c) of the SM~\cite{supple}).   With Sn deposition, the anti-site defects  are  enhanced, possibly because of the propensity for chemical bonding aided by the kinetic energy of the impinging Sn atoms (equal to 1073~K) and the closeness of their atomic diameters (0.278 nm, 0.276 nm, 0.296 nm, and 0.278 nm for Sn, Te , Bi, and Sb, respectively~\cite{Cordero2008}).

	\subsection{Band dispersion and Fermi surface from ARPES}
	\begin{figure*}[t!] 
	\includegraphics[width=\linewidth,keepaspectratio,trim={0.5cm 0.7cm 9.5cm 0cm},clip] {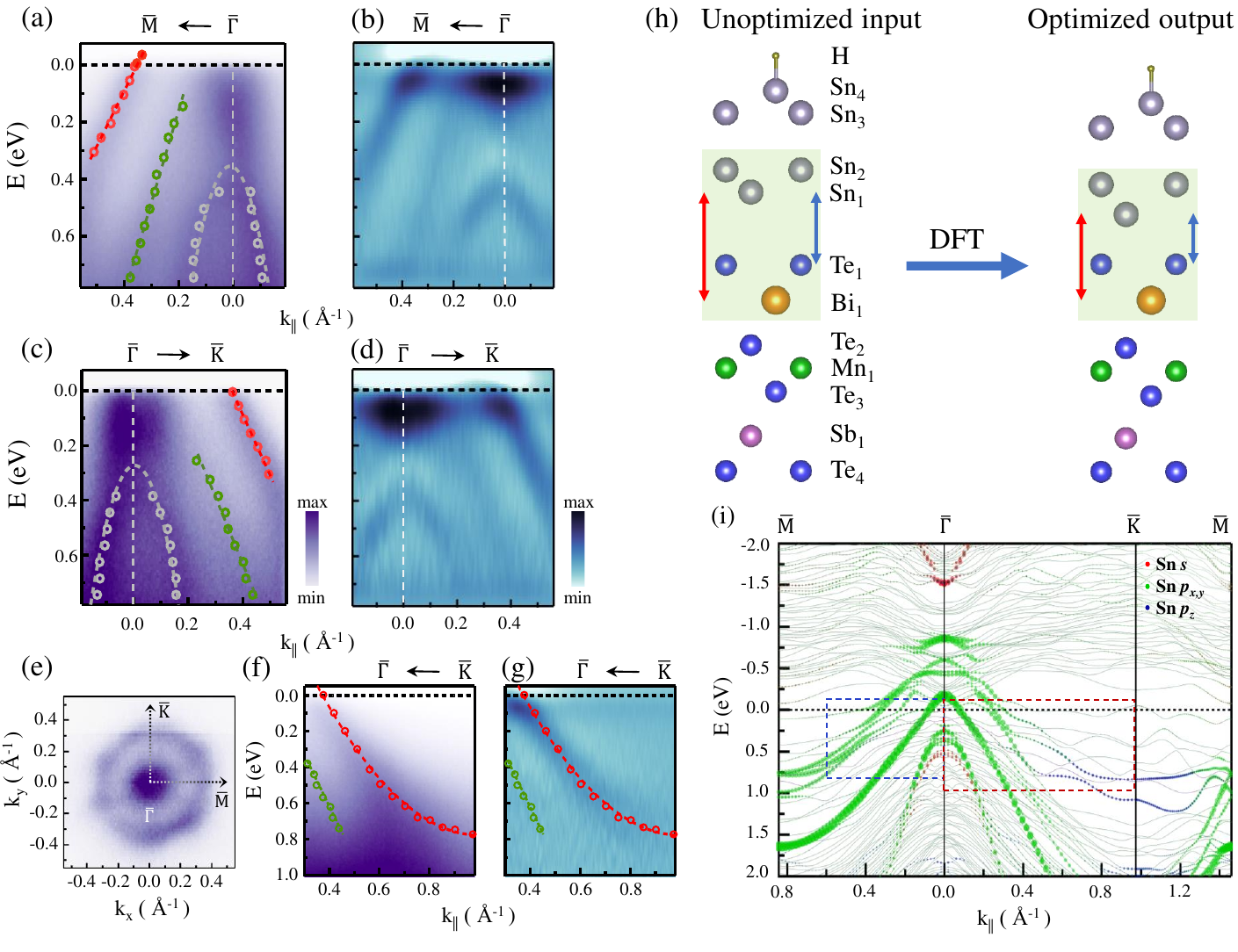}
	\caption{%\textbf {ARPES and STM of stanene.}  
		 ARPES intensity plots of \srb{predominantly monolayer stanene} from $\overline{\Gamma}$ towards (a) $\overline{M}$ and (c) $\overline{K}$ directions. The BZ is shown in the inset of Fig.~\ref{arpes_All}(a).   Open circles (red: outer band, green: inner band and gray: inverted parabolic band) represent the   positions of  the bands obtained by fitting the MDCs. The dashed curves  are obtained by fitting the open circles.   (b, d) Second derivative image of  panels \textbf{a} and \textbf{c}, respectively.  \srb{ (e) The Fermi surface of  stanene obtained from ARPES after averaging over $\pm$15 meV. (f) Raw and (g) second derivative ARPES intensity plot toward the $\overline{K}$-$\overline{\Gamma}$  direction. \srbn{In panel \textbf{f}, the red circles and the dashed curve are obtained from fitting %as in panel \textbf{c} 
		 	and these are overlaid on panel \textbf{g}.}   (h) The unoptimized input (left panel)  and the optimized output (right panel) structures of  \textbf{``S+BL+H"}  DFT calculation for monolayer stanene on MBST including  BL (green shading) and H passivation (topmost layer).  Both the structures have been drawn in the same scale using the  VESTA program~\cite{Momma2011}.The labeled atoms are represented by size-scaled circles of various colors. The blue (red) double arrow indicates the Sn$_1$-Te$_1$ (Sn$_1$-Bi$_1$) layer separation. (i)  The  $E-k_{||}$ band dispersion calculated including SOC for  \textbf{``S+BL+H"}  along $\overline M$-$\overline{\Gamma}$-$\overline K$-$\overline M$ direction.  The bands of Sn $p_{x,y}$, $p_{z}$, and $s$ character are shown.  Blue (red) dashed rectangle shows the  range measured in ARPES  towards $\overline{M}$ in panel  \textbf{a} ($\overline{K}$ in panels \textbf{c, f}).}} 
	\label{arpes_sta}	
\end{figure*}  

%	\subsubsection{Bilayer stanene}
\srb{The ARPES intensity plots of predominantly monolayer stanene that almost completely covers the BL at \t= 4.8 min are depicted in Figs.~\ref{arpes_sta}(a) to \ref{arpes_sta}(d). Some monolayer regions are covered by an additional stanene layer, i.e.,  bilayer stanene as a result of SK growth, as demonstrated by STM.} These measurements have been performed at 80~K after deposition at room temperature with statistics better than that of Fig.~\ref{arpes_All}(h).  %Based on STM images in Figs.~\ref{stmseries}(e,g), at \t=4.8 min, 
 %Thus, the ARPES data that, unlike STM, is recorded over macroscopic region ($\sim$1 mm) of the crystal, would have signature of both bilayer and monolayer stanene.} 

\vskip 5mm\noindent \textbf{\underline {Bands around the $\overline{\Gamma}$ point and the Fermi surface}}:~%towards $\overline{\Gamma}$-$\overline{M}$ and $\overline{\Gamma}$-$\overline{K}$}}:~~ 
Figure~\ref{arpes_sta}(a)  shows two bands, outer and inner,  towards  the $\overline{\Gamma}$-$\overline{M}$ direction. Their dispersion is hole-like, as shown by  red open circles for the outer band. These circles correspond to the maximum intensity  of  the momentum distribution curves (MDCs) obtained from Fig.~\ref{arpes_sta}(a) at different $E$  (Fig.~S21 of the SM~\cite{supple} shows how the maxima of the MDCs are  identified by fitting).  The red dashed line, which is a linear fit to the red circles, shows that the dispersion is a close approximation to linearity.  The inner hole band  (green dashed line),  determined in a similar way, %   shows a nearly linear 
	~disperses   down to  0.8 eV at  0.4 \a.  An  inverted parabolic band is also observed that disperses from a maximum $E$ of  $\sim$0.4 eV at  $\overline{\Gamma}$ to  0.8 eV at  $\sim$0.2~\a.   A second derivative ARPES intensity plot in Fig.~\ref{arpes_sta}(b) shows these bands prominently.

	To  probe whether the outer  band  crosses \ef, we have deposited potassium (K), which provides electrons to the surface that would fill up this band~\cite{Neupane2012,Zhu2015}. Thus, the band positions  above \ef\, can be determined. As  shown by the red filled circles in Fig.~\ref{arpes_sta}(a)  (the corresponding ARPES intensity plots shown in  Fig.~S22 of the SM~\cite{supple}), we find that the outer band crosses \ef. %, which is in agreement with DFT [Fig.~\ref{arpes_sta}(e)].  

	 The outer and inner stanene bands towards the $\overline{\Gamma}$-$\overline{K}$ direction  are observed in  Figs.~\ref{arpes_sta}(c) and \ref{arpes_sta}(d), which are highlighted by open circles and the dashed curves. ARPES data in an extended range   ($\overline{M}$-$\overline{\Gamma}$-$\overline{K}$) taken at room temperature are shown in Fig.~S23 of SM~\cite{supple}, with the bands obtained above overlaid.  %The outer band crosses \ef\, at a slightly larger Fermi vector  %($k_F$) of 0.4~\a\, ~compared to  $\overline{\Gamma}$-$\overline{M}$. % where $k_F$= 0.35~\a. srbtosrb: I see that these are similar, checl with SB
	~The Fermi surface shown in  Fig.~\ref{arpes_sta}(e) exhibits  hexagonal symmetry, where  $k_F$ is 0.4~\a\, towards $\overline{K}$ compared to  0.35~\a  towards $\overline{M}$. Such hexagonal Fermi surface of stanene has been  observed   on \bt~\cite{Li2020} and Pd$_2$Sn~\cite{Yuhara2021}. A series of $k_x$-$k_y$ isosurface plots for $E$ varying from 0.1 to 0.4 eV  in  Fig.~S24 of the SM~\cite{supple} also exhibits the hexagonal symmetry. Interestingly,  segmentation of the isosurface plots indicate formation of nodes at $\overline{K}$ and $\overline{M}$  that could possibly be related to  band crossings.  % non-trivial??
	
	Note that the Fermi velocities  of stanene (estimated from the slope of the outer band)   towards $\overline{\Gamma}$-$\overline{M}$ % 3.2$\times$10$^{5}$~m/s., page ?? of SB book
	and $\overline{\Gamma}$-$\overline{K}$ %  3.2$\times$10$^{5}$~m/s., page ?? of SB book
	~ are similar, i.e., 3.2$\pm$0.02$\times$10$^{5}$~m/s. % and 3$\pm$ 0.01$\times$10$^{5}$~m/s,
	~ This value is somewhat larger compared to stanene on \bt\, (2.79$\times$10$^{5}$~m/s)~\cite{Li2020} or Sb$_2$Te$_3$ (2.83$\times$10$^{5}$~m/s)~\cite{Li_stanene_Bi2Te3_2020},  indicating that stanene on MBST may be useful in high-speed electronic devices. % indicating nearly similar carrier mobility for stanene on MBST that points to its usefulness in high speed electronic devices. 
	
	%The Sn $p_z$ related bands from DFT calculation appear near \ef\,  
	%	\noindent \textbf{\underline {Bandgap at the $\overline{K}$ point}}:
	\vskip 5mm\noindent \textbf{\underline {Bands around the $\overline{K}$ point}}:	\srb{Figures~\ref{arpes_sta}(f,g) show the outer hole band   up to the $\overline{K}$ point. This band  crosses the $\overline{K}$ point at $\sim$0.8 eV, as shown by the red circles and dashed curve obtained from fitting the  raw data in panel \textbf{f}.   No other bands are observed near \ef, thus indicating a bandgap of 0.8 eV at the  $\overline{K}$ point. This behavior is in contrast to  freestanding stanene,  where this band disperses towards \ef\ due to the $\pi$-$\pi$ bonding between the unsaturated Sn $p_z$ orbitals and opens a nontrivial band gap of 0.1 eV~\cite{XuBinghai2013,Matthes2013}. }
	
	The above discussed deviation of the ARPES band dispersion at the $\overline{K}$ point has  been reported earlier for  stanene grown on telluride substrates~\cite{Zhu2015,Liao2018,Zang2018}, where an experimental band was likewise shifted to  higher $E$ opening up a large bandgap.  On the contrary, the DFT calculations for stanene including the substrate %~\cite{Liao2018,Zhao_Sn_Bi_2022} 
	showed absence of the gap, unless hydrogen passivation  was considered~\cite{Liao2018,Zhao_Sn_Bi_2022,Zhu2015,Zang2018}. It was argued that hydrogen  passivates  the reactive Sn $p_z$ orbital, and this is possible  in ultra high vacuum chambers made of stainless steel because hydrogen is the dominant residual gas~\cite{Redhead2003,Avdiaj2012}.
	\srb{Residual gas analysis conducted with a quadrupole mass spectrometer indicates that hydrogen is the most prevalent gas in our chamber (Fig.~S25 of SM~\cite{supple}).  This mass spectrum is comparable to that of Ref.~\onlinecite{Zhao_Sn_Bi_2022}, in which hydrogen passivation of stanene was demonstrated and the chamber pressure was nearly identical to ours. The probability of hydrogen passivation is evident when we %consider that the measurements are conducted over an extended period of hours and 
		estimate  the number of hydrogen molecules at a pressure of 1$\times$10$^{-10}$ mbar to be $\sim$2.5$\times$10$^{12}$/m$^3$.%**srbtosb/pb: check this number that I got
		 ~Additionally, being the lightest gas, hydrogen has the highest molecular speed %(1927 m/s at room temperature) 
		~among all the gases.}

%**srbtosrb: give the 1ML+BL-H DFT, talk about large buckling that increases the pxpy and pz interaction moving the bands up and down! (Done)

\subsection{DFT calculation for stanene on MBST}
\noindent \textbf{\underline {\srb{Input structure for DFT calculation}}}:~\srb{The left panel of  Fig.~\ref{arpes_sta}(h) shows the  input structure for DFT calculation, where buckled monolayer stanene  is represented by an upper sublattice (Sn$_4$) and a lower sublattice (Sn$_3$). The BL is modeled based on the information about its composition and symmetry obtained from XPS, LEED, and STM; random anti-site disorder is, however, not considered. The BL is represented by  two Sn sublattices similar to stanene (Sn$_2$ and Sn$_1$, where the upper Sn$_2$  sublattice is at the fcc site) and the top Te  (Te$_1$) and  Bi (Bi$_1$) layers of MBST, as shown by the green shading. The Sn layers of BL and stanene resemble the $\alpha$-Sn structure in the (111) direction. %  The starting separation between Sn$_1$ and Te$_1$ layers is taken to be 0.365 nm, which is reported for stanene on PbTe, where BL formation is not reported~\cite{Zang2018}. % Random antisite defects are not taken into account in our DFT calculation, however it is well known that disorder broadens and weakens the bands.  Monolayer stanene  comprising of two sublattices (Sn$_1$ and Sn$_2$) was taken  as per the $\alpha$-Sn structure in the (111) direction. %The separation between stanene and the BL (i.e. Sn2-Sn3 distance) is taken to be 0.285 nm.   the  upper (lower) Sn sublattice is located at the hollow (face centered cubic) site of the substrate.
	 ~Hydrogen (H) passivation is considered in our calculation for the reasons discussed at the end of the previous section (III.E). The H atom is placed at the on top position at a distance of  0.176 nm  from Sn$_4$ [see Fig.~S26(a) for all the interlayer separations]. The substrate MBST(0001) is represented by its unit cell (Fig.~S3(a) of SM~\cite{supple}) with three septuple layers,  where 2 of the 6 Bi atoms were replaced by Sb so that the doping level is 33\%, which is close to the experiment. The lattice constant is taken to be 0.430 nm based on  x-ray crystallography data~\cite{Yan2019}.  The atom positions  have been  fully relaxed  
	 for  stanene, BL, H, and the top septuple of MBST,   whereas the two bottom septuples were kept fixed.}
%~\srb{The lattice constant of stanene was taken to be 0.46 nm based on our STM measurement.} %For hydrogen passivation calculations,  charge of H taken to be 1 for calculating the potential. 
%Outer shell configurations:

%Sn:  [Kr].4d10.5s2.5p2 
%Te: [Kr].4d10.5s2.5p4 
%Bi:  [Xe].4f14.5d10.6s2.6p3
%Sb:[Kr].4d10.5s2.5p3
\vskip 5mm
\noindent \textbf{\underline {\srb{Optimized output structure from DFT }}}:~\srb{The   output structure  in the right panel of Fig.~\ref{arpes_sta}(h)  shows  significant change in separation of the layers representing the BL (green shading) (also see Fig.~S26(b) of the SM~\cite{supple}).  Particularly,  Sn$_1$  shifts down  and the  inter layer separation  between the Sn$_1$ and Te$_1$ layers (Sn$_1$-Te$_1$ distance along the $z$ direction shown by blue double arrow)  decreases substantially to 0.25 nm  from the starting value of 0.365 nm (which is the separation  for stanene on PbTe after optimization, where BL is not present~\cite{Zang2018}) .   Consequently, the   distance between Sn$_1$ and Te$_1$  atoms considering their ($x$, $y$, $z$) coordinates % interatomic bond 
	~reduces to 0.353 nm from 0.44 nm before optimization. This distance is close to  that  of  SnTe  (0.316 nm)~\cite{Lefebvre1998}. Thus, the formation of  a chemical bond between Sn$_1$ and Te$_1$ can be suggested. The Sn$_2$-Te$_1$ distance also decrease from  0.472 nm to 0.395 nm.  
	~Similar to the Te$_1$ layer, %the Bi$_1$ layer also shifts up closer to  Sn$_1$ such that 
	 ~the Sn$_1$-Bi$_1$ layer distance (red double arrow) is reduced by 0.111 nm (from 0.54 to 0.429 nm).  %Sn1 and Bi1 bond distance for S+BL+H:1. Unoptimized input structure: 0.593 nm; 2. optimized output structure: 0.496 nm
	 ~A substantial decrease in the interlayer separations between Sn and the top two atomic layers of the MBST substrate, %indicating the formation of chemical bonds, 
	 ~is the signature of BL formation from DFT. Due to this, the separation between the  Bi$_1$ and the  underlying Te$_2$ layer increases slightly  from 0.223 to 0.238 nm.  The Te$_2$-Mn$_1$ layer separation remain almost  unchanged, %and neither of these layers shifts up, 
	 ~indicating that these layers are not part of the BL. } %total BL height: 5,74 optimized versus 6.47 input 	

 \begin{figure*}[t!] 
	\includegraphics[width=0.9\linewidth,keepaspectratio,trim={0.7cm 12cm 2cm 0cm},clip]{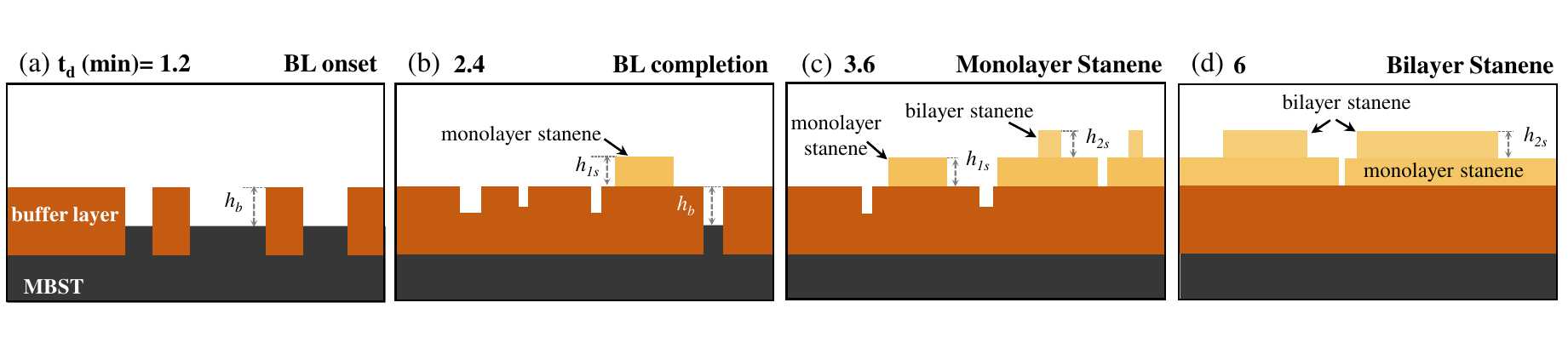} %{STM_growth_evolution_v4}
	\caption{% \textbf{Schematic diagrams.} 
		(a-d) A set of schematic diagrams  showing the growth of monolayer and bilayer stanene  (yellow) on the BL  (reddish brown). $h_b$, $h_{1s}$, $h_{2s}$ represent the heights of the BL, monolayer stanene and the bilayer stanene, respectively. The MBST substrate is shown in black color. }
	\label{schematics}
\end{figure*}

	\srb{After optimization, stanene shows a buckling height of 0.124 nm given by Sn$_4$-Sn$_3$ separation, which  is in good agreement with STM (0.1$\pm$0.015 nm).  Additionally, the separation between stanene and the BL, i.e.,  the Sn$_3$-Sn$_2$ separation (0.284 nm), remains almost unchanged.  For comparison with the other DFT calculation discussed later, we annotate this  calculation as \textbf{``S+BL+H"}, where both  BL and H passivation are considered (``\textbf{S}" stands for monolayer stanene).}
\vskip 5mm
\noindent \textbf{\underline {\srb{	DFT band structure for  \textbf{S+BL+H} calculation}}}:~\srb{Figure~\ref{arpes_sta}(i) shows the stanene related bands for the above discussed structure. Two adjacent  bands of Sn $p_z$ orbital character are found to traverse the $\overline{K}$ point at $E$= 0.85 and 1.1 eV. Thus, the bandgap at the $\overline{K}$ point is measured to be 0.85 eV, which closely matches the bandgap of 0.8 eV from ARPES [Figs.~\ref{arpes_sta}(f) and \ref{arpes_sta}(g)]. These bands disperse up towards \ef\ and approach the $\overline{\Gamma}$ point, and constitute the outer hole band that is observed in ARPES [Figs.~\ref{arpes_sta}(c, d, f, g)].  The Sn $p_{x,y}$ orbital character (green circles) of the bands is apparent below 0.5~\a. %These bands merge below 0.3~\a and form the outer hole band that is observed in ARPES around $\overline{\Gamma}$ in Figs.~\ref{arpes_sta}(c,d,f). 
	  ~The outer hole band of  Sn $p_{x,y}$ character  is also observed  towards the $\overline{M}$ point, which shows a minimum around 1 eV at $\overline{M}$.} %crosses \ef\, around 0.3\a. 
~\srb{DFT also exhibits the inner hole band found in ARPES about $\overline{\Gamma}$. It is of   $p_{x,y}$ character  and crosses \ef\ from both $\overline{M}$ and $\overline{K}$ directions at 0.07~\a\ and shows a maximum at 0.2 eV above \ef.  It shows a minimum at  1.6 eV at $\overline{M}$ and  disperses  to $>$2 eV towards $\overline{K}$. }%Note that nearly linear dispersion observed experimentally  in Figs.~\ref{arpes_sta}(a,c) is also reproduced in DFT in the relevant $E$-$k$ range shown by a blue rectangle in Fig.~\ref{arpes_sta}(h).   }

\srb{Besides the outer and inner bands,  the inverted parabolic band observed in ARPES is also observed in both directions, with a maximum at $\overline{\Gamma}$ at $\sim$0.4~eV.    It has primarily    $p_{x,y}$ character with approximately equal contributions from the  $p_x$ and $p_y$ orbitals,  with some admixture of  $s$ character (Fig.~S27 of SM~\cite{supple}).   The outer and inner hole bands  also have nearly similar  $p_x$ and $p_y$ contributions.  A primarily  Sn $s$ band appears above \ef\ with a minimum at $E$= -1.5 eV around the $\overline{\Gamma}$ point. Stanene on MBST is found to be metallic, where both the inner and outer  bands cross \ef\ around $\overline{\Gamma}$ and give rise to the outer hexagonal and inner circular pocket of the Fermi surface, respectively  [Fig.~\ref{arpes_sta}(e)]. }

\srb  {The preceding discussion demonstrates a good agreement between ARPES [Figs.~\ref{arpes_sta}(a) to \ref{arpes_sta}(g)] and DFT [Fig.~\ref{arpes_sta}(i)]. The bandgap at $\overline{K}$ is reproduced, and all the bands appear in the latter. However, unlike freestanding stanene, the Dirac cone at the $\overline{K}$ point is absent due to the  bandgap,  and band inversion is not observed. Therefore, the non-trivial band structure is not observed for stanene on MBST, as is also the case for stanene on \bt~\cite{Zhu2015} and PbTe~\cite{Zang2018}. }   

%Additionally, the optimized  output structure   shows that the Sn4-Sn3 distance between the upper and lower sublattices of stanene, which is the buckling height, is  0.124 nm. This is in reasonable agreement with that obtained from STM (0.1$\pm$0.01 nm) considering that the DFT has been performed with the stanene lattice matched with the substrate i.e. with lattice constant of 0.43 nm.  A calculation  with lattice constant of 0.46 nm, although resulting in higher energy, could be converged, where the buckling was found to be 0.103 nm, which is in excellent agreement with the experimental value. This shows that buckling  increases if stanene is  compressed.  
 \vskip 5mm
\noindent \textbf{\underline {\srb{\textbf{S+BL} DFT calculation}}}:~\srb{To examine the role of H passivation, %\textit{vis-a-vis} that of BL on the stanene bands, we have performed 
	~another calculation has been performed with the input structure as in the left panel of Fig.~\ref{arpes_sta}(h), but with the H layer removed. It is referred to as the \textbf{``S+BL"} calculation. The resulting optimized  structure in  Fig.~S28(a) of SM~\cite{supple}  shows BL formation with reduced Sn$_1$-Te$_1$ (0.25 nm) and Sn$_1$-Bi$_1$ (0.435 nm) separations. %The Sn$_3$-Sn$_2$ separation between BL and  stanene is 0.288 nm. 
	~These values are close to the  \textbf{S+BL+H} calculation (Fig.~S26(b) of SM~\cite{supple}). Therefore, the absence of H passivation does not influence the BL formation. On the other hand, stanene demonstrates a notable increase of $\sim$28\% to 0.159 nm in the buckling height, which is  observed from the Sn$_4$-Sn$_3$ separation.	  This enhances the hybridization between the  unpassivated $p_z$  and the $p_{x,y}$ orbitals~\cite{XuBinghai2013} and  influences the  band structure (Fig.~S28(b) of SM~\cite{supple}). The maximum of the inner (outer) hole band of Sn $p_{x,y}$ character moves up in $E$ by about 0.7 (0.5)  eV to above \ef. The inverted parabolic band that is  of $p_{x,y}$ character  shows a $p_z$ contribution near the maximum due to the enhanced hybridization. Here too, the maximum  is  shifted %by 0.7 eV  
	~to $E$= -0.4 eV. % above \ef. 
	 ~In contrast, the $p_z$ bands around  $\overline{K}$ are shifted  to $E$= 0.6 eV. % and a bandgap at the $\overline{K}$ opens up. %Note that the bandgap opens up in absence of  H passivation. %although it is smaller compared to that observed in experiment (0.85 eV).  
	~Thus, the bands of the $p_{x,y}$ ($p_{z}$) character are shifted up (down) as a result of the stabilization of the stanene structure with larger buckling and the resulting increase in the $p_{x,y}$-$p_{z}$ hydridization.} %between the unpassivated $p_z$ orbitals and the $p_{x,y}$ orbitals 
%	resulting in a hybridization gap. 
 %The Sn $p_{x,y}$ orbitals of the BL also has larger hybridization  with the $p_z$ orbitals of stanene. %the lower stanene sublattice. 
	%This seems plausible because in absence of the BL  [Fig.~S28(a), the \textbf{``SorBL-H"} calculation],**srb: well BL is here in BL-H calculation, seems like wrong logic, delete!!** the buckling height is 0.14 nm compared to 0.16 nm here with BL. So, Sn-Sn interaction between Sn of the BL and that of stanene could slightly decrease the buckling and aid in opening up the bandgap at the $\overline{K}$ point. 
\vskip 5mm
\noindent \textbf{\underline {\srb{ \textbf{%``S+H" that transforms to 
				BL+H DFT calculation % or``SorBL+H"
}}}}:~\srb{A calculation conducted using a monolayer stanene structure that is passivated with hydrogen [Fig.~S29(a)] results in an optimized structure where the distances between Sn$_1$-Te$_1$ (0.25 nm) and Sn$_1$-Bi$_1$ (0.43 nm) significantly decrease. This indicates that the stanene transforms to the BL structure (Fig.~S29(b) of the Supplementary Material~\cite{supple}). %A  calculation performed with an input structure of monolayer stanene only with H passivation [Fig.~S29(a)]  transforms to an  optimized structure with a significant decrease in Sn$_1$-Te$_1$ (0.25 nm) and Sn$_1$-Bi$_1$ (0.43 nm) distances that are signatures of BL formation [Fig.~S29(b)  of SM~\cite{supple}]. %without the BL   %is referred to as \textbf{``S+H"}, do not mention this as it will cause confusion, we stick to the output str for naming as the band is for that, other wise in the S+B+H calc, the input needs to be named as 2S+H, also remove from figure 
 So, the optimized structure can be regarded as the BL without a stanene (\textbf{S}) layer,  and is termed  as the \textbf{``BL+H"} calculation.  This calculation shows the importance of the BL, in the absence of which  stanene  will  exhibit chemical bonding with the top two layers of the substrate and the stanene related bands might not be observed. Indeed,  Fig.~S29(c) of SM~\cite{supple} shows that the stanene related inner and parabolic bands are absent. Nevertheless, a Sn related band resembling the outer band  is observed in theory. However,   the random anti-site disorder present in the BL, which is not considered in our calculation, is expected to broaden and thus weaken this band. So, this band is faint in the raw data in Fig.~\ref{arpes_All}(f), but can be identified in the second derivative plot (white arrows in  Fig.~S4(f) of SM~\cite{supple}) towards the  $\overline{M}$ direction for \t= 2.4 min, when the BL formation is nearly completed and there is minimal stanene formation. } %not relevant: This band becomes more prominent with monolayer stanene formation, as shown in Figs.~\ref{arpes_All}(h) and \ref{arpes_sta}(a-d).} 

\srb{Thus, based on the three DFT calculations discussed above (\textbf{S+BL+H}, \textbf{S+BL}, and \textbf{BL+H}), it can be concluded that the stanene band structure calculated with both BL and H passivation yields good agreement with ARPES, as seen from the results of  the \textbf{S+BL+H} calculation presented in Figs.~\ref{arpes_sta}(h) and \ref{arpes_sta}(i).}
\begin{comment}srbtosrb: leave this out, not important to present context and also DFT discussions becoming too extended
\noindent \textbf{\underline {\srb{DFT band structure for  \textbf{``S-H" %or``SorBL-H" **srb: the SorBL nomenclature is confusing
}}}}:~\srb{The DFT calculation discussed above %(\textbf{``S+H"}) 
~include  the H layer. To decipher the role of H, we have performed a  \textbf{``S-H"} calculation, i.e.,  without H. %, and this is annotated as \textbf{``S-H"}. 
~The optimized structure  in Fig.~S28(a) of SM~\cite{supple} shows the signatures of BL formation from decreased Sn$_2$-Te$_1$ (0.275 nm) and Sn$_2$-Bi$_1$ (0.46 nm) distances, and thus this calculation can be termed as \textbf{``BL-H"}.  The band dispersion shows  **(description of bands)**.  The bandgap  at the $\overline{K}$ point is absent  in absence of H passivation  [Fig.~S28(b)  of SM~\cite{supple}].}
\end{comment}

%**srbtopb: random height modulation is indicatibe of H adsorption: Zhou Nmat2015
	 
%	2MLwithoutH= 1ML withBL. If this shows gap at K point, then H passivation is important. If it shows bands dispersing to EF, BL has no effect on it. If G bands not affected, then we can say: "The two hole bands represent a characteristic feature of stanene[16] which is insensitive to substrate	and decoration because of the intactness of in-plane Sn  " from Zang, could be used somewhere.
	
%transfer of electrons from Sn p orbital to Te p /Bi ? orbitals, so characteristic stn bands not observed for only BL calculation.	
	
\section{Conclusion}
In conclusion, using a combination of complementary analytical techniques and density functional theory (DFT), we report  the formation of  stanene on a magnetic topological insulator (MBST), whose growth is aided by the formation of an  underlying BL.  The whole sequence of growth as a function of Sn deposition time (\t) is depicted in Fig.~\ref{schematics} through a series of schematic representations. For the initial depositions,  angle resolved photoemission spectroscopy (ARPES) shows a gradual shift of  the  MBST valence band, indicating electron transfer to the substrate. This is the onset of the BL formation  [Fig.~\ref{schematics}(a)], as shown also by scanning electron microscopy (STM) and core level spectroscopy. Stanene formation starts on the BL  [Fig.~\ref{schematics}(b)], the latter having a thickness  of 0.9 nm and a composition  of  Sn:Te:Bi/Sb $\approx$ 2:1:1. Low energy electron diffraction (LEED) and STM show that the BL has an ordered crystal lattice, albeit with random anti-site defects.  

\srb{The height profile analysis of the atomic resolution STM images provides evidence for both monolayer and bilayer stanene. The stanene layers develop  in the form of islands, exhibiting a Stranski-Krastanov growth mode with a step height of   0.35$\pm$0.03 nm  [Figs.~\ref{schematics}(c) and \ref{schematics}(d)]. The buckling height is estimated from STM to be 0.1$\pm$0.015 nm.  The LEED intensity profiles indicate that the lattice constant of stanene is $\sim$3.1\% larger than that of MBST, which aligns with the findings from STM. The BL bridges this disparity and provides a platform for stanene growth.} A stanene-related component is  detected in the Sn $d$ core level spectra in addition to a BL-related component.

 ARPES intensity plots and DFT calculations for  monolayer stanene (the \textbf{S+BL+H} calculation) show hole-like inner and outer bands  and an inverted parabolic band  around the $\overline{\Gamma}$ point of primarily Sn $p_{x,y}$ orbital character. \srb{The outer band traverses the \ef\ showing that stanene is metallic with a Fermi surface of hexagonal symmetry. In contrast, a  bandgap of 0.8 eV is observed at the  $\overline{K}$ point, where the outer band is of Sn $p_z$ orbital character.  We find that DFT band structure calculations show good agreement with the ARPES  only when the BL and H passivation are considered in the calculation. From this calculation, all the bands %around the $\overline{\Gamma}$ point 
 	~are identified, and a  nearly similar bandgap at the $\overline{K}$ point is obtained. On the other hand, if either  BL or H passivation are not considered in the calculation, the agreement is unsatisfactory. }%Evidence of  BL formation from DFT is obtained from the decrease in the distances between Sn and the top two layers of  MBST. 

Formation of  stanene   opens up the  possibility of  superconductivity~\cite{Liao2018,Zhao_Sn_Bi_2022}  and its ensuing proximity effect with MBST.  %the substrate 
%~MBST  that is antiferromagnetic below 24~K. 
~\srbg{The proximity effect could be tuned by varying the thickness of the BL through modification of  the growth conditions.   The BL could also possibly  become superconducting, as there are  reports in literature about superconductivity in  Sn$_{0.57}$Sb$_{0.43}$ with $T_C$= 1.45 K~\cite{Liu2019} and Sn$_{1-\delta}$Te with $T_C$$<$\,0.3 K~\cite{Erickson2009}, both having rhombohedral structure as is the case here.} 
\srbn{Despite the presence of the BL, the magnetic interaction  due to the Mn layer of the substrate could  potentially impact the properties of stanene because  the separation of the latter from the  Mn layer ($\sim$1.3 nm, Fig.~S26(b) of SM~\cite{supple}) is smaller than the magnetic interaction range ($\geq$3.36 nm~\cite{text1,Klimovskikh2020}).} %On the other hand, in spite of the BL, since the separation between stanene and the Mn layer %(1.336 between Sn4 and Mn1 in Fig. S26b)
	 % ~is smaller compared to the magnetic interaction range of  $\geq$3.36 nm in the MBT series~\cite{text1}, magnetic interaction could influence the properties of stanene.} 
 %text1: MBT and higher n members of the series MnBi$_2$Te$_4$(Bi$_2$Te$_3$)$_{\rm n}$ exhibit antiferromagnetic interaction between Mn layers up to n=2. So, the magnetic interaction range in this series is expected to be more than or equal to the Mn layer separation (3.36 nm for n= 2)\cite{Klimovskikh2020}.
 % [I.I. Klimovskikh et al. npj Quantum Mater. 5, 54, 2020].   
~ Our work establishes the formation of stanene on a magnetic topological insulator, which may lead to an interesting interplay between  band topology, superconductivity, and magnetism. \\

\vskip 10mm
\noindent\textbf{\large Acknowledgments}\\
The Computer division of Raja Ramanna Centre for Advanced Technology  is thanked for installing the DFT codes and providing support. 	R.B. would like to acknowledge the partial support from the Start-up Research Grant (SRG), SERB, Govt. of India, file no. SRG/2022/000552-G. A.P. acknowledges the  Israel Council for Higher Education for a postdoctoral fellowship through the Study in Israel program. M.H. thanks the Leona M. and Harry B. Helmsley Charitable Trust grants 2018PG-ISL006 and 2112-04911 for support.  M.B. and S.R.B. are thankful for the support from Science and Engineering Board through a CRG project (CRG/2023/001719). We are thankful to  A.~Gloskovskii for support during the HAXPES measurement. The authors acknowledge the India@DESY collaboration and the CERIC-ERIC Consortium for financial support and access to  Petra-III and Elettra synchrotron facilities, respectively. \\
\noindent $^*$electronic address: barmansr@gmail.com

%\clearpage
%\bibliographystyle{apsrev4-2_with_titile}
%\bibliography{references_v23}% Produces the bibliography via BibTeX.
\providecommand{\noopsort}[1]{}\providecommand{\singleletter}[1]{#1}%

\clearpage
\begin{figure} 
	\includegraphics[page=1, width=1.3\linewidth,keepaspectratio,trim={2cm 0 -1cm 0},clip]{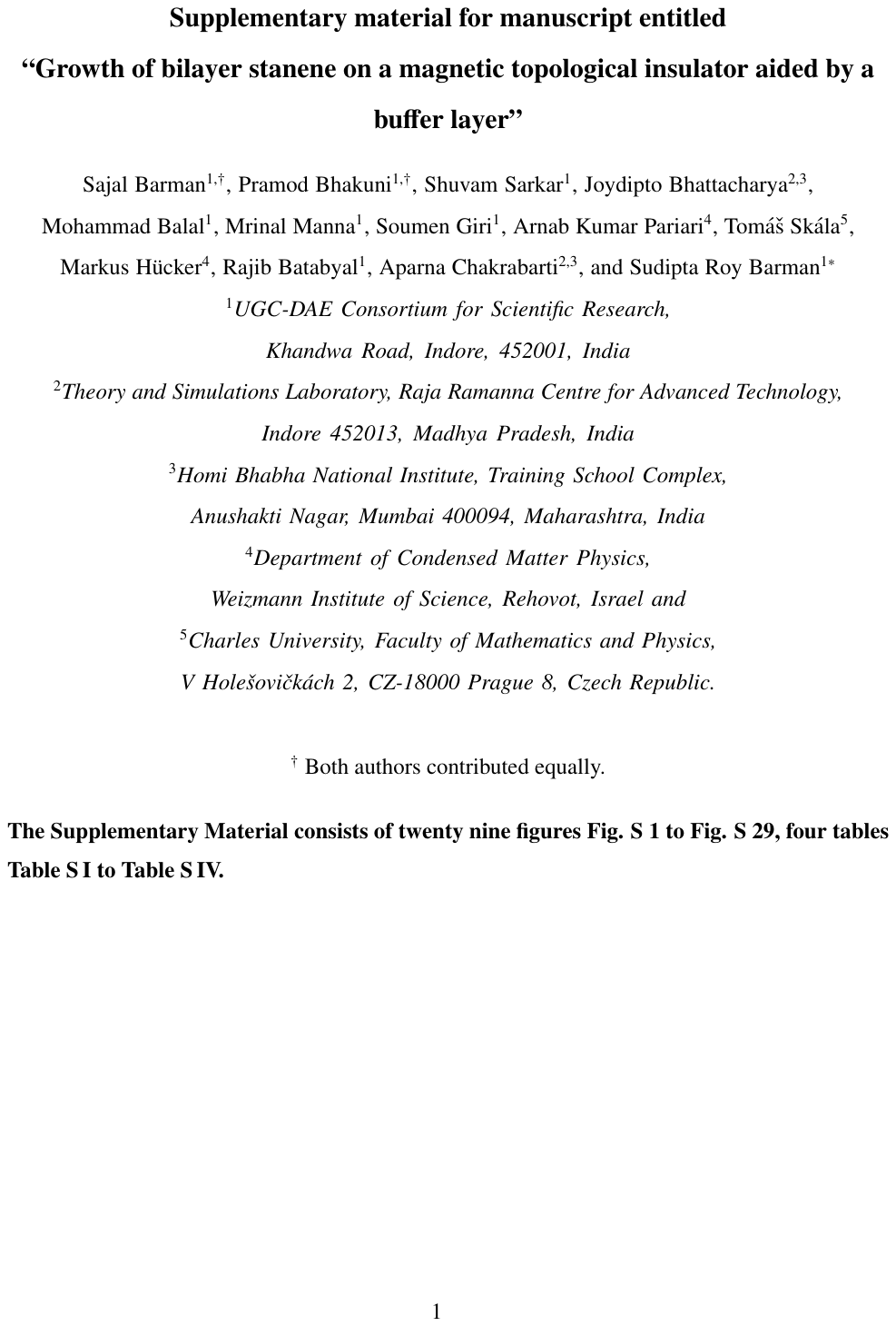} 
\end{figure}
\newpage
\begin{figure} 
	\includegraphics[page=2,width=1.3\linewidth,keepaspectratio,trim={2cm 0 -1cm 0},clip]{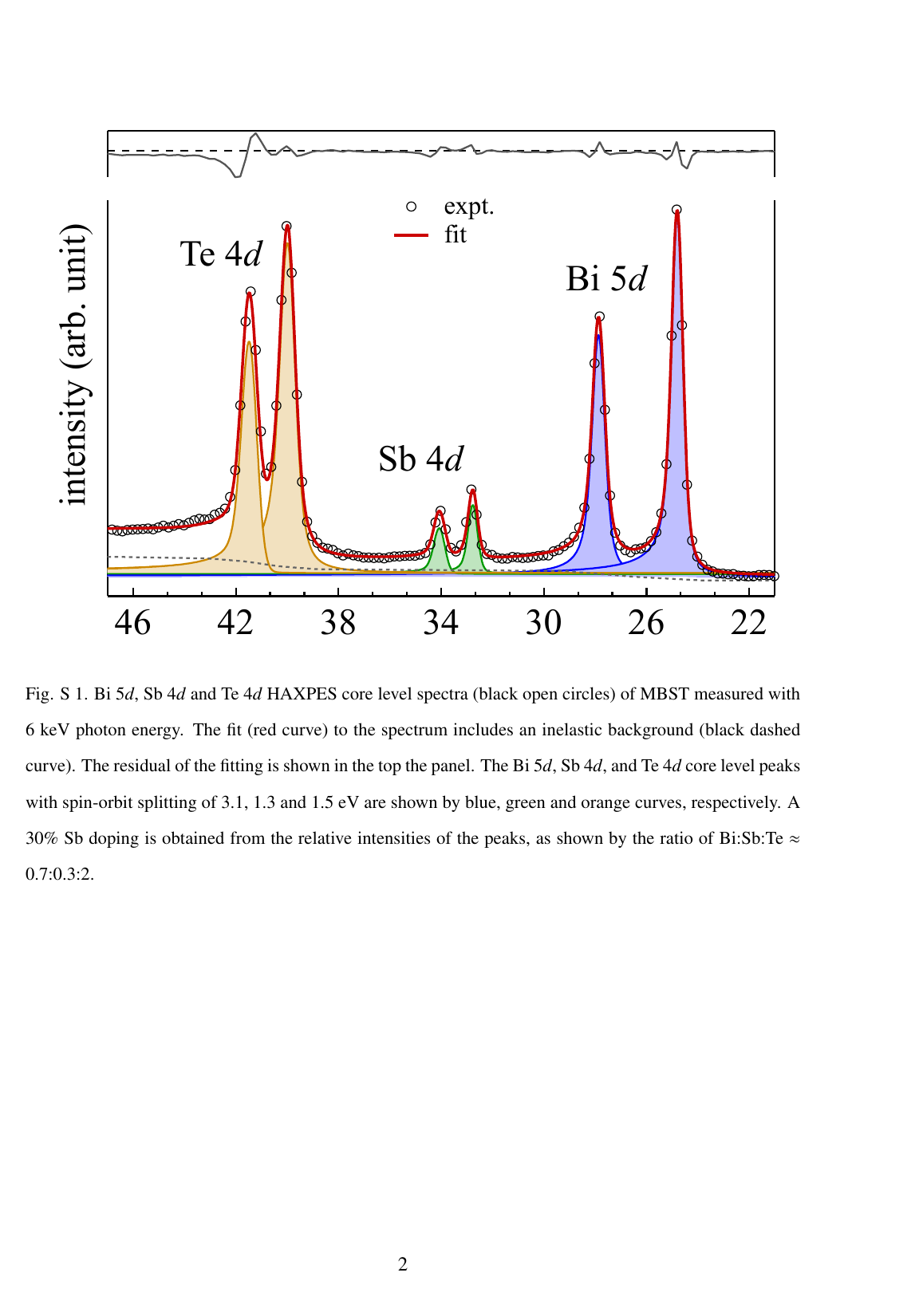} 
\end{figure}
\newpage
\begin{figure} 
	\includegraphics[page=3,width=1.3\linewidth,keepaspectratio,trim={2cm 0 -1cm 0},clip]{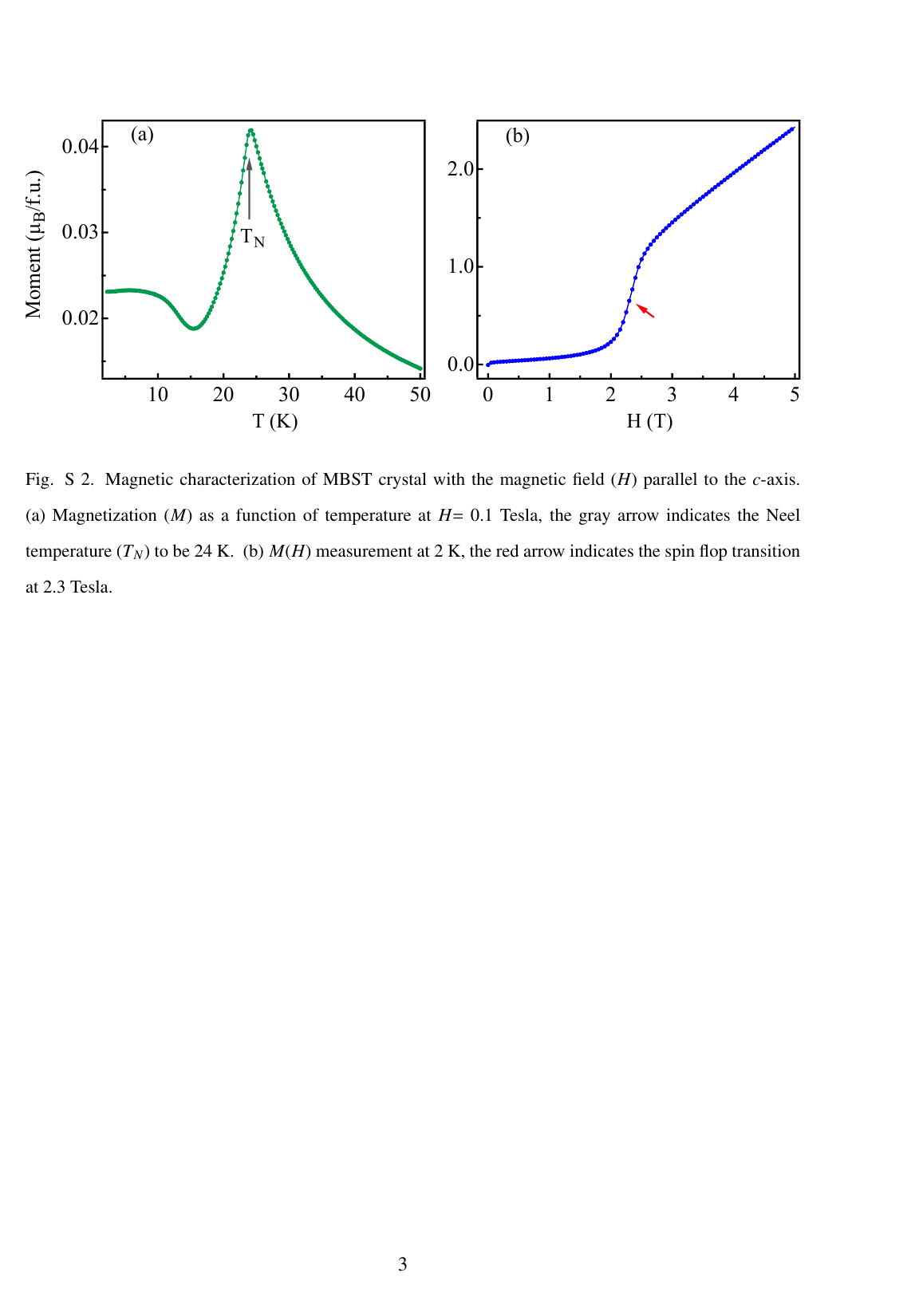} 
\end{figure}
\begin{figure} 
	\includegraphics[page=4,width=1.3\linewidth,keepaspectratio,trim={2cm 0 -1cm 0},clip]{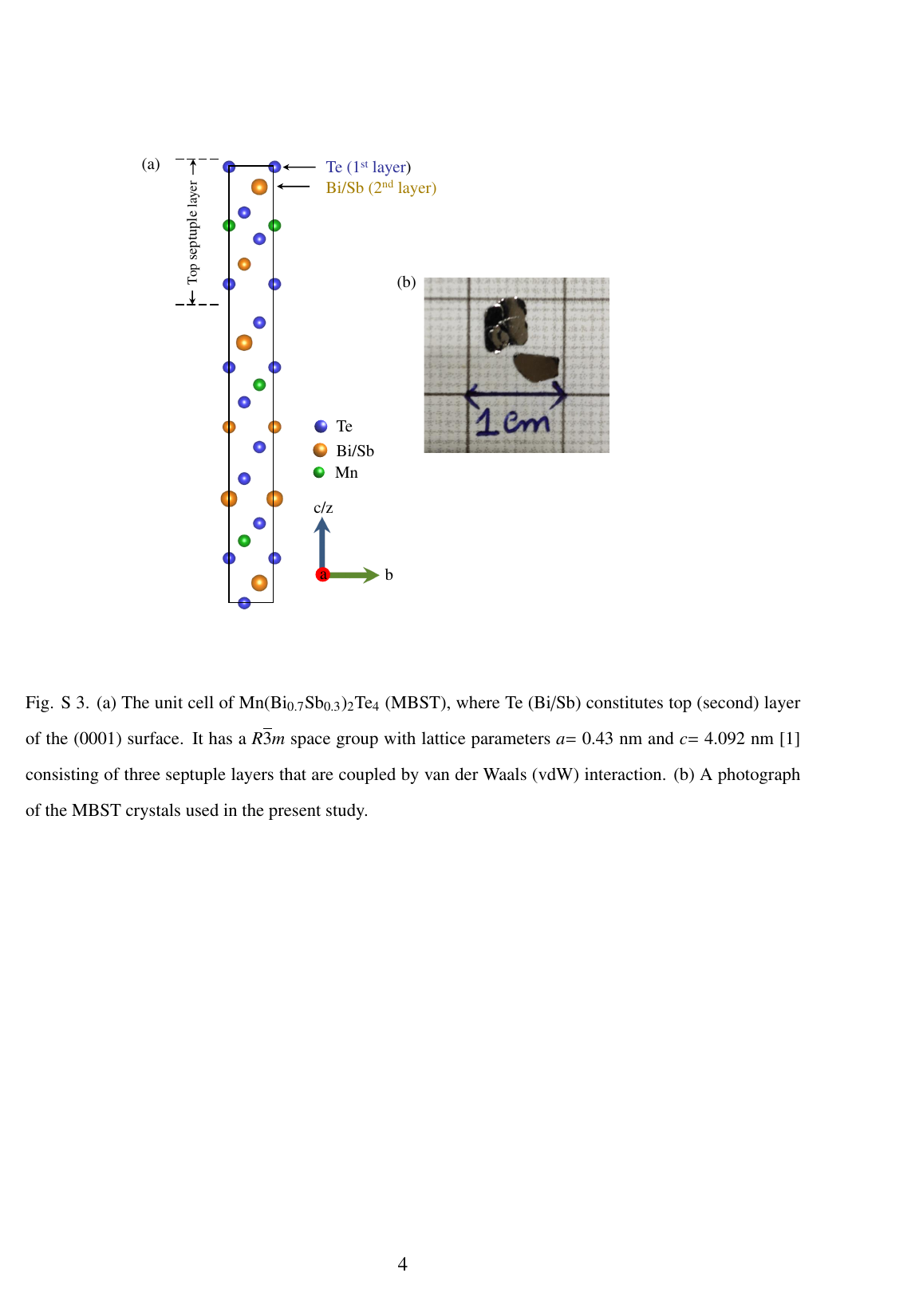} 
\end{figure}
\newpage
\begin{figure}
	\includegraphics[page=5,width=1.3\linewidth,keepaspectratio,trim={2cm 0 -1cm 0},clip]{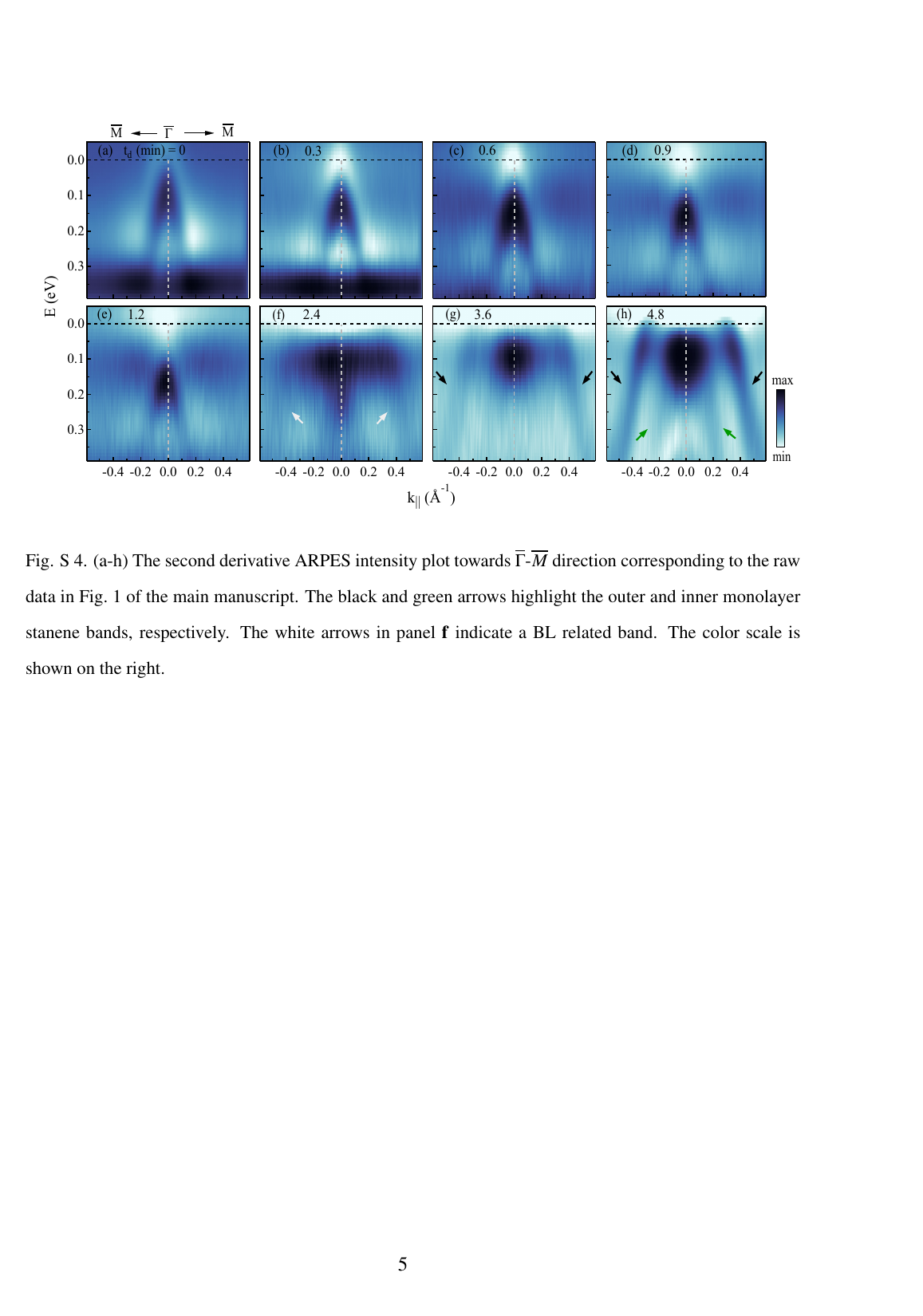} 
\end{figure}
\newpage
\begin{figure}
	\includegraphics[page=6,width=1.3\linewidth,keepaspectratio,trim={2cm 0 -1cm 0},clip]{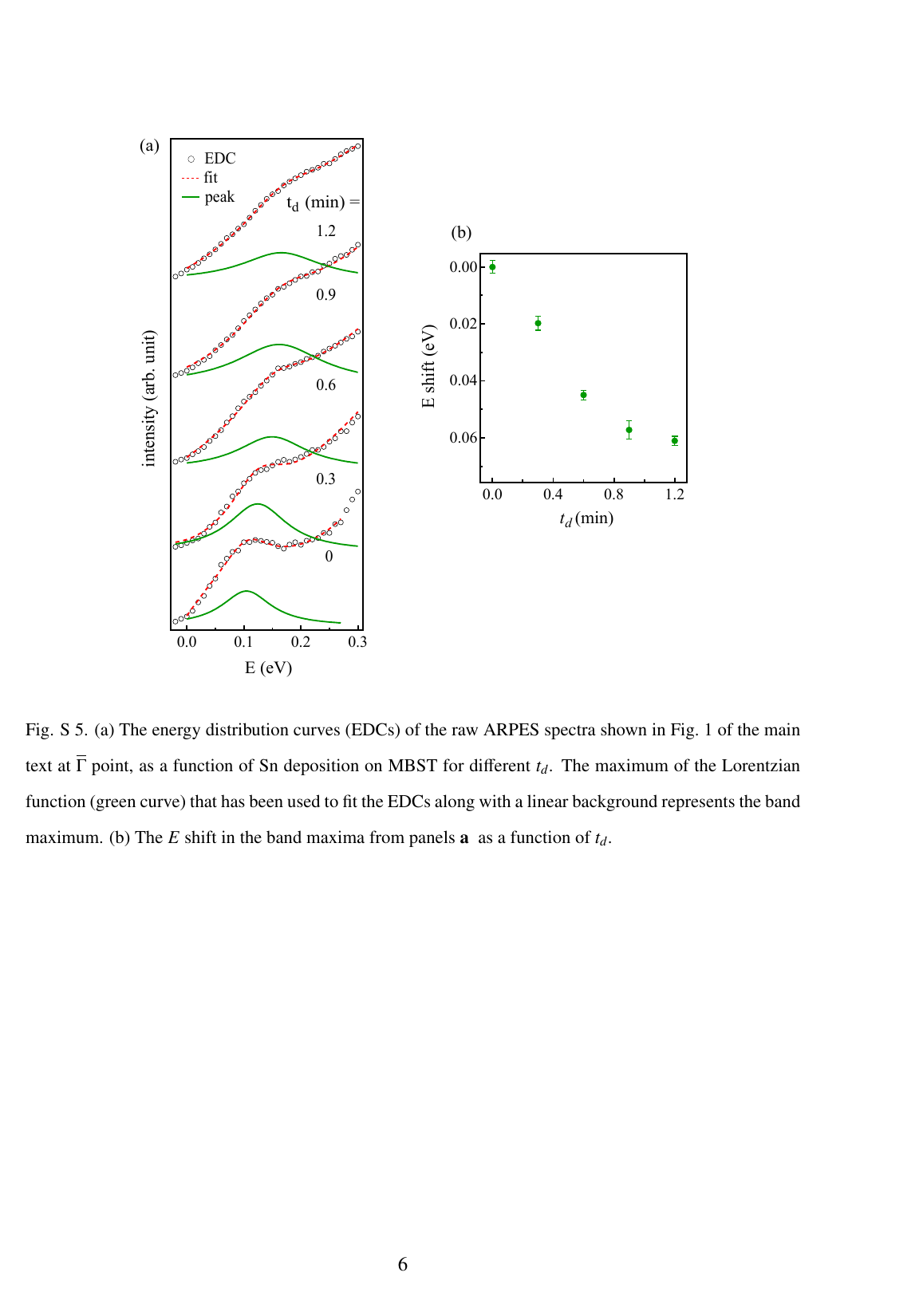} 
\end{figure}
\newpage
\begin{figure}
	\includegraphics[page=7,width=1.3\linewidth,keepaspectratio,trim={2cm 0 -1cm 0},clip]{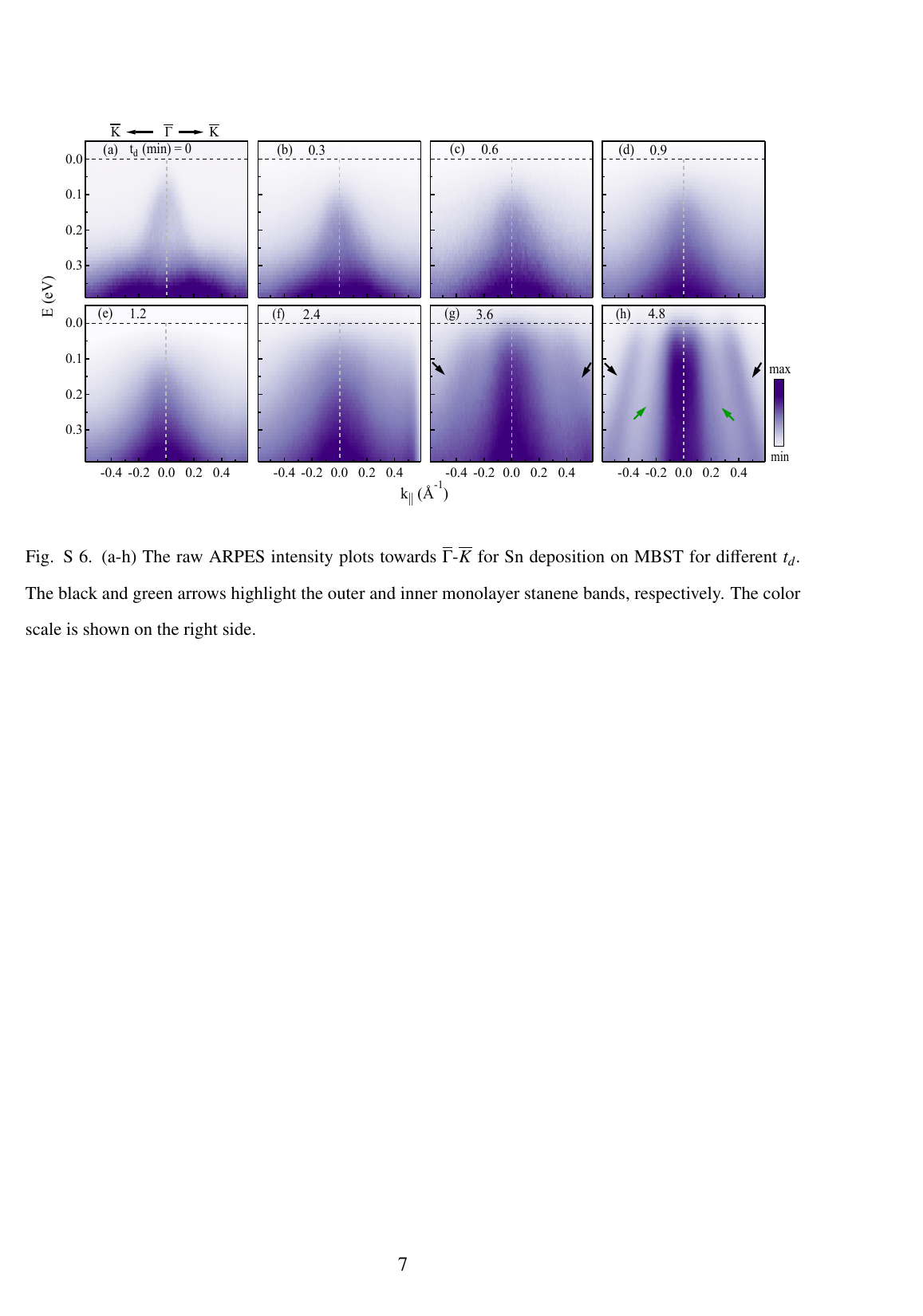} 
\end{figure}
\newpage
\begin{figure}
	\includegraphics[page=8,width=1.3\linewidth,keepaspectratio,trim={2cm 0 -1cm 0},clip]{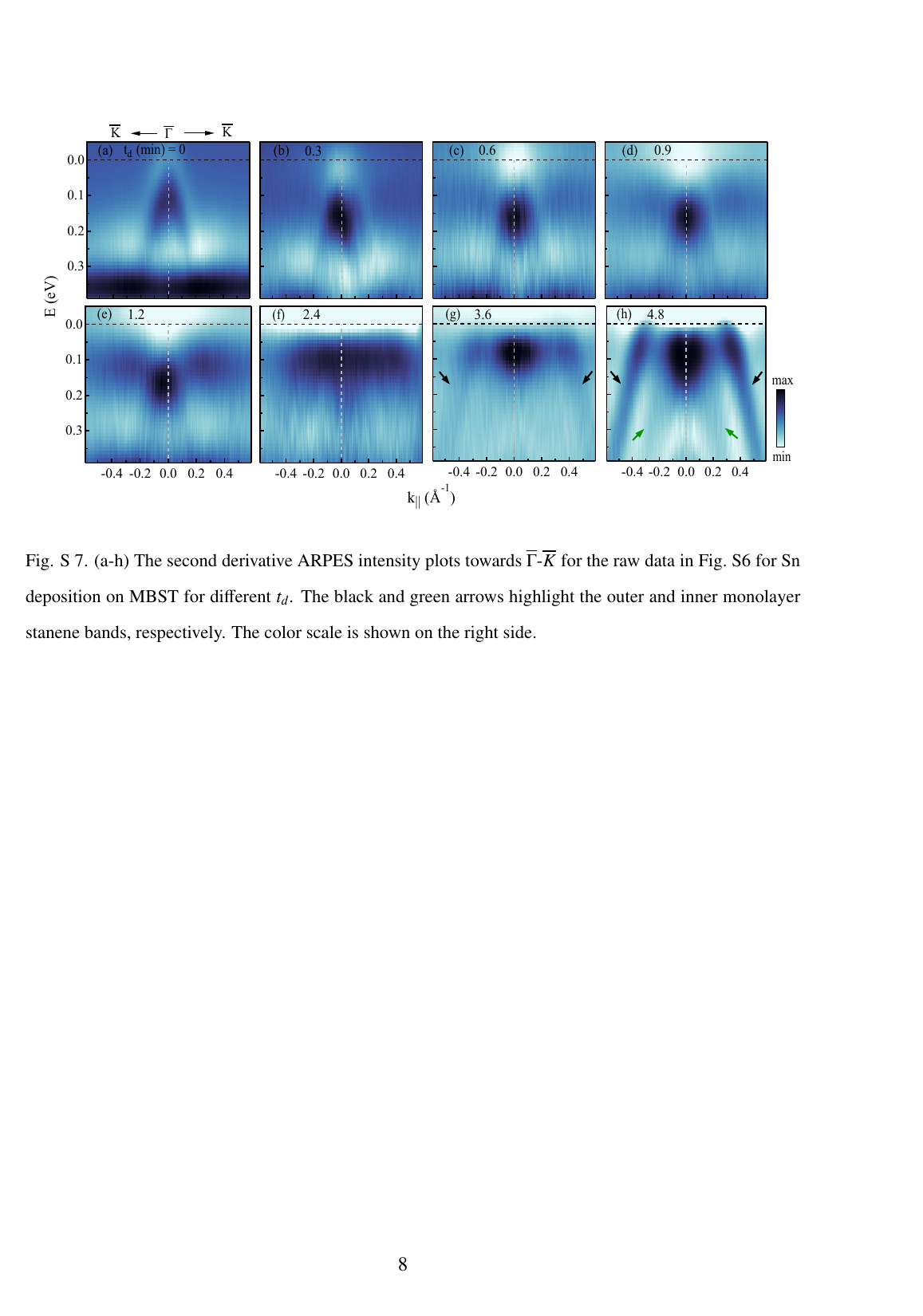} 
\end{figure}
\newpage
\begin{figure}
	\includegraphics[page=9,width=1.3\linewidth,keepaspectratio,trim={2cm 0 -1cm 0},clip]{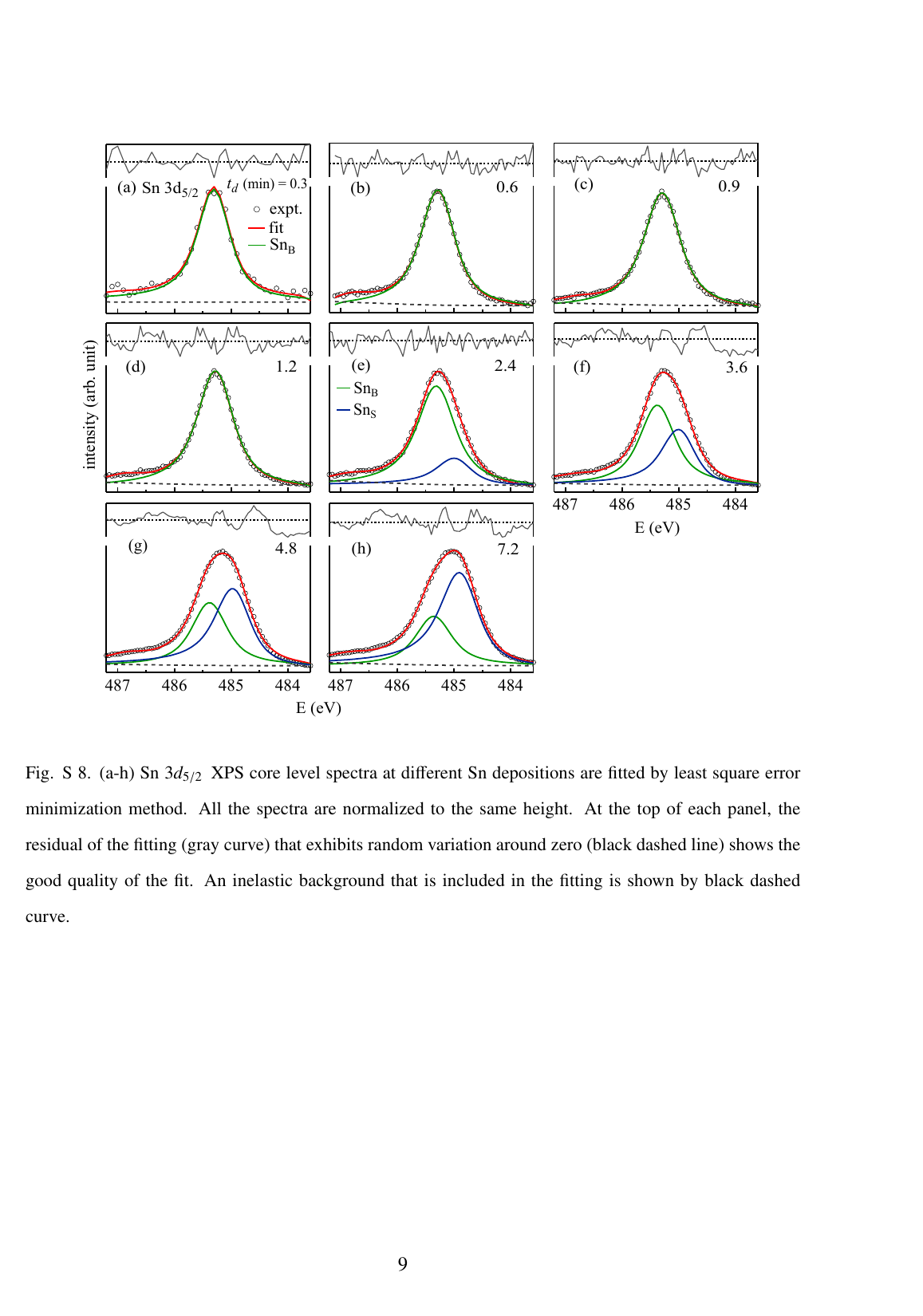} 
\end{figure}
\begin{figure}
	\includegraphics[page=10,width=1.3\linewidth,keepaspectratio,trim={2cm 0 -1cm 0},clip]{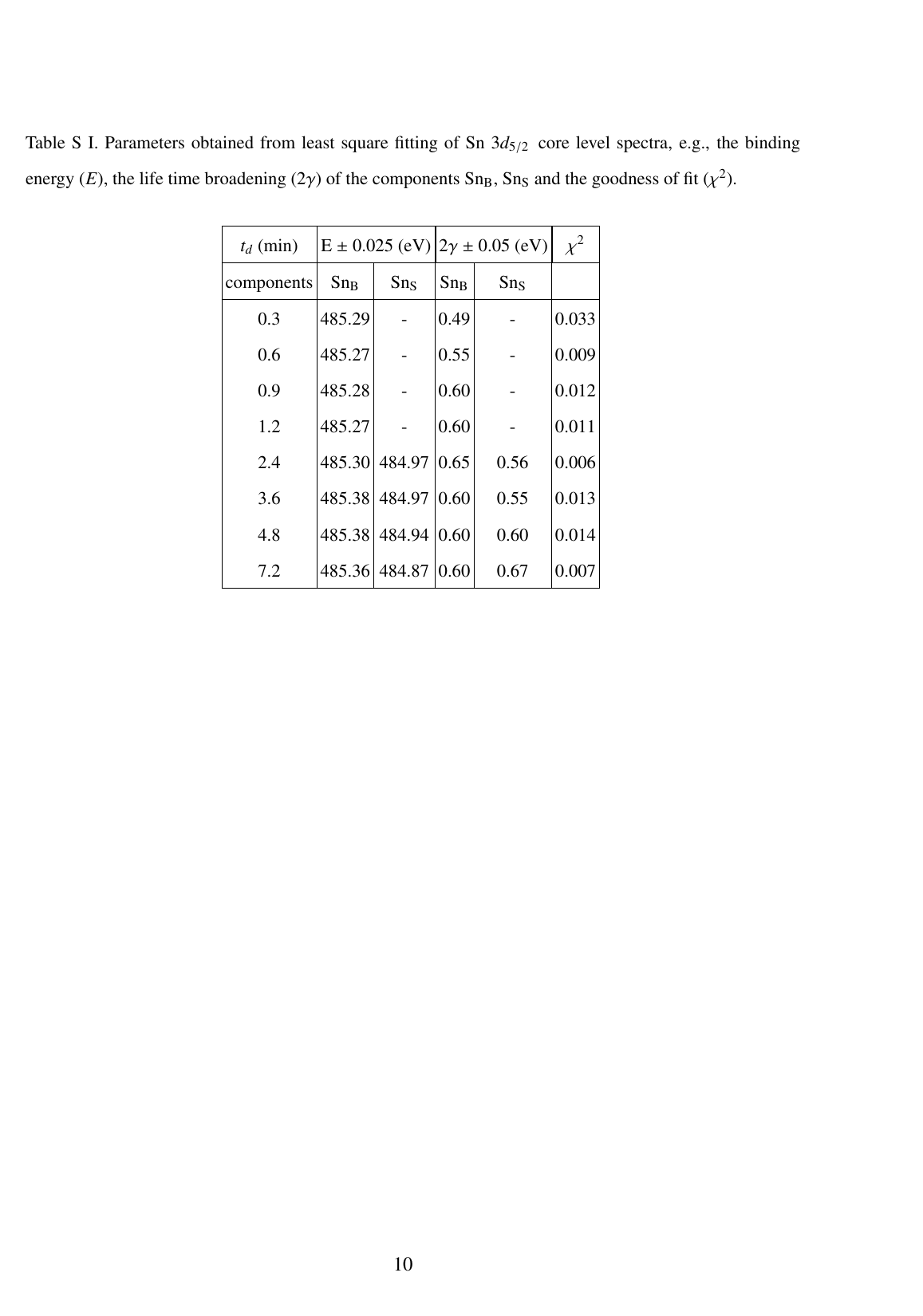} 
\end{figure}
\newpage
\begin{figure}
	\includegraphics[page=11,width=1.3\linewidth,keepaspectratio,trim={2cm 0 -1cm 0},clip]{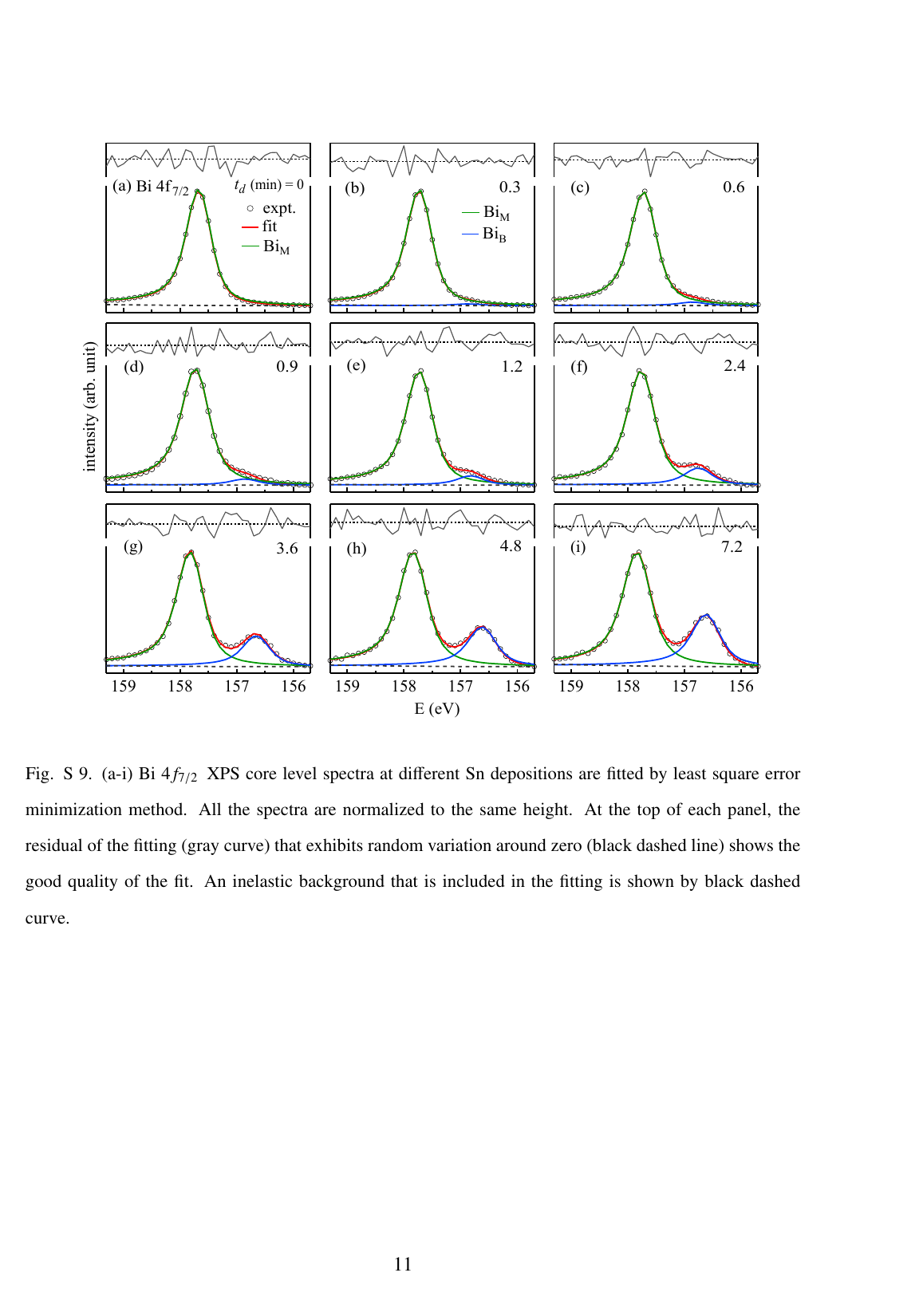} 
\end{figure}
\newpage
\begin{figure}
	\includegraphics[page=12,width=1.3\linewidth,keepaspectratio,trim={2cm 0 -1cm 0},clip]{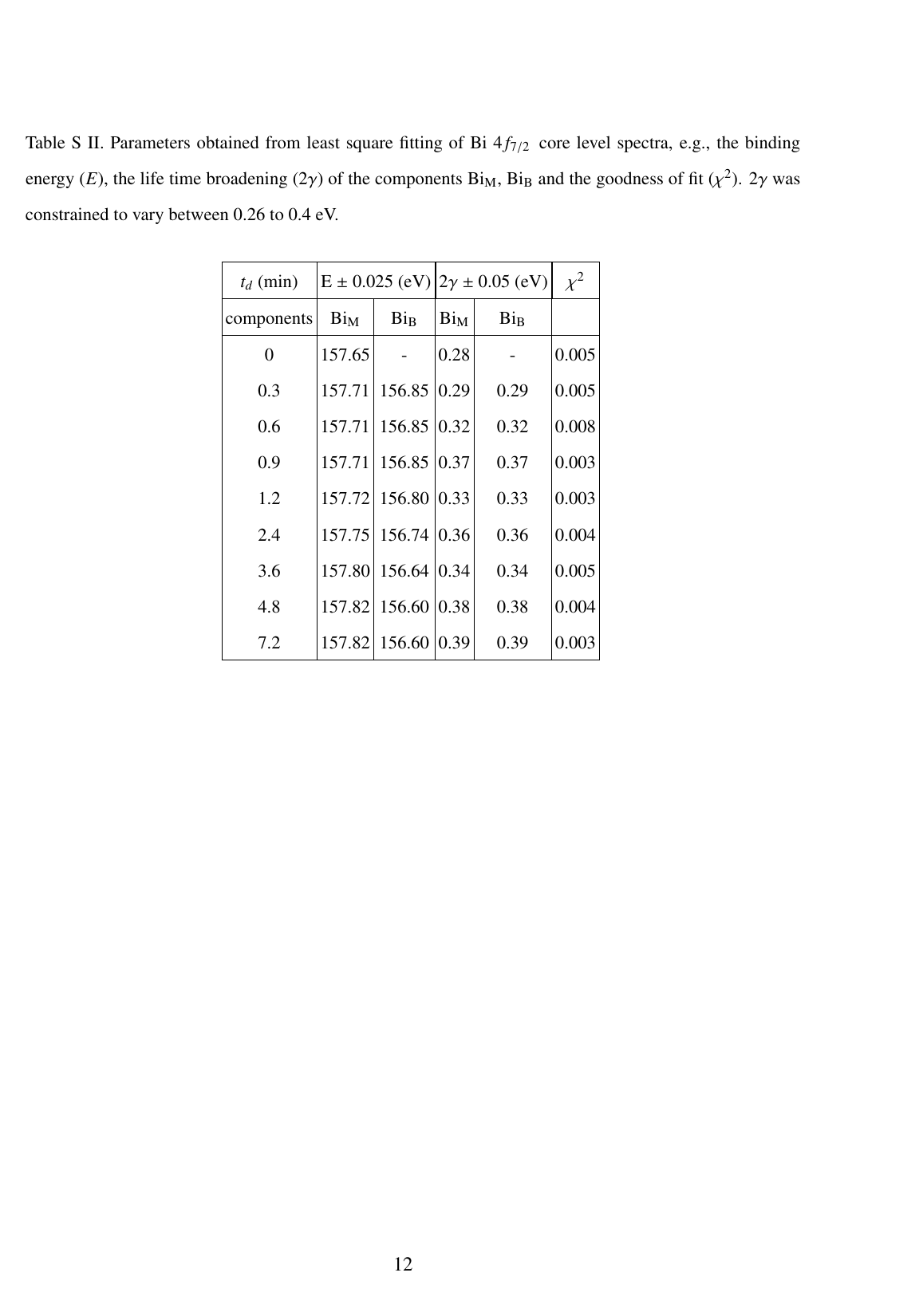} 
\end{figure}
\begin{figure}
	\includegraphics[page=13,width=1.3\linewidth,keepaspectratio,trim={2cm 0 -1cm 0},clip]{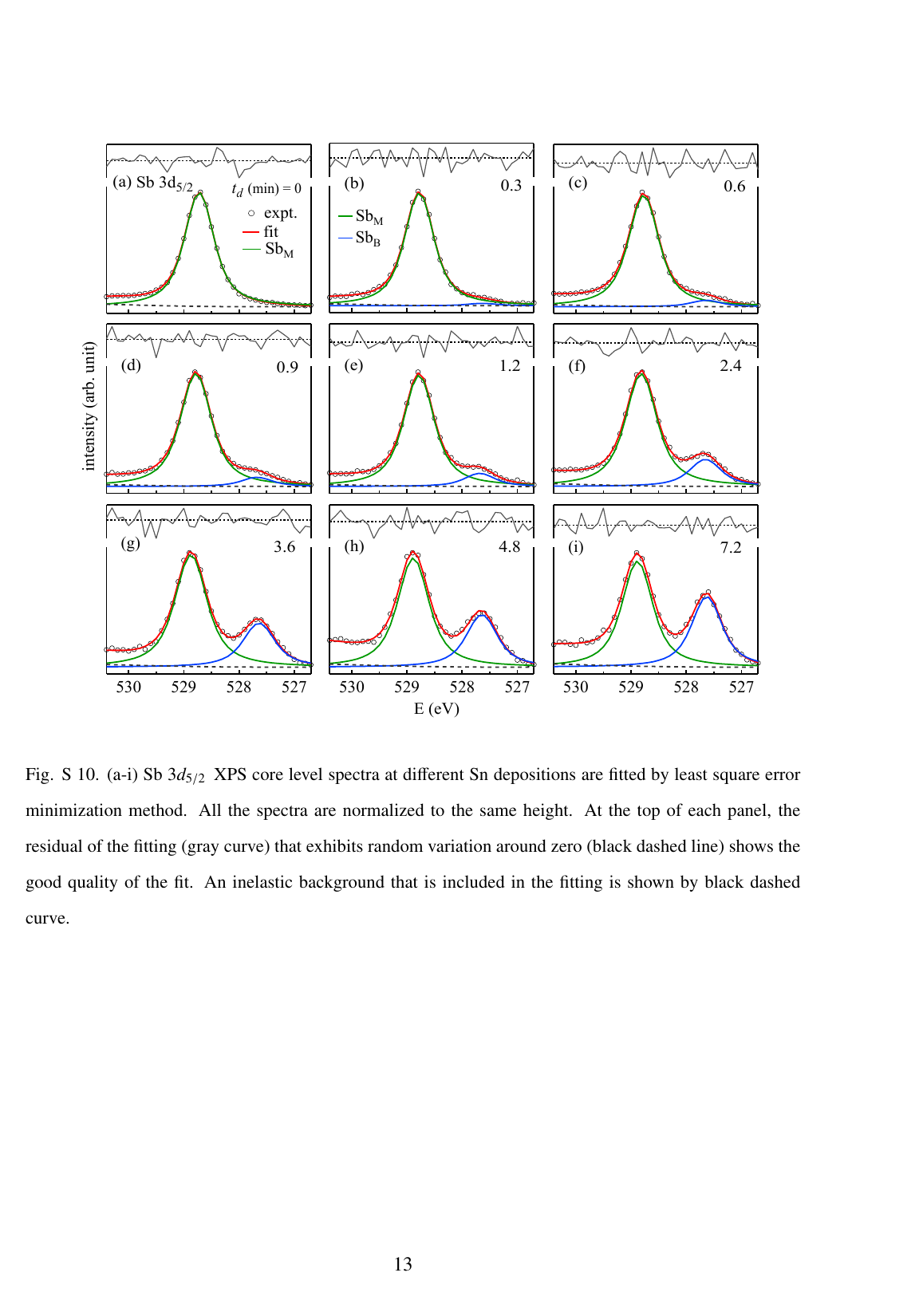} 
\end{figure}\newpage
\begin{figure}
	\includegraphics[page=14,width=1.3\linewidth,keepaspectratio,trim={2cm 0 -1cm 0},clip]{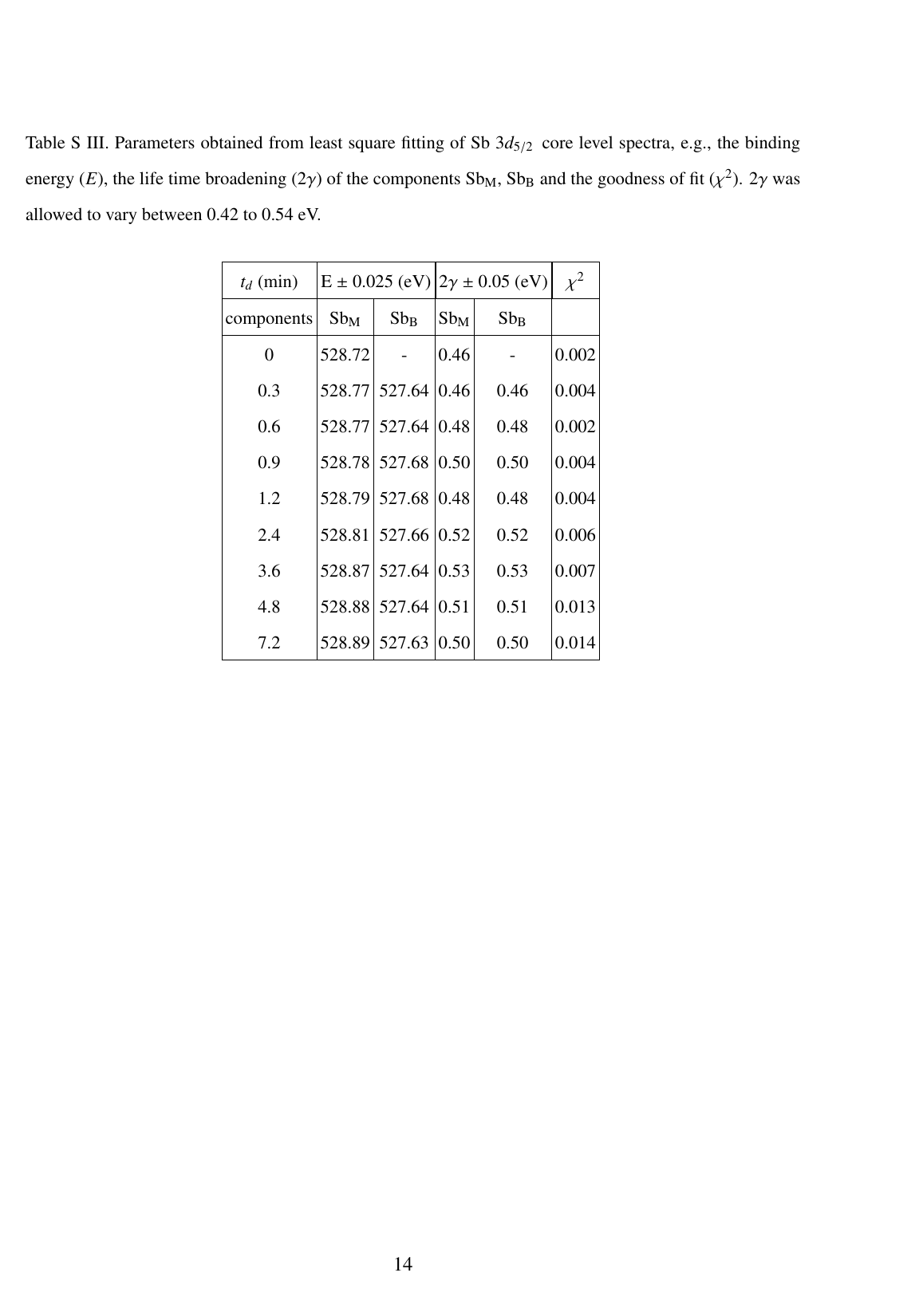} 
\end{figure}\newpage
\begin{figure}
	\includegraphics[page=15,width=1.3\linewidth,keepaspectratio,trim={2cm 0 -1cm 0},clip]{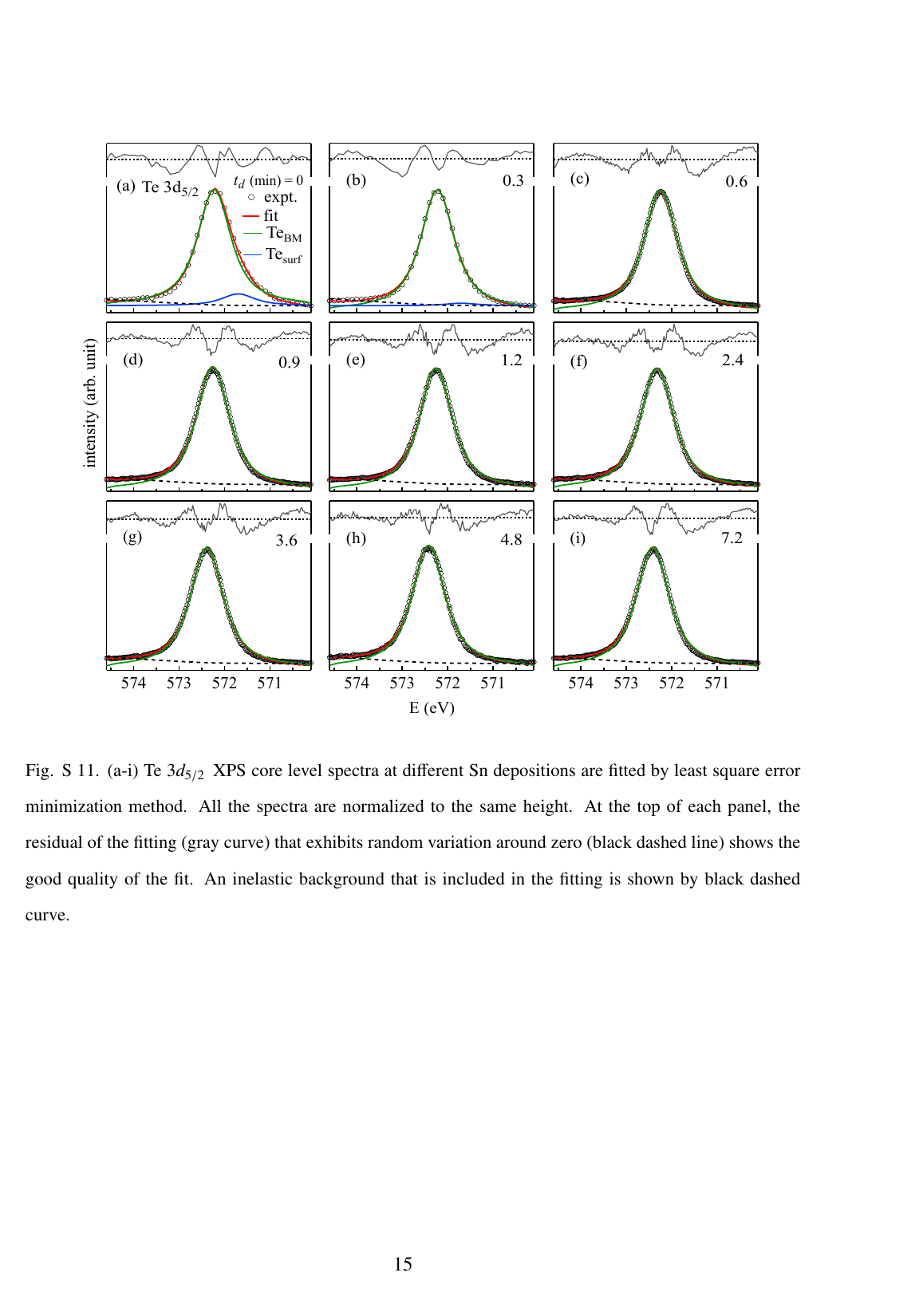} 
\end{figure}\newpage
\begin{figure}
	\includegraphics[page=16,width=1.3\linewidth,keepaspectratio,trim={2cm 0 -1cm 0},clip]{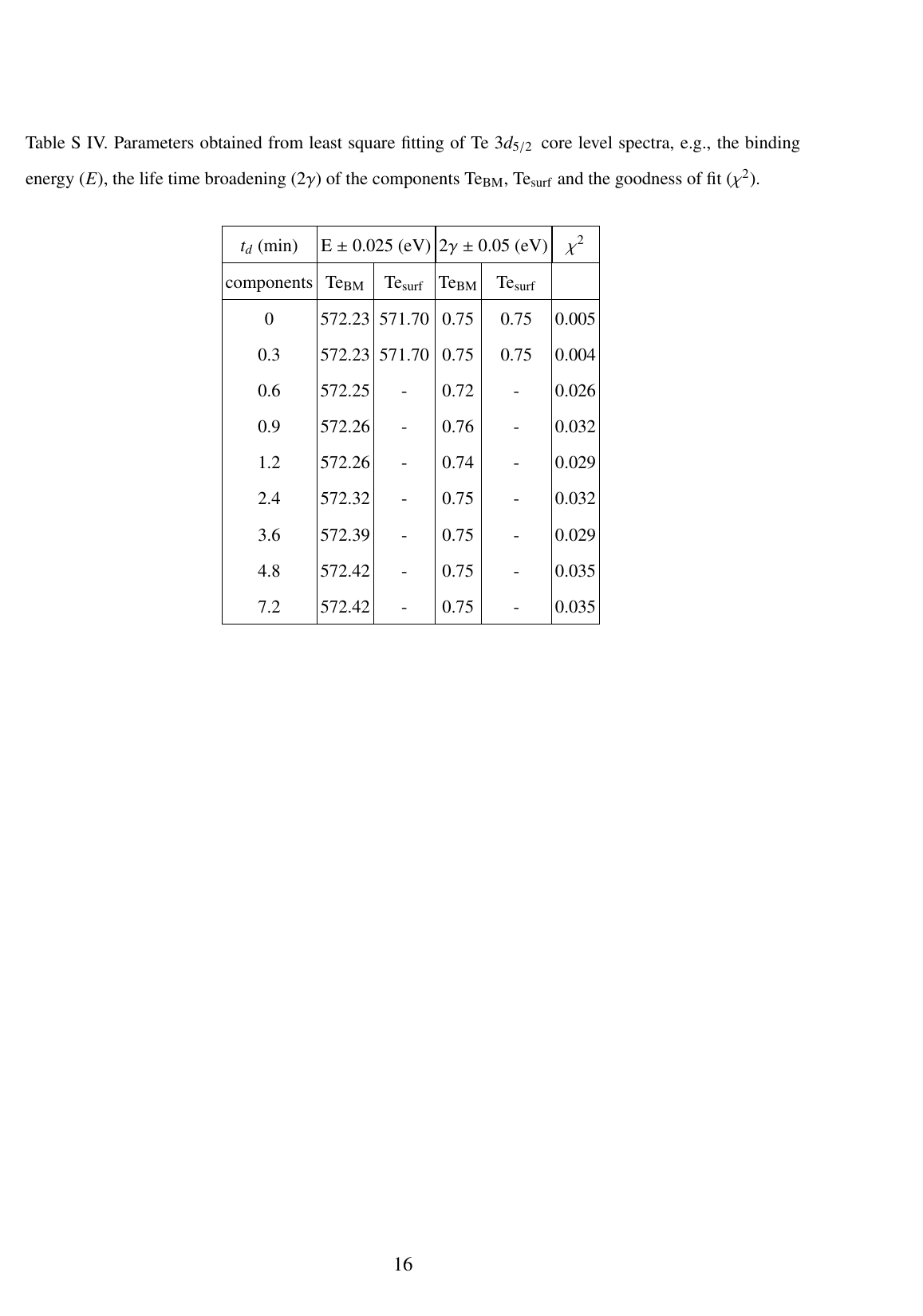} 
\end{figure}\newpage
\begin{figure}
\includegraphics[page=17,width=1.3\linewidth,keepaspectratio,trim={2cm 0 -1cm 0},clip]{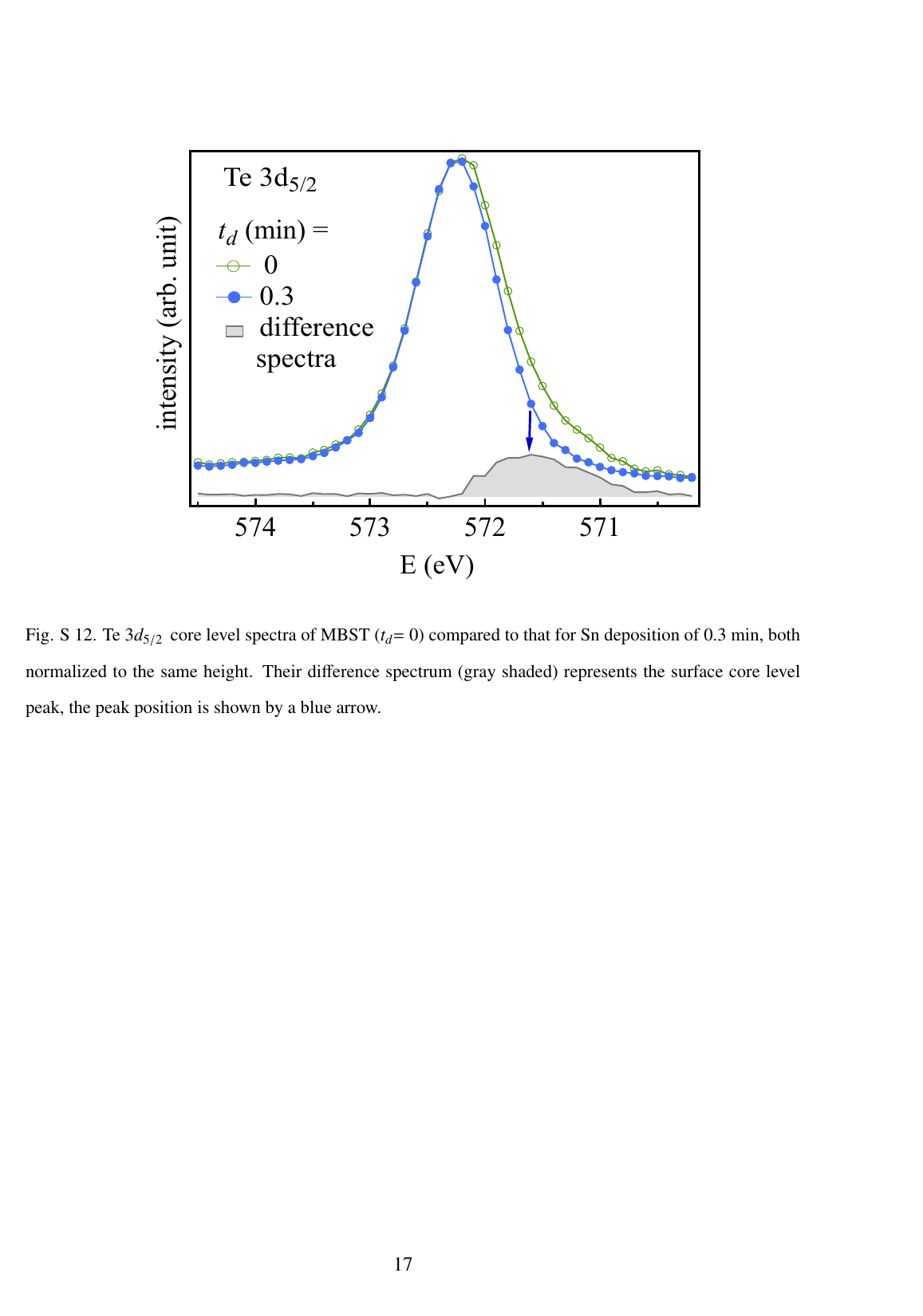} 
\end{figure}
%\newpage
\begin{figure}
	\includegraphics[page=18,width=1.3\linewidth,keepaspectratio,trim={2cm 0 -1cm 0},clip]{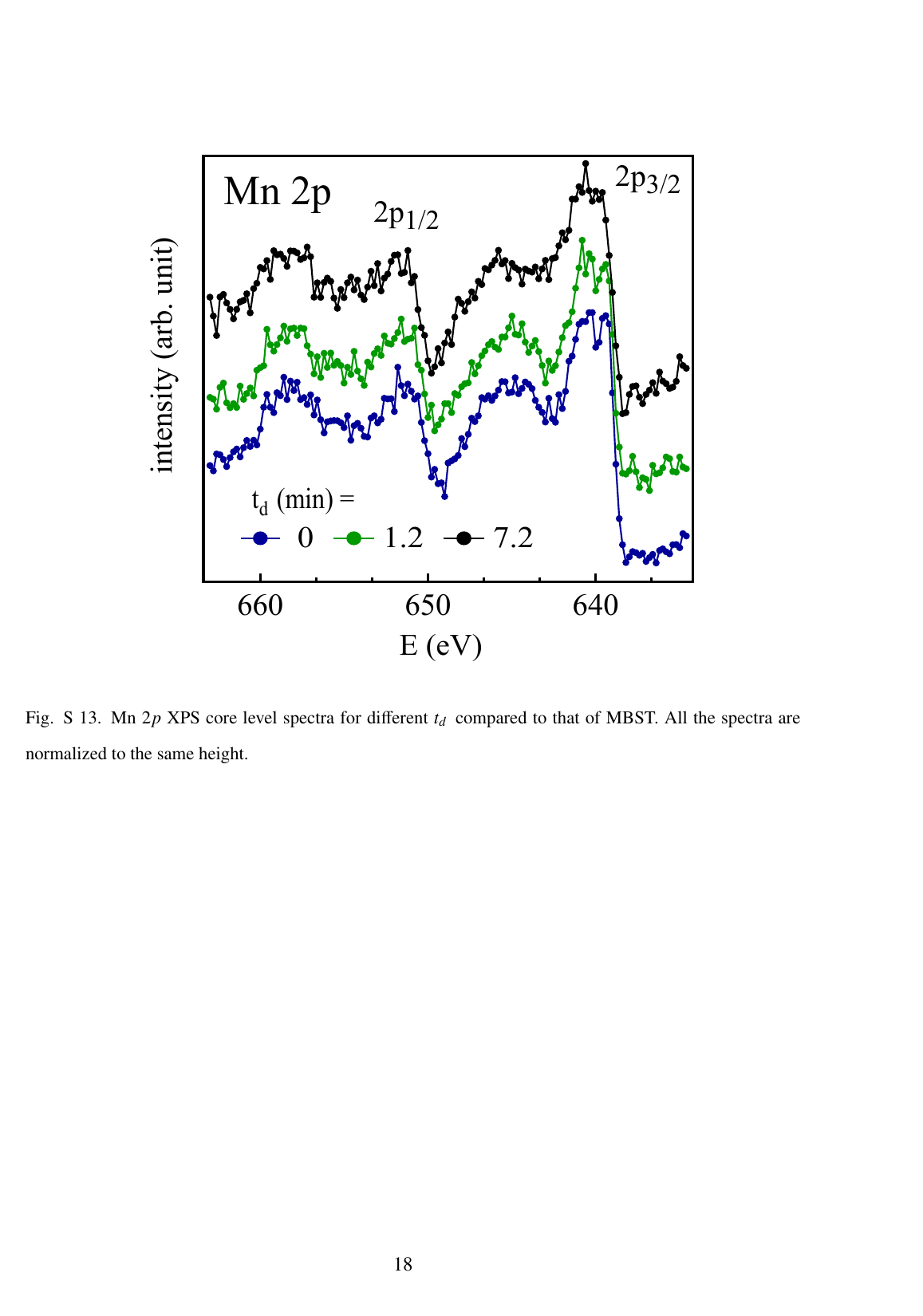} 
\end{figure}
%\newpage
\begin{figure}
	\includegraphics[page=19,width=1.3\linewidth,keepaspectratio,trim={2cm 0 -1cm 1cm},clip]{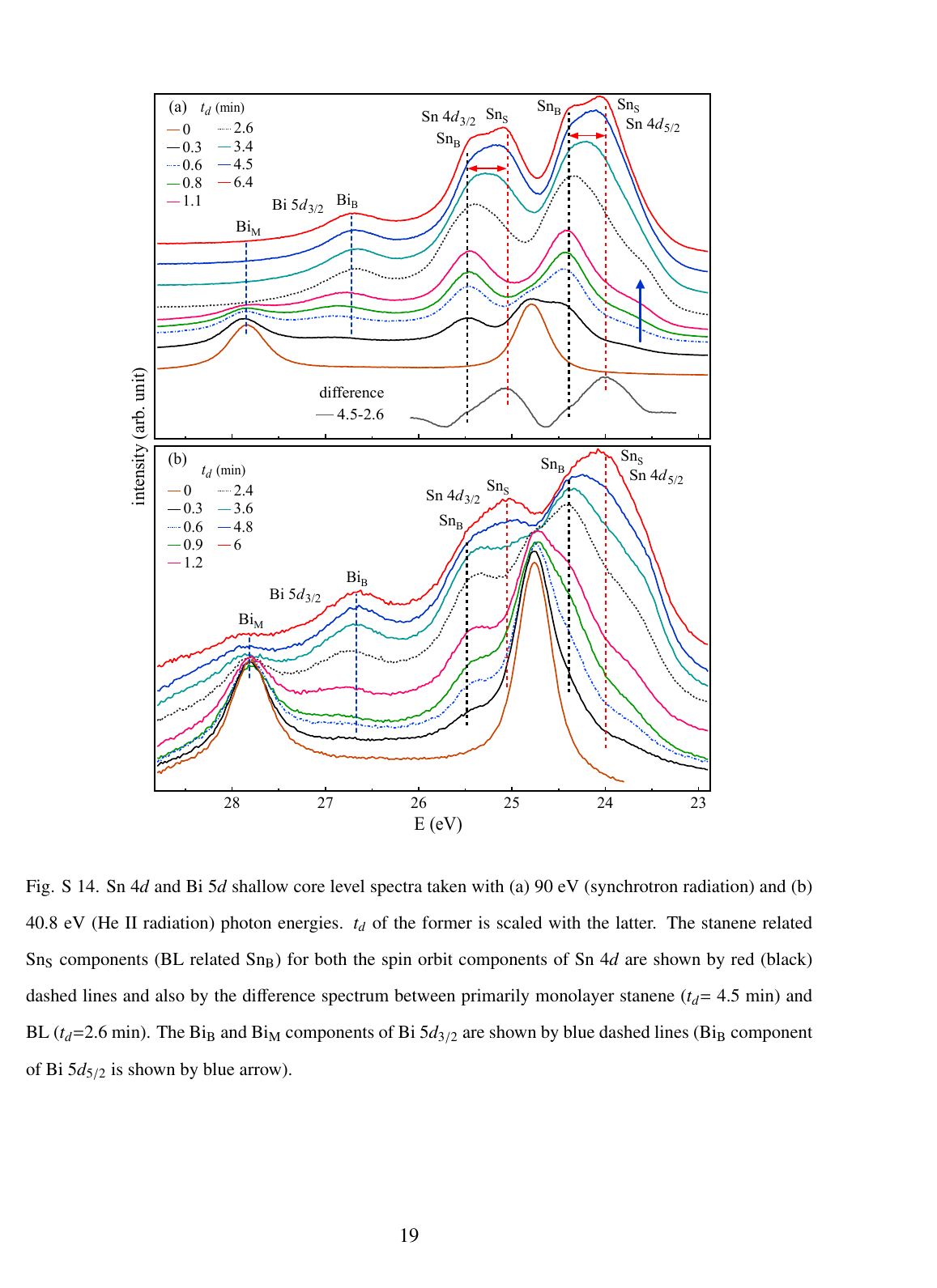} 
\end{figure}
\newpage
\begin{figure}
	\includegraphics[page=20,width=1.3\linewidth,keepaspectratio,trim={2cm 0 -1cm 1cm},clip]{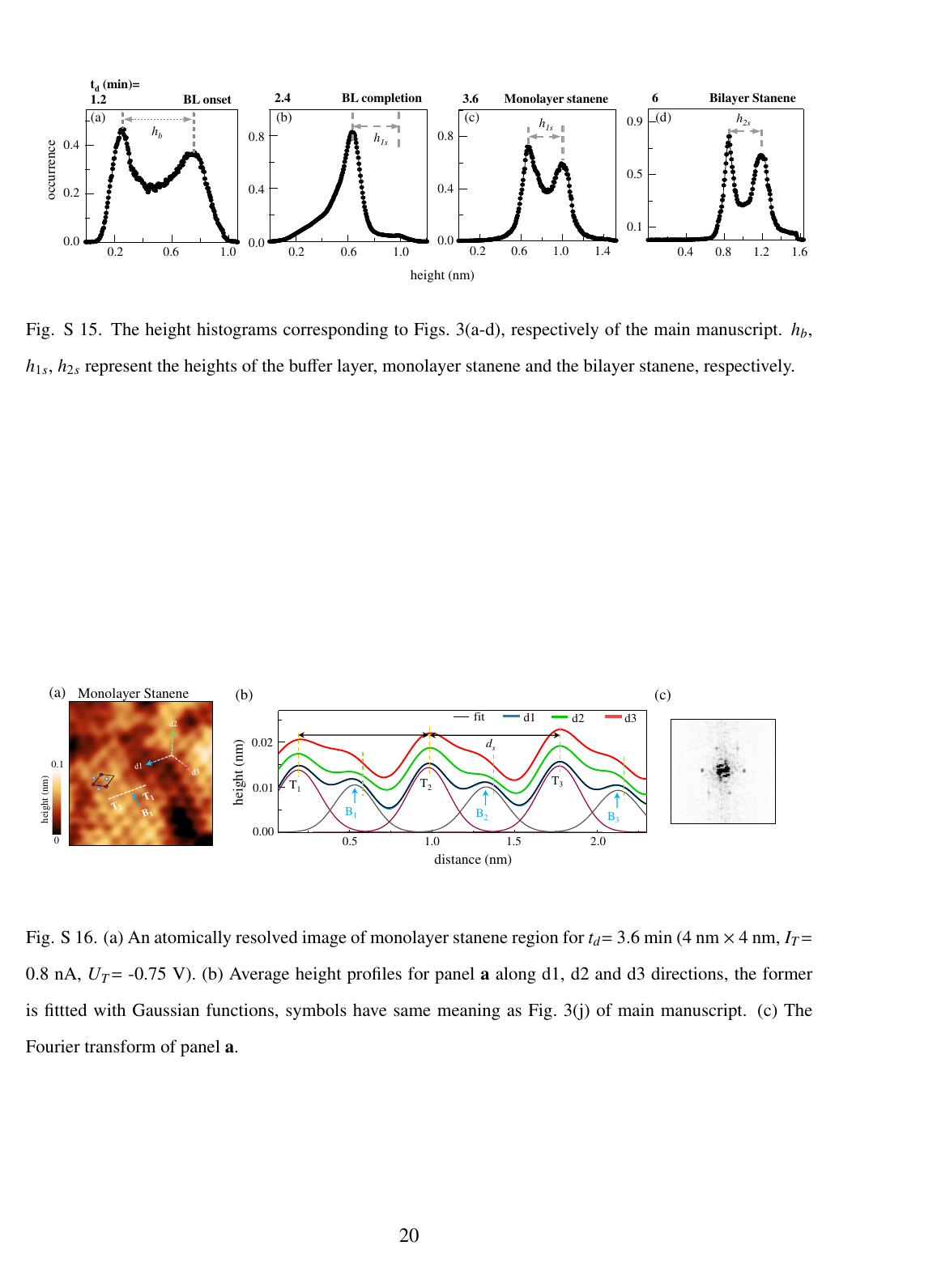} 
\end{figure}\newpage
\begin{figure}
	\includegraphics[page=21,width=1.3\linewidth,keepaspectratio,trim={2cm 0 -1cm 0},clip]{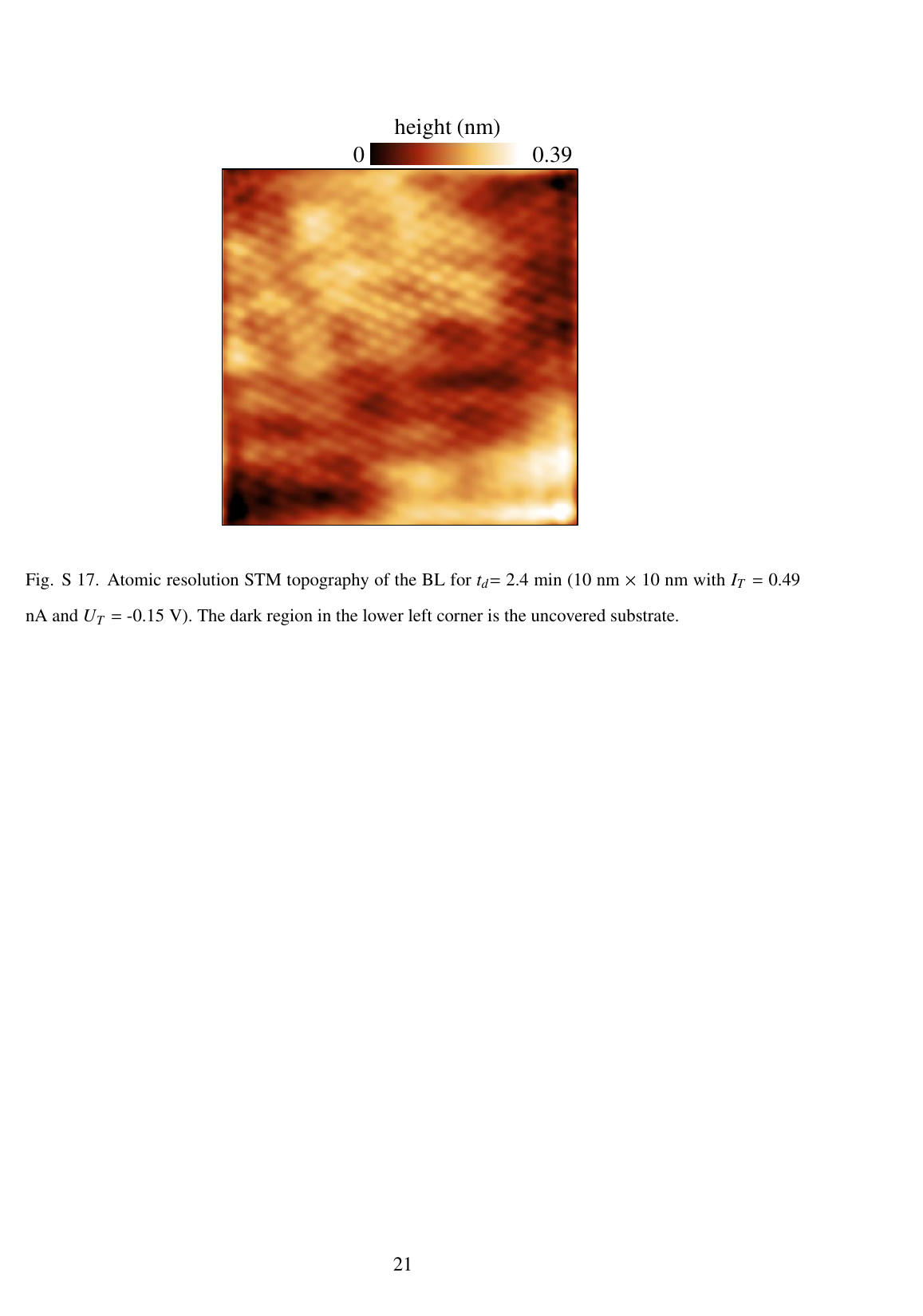} 
\end{figure}\newpage
\begin{figure}
	\includegraphics[page=22,width=1.3\linewidth,keepaspectratio,trim={2cm 0 -1cm 0},clip]{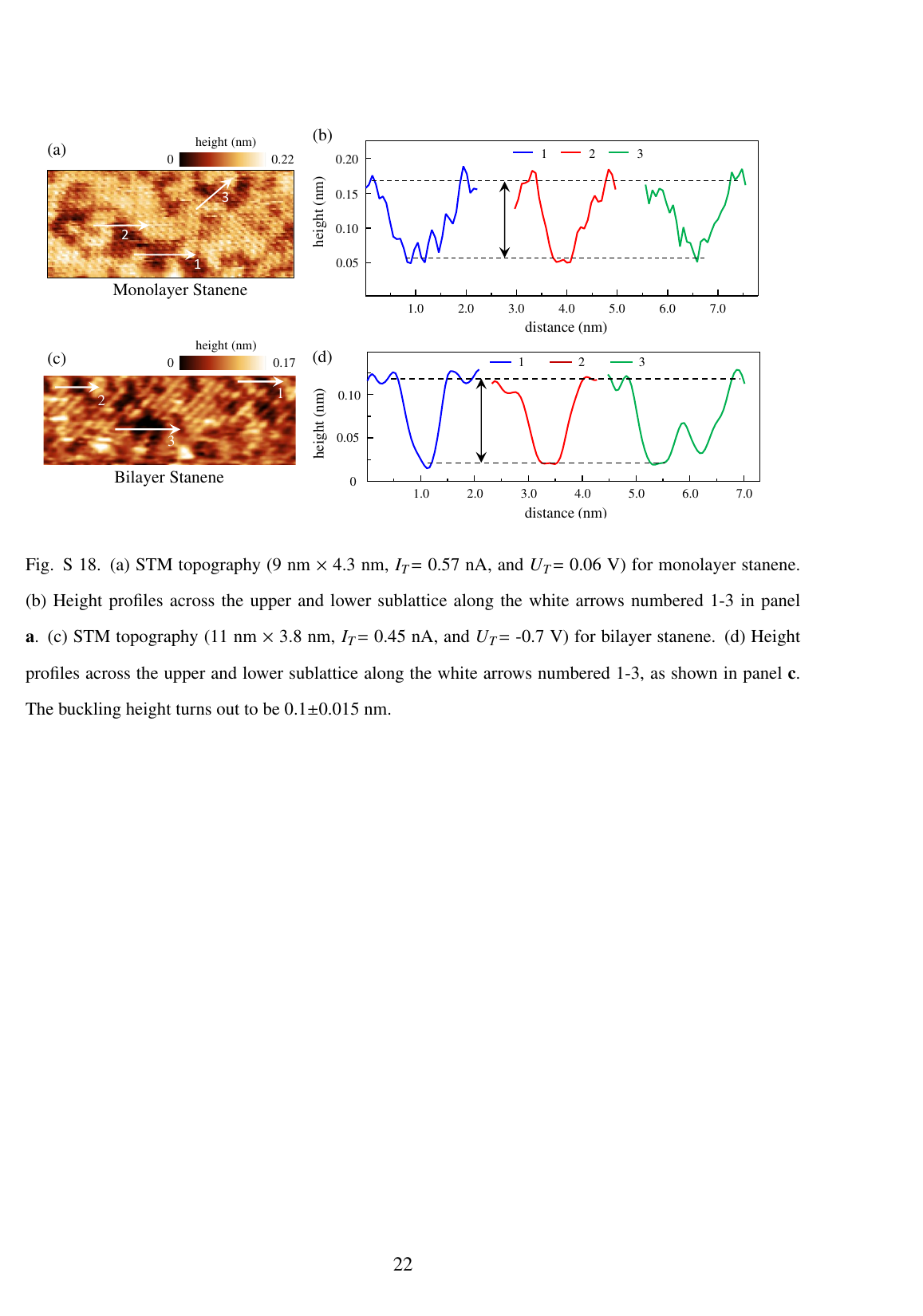} 
\end{figure}\newpage
\begin{figure}
	\includegraphics[page=23,width=1.3\linewidth,keepaspectratio,trim={2cm 0 -1cm 0},clip]{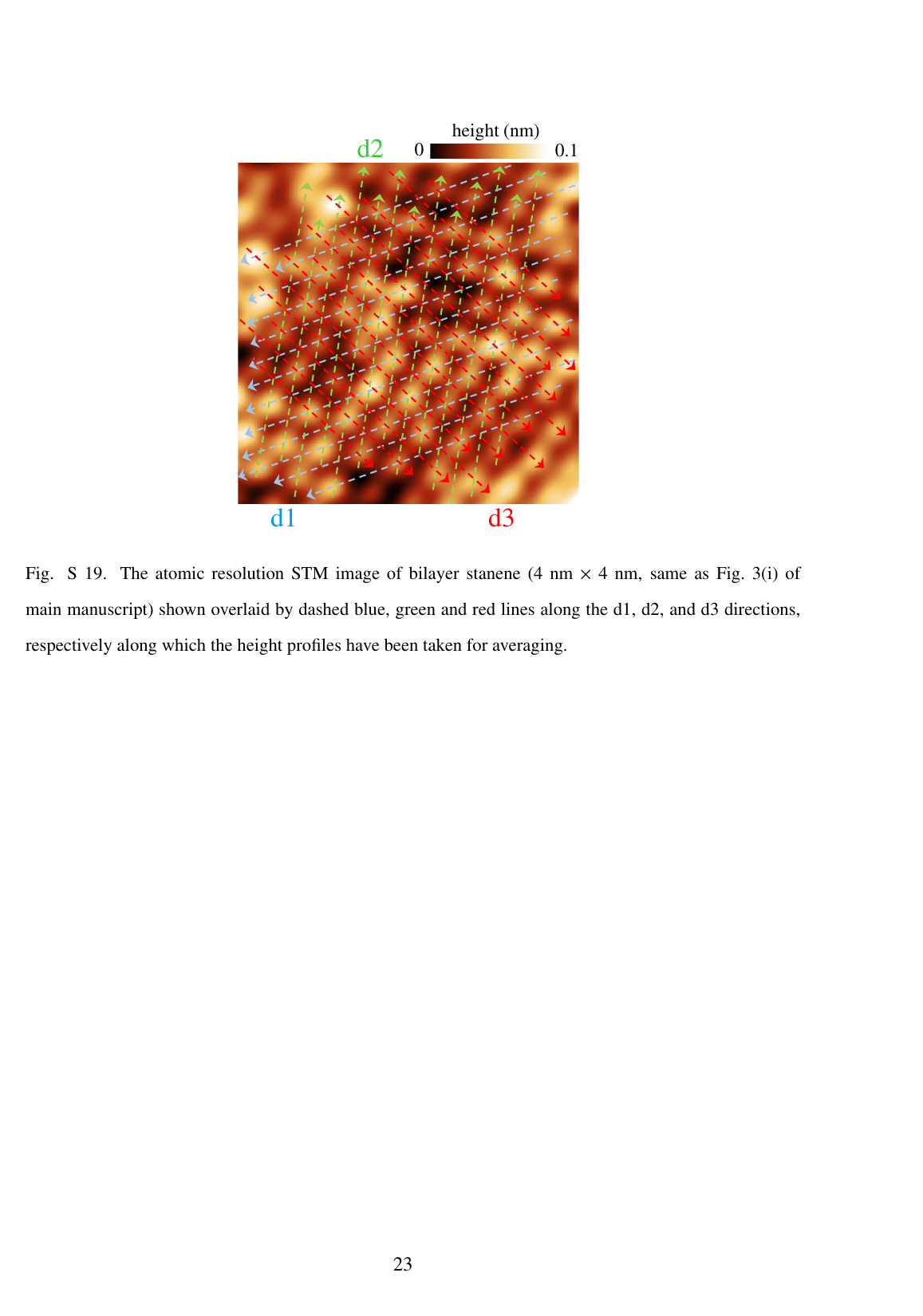} 
\end{figure}\newpage
\begin{figure}
	\includegraphics[page=24,width=1.3\linewidth,keepaspectratio,trim={2cm 0 -1cm 2cm},clip]{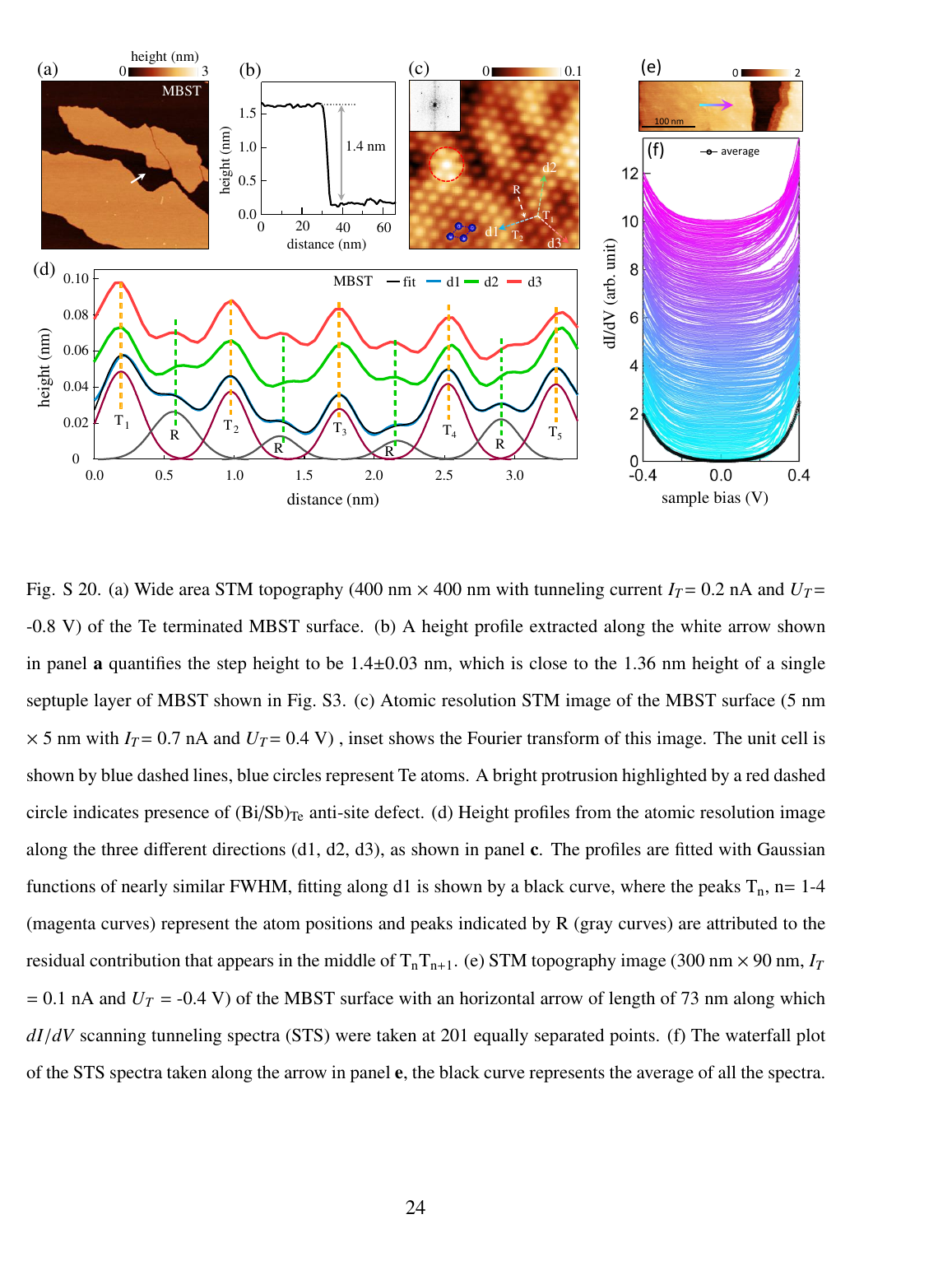} 
\end{figure}\newpage
\begin{figure}
	\includegraphics[page=25,width=1.3\linewidth,keepaspectratio,trim={2cm 0 -1cm 0},clip]{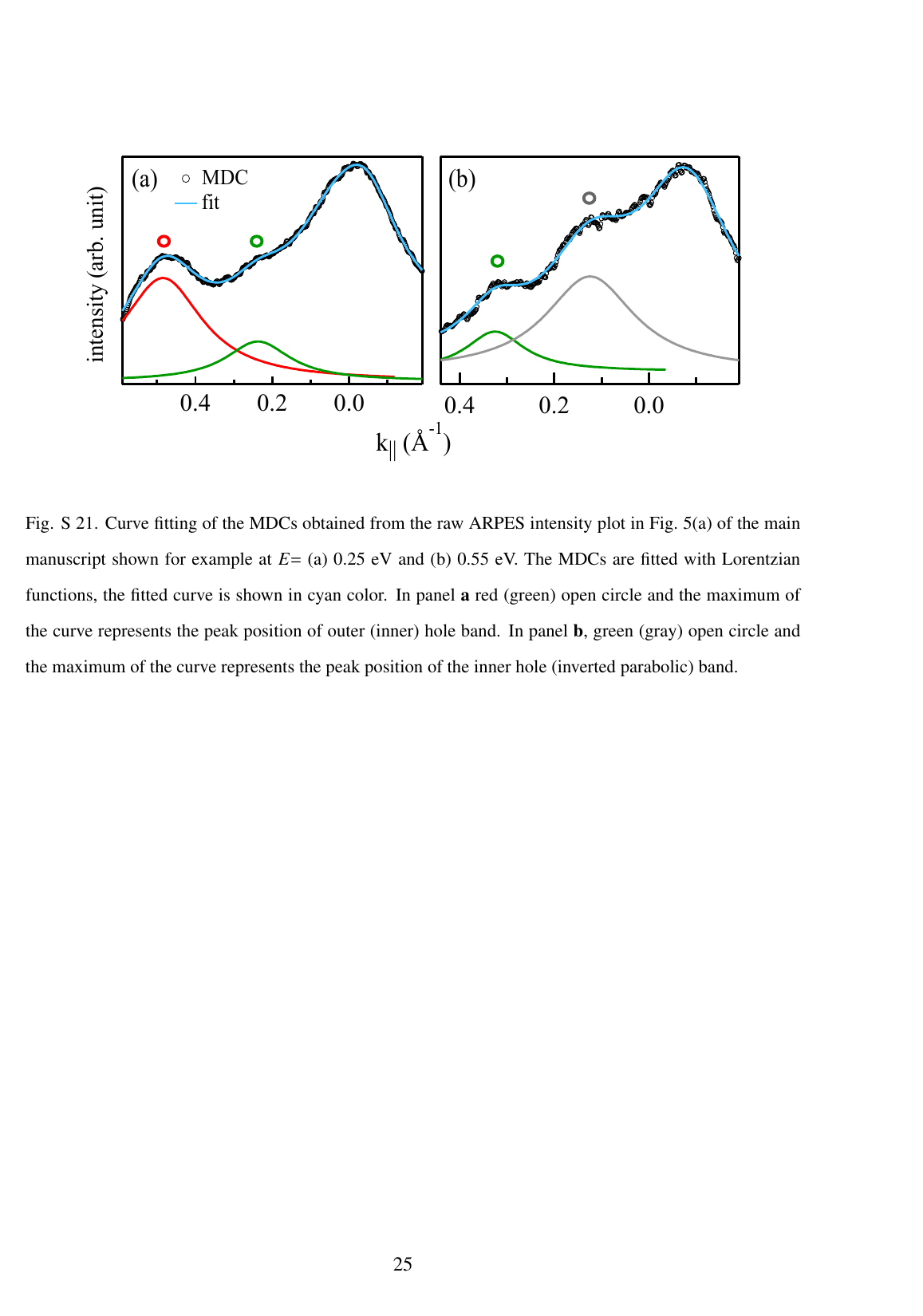} 
\end{figure}\newpage
\begin{figure}
	\includegraphics[page=26,width=1.3\linewidth,keepaspectratio,trim={2cm 2cm -1cm 2.5cm},clip]{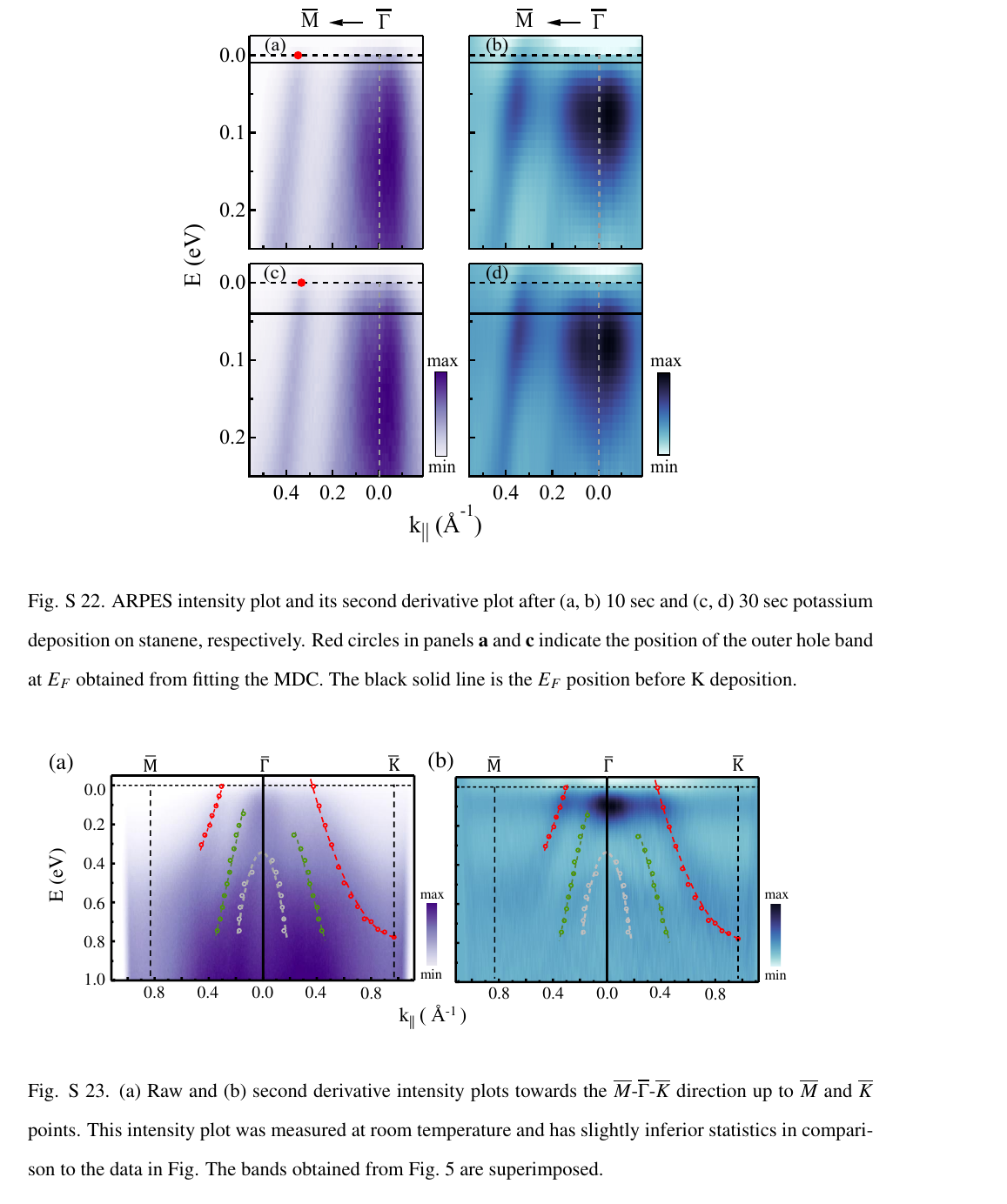} 
\end{figure}\newpage
\begin{figure}
	\includegraphics[page=27,width=1.3\linewidth,keepaspectratio,trim={2cm 0 -1cm 1.5cm},clip]{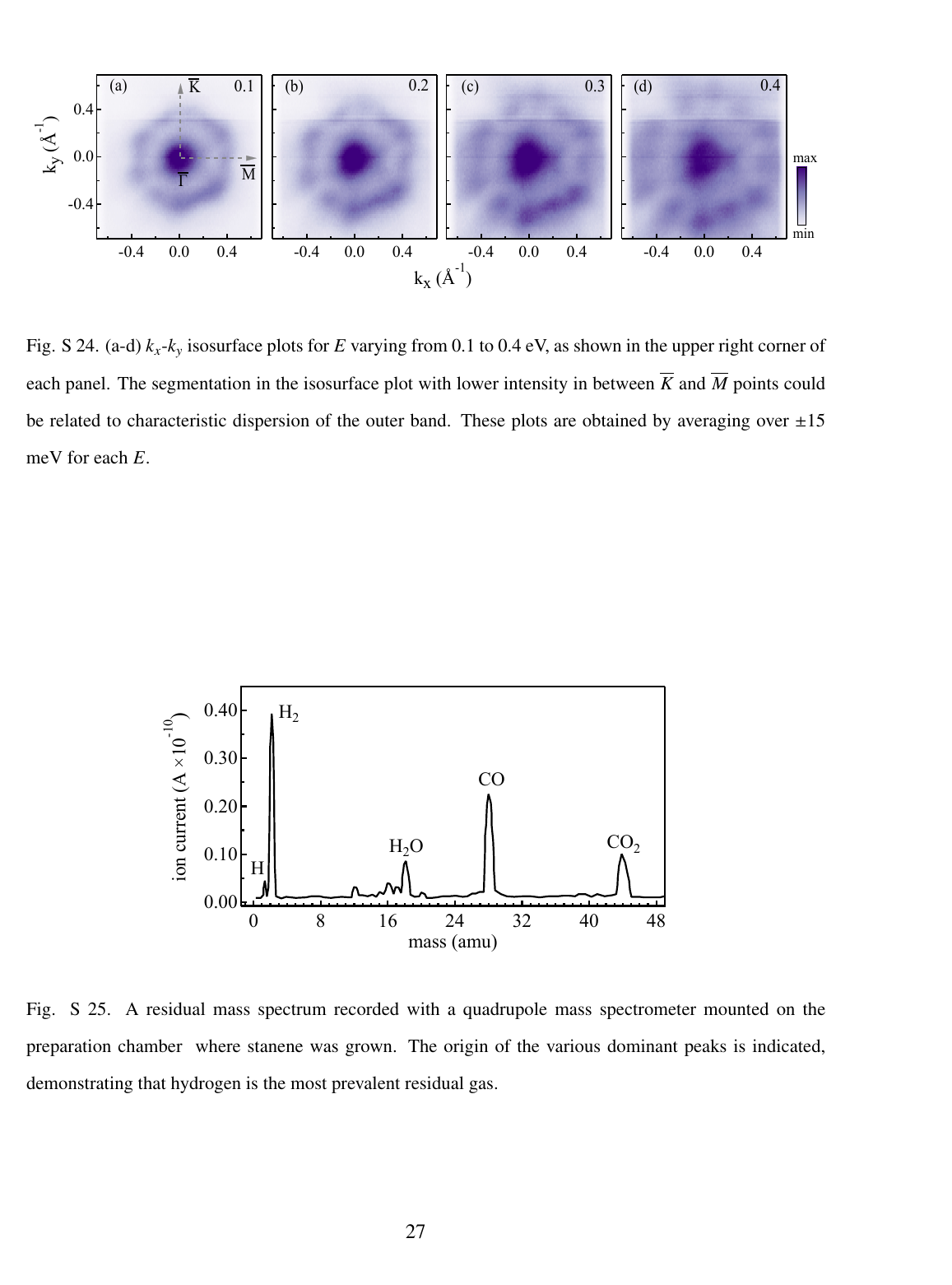} 
\end{figure}\newpage
\begin{figure}
	\includegraphics[page=28,width=1.2\linewidth,keepaspectratio,trim={2cm .5cm -1cm 0.2cm},clip]{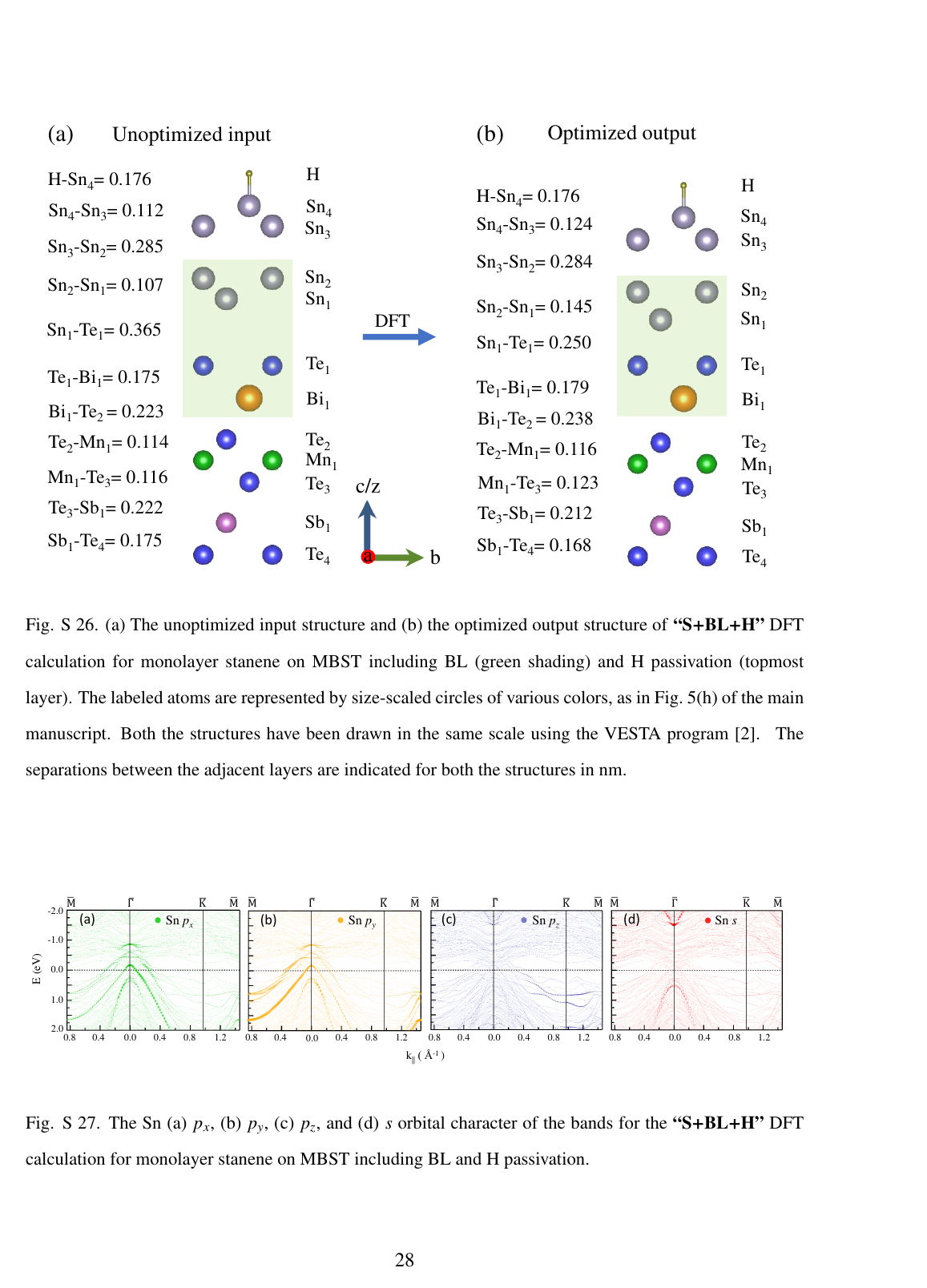} 
\end{figure}\newpage
\begin{figure}
	\includegraphics[page=29,width=1.3\linewidth,keepaspectratio,trim={2cm 0 -1cm 0},clip]{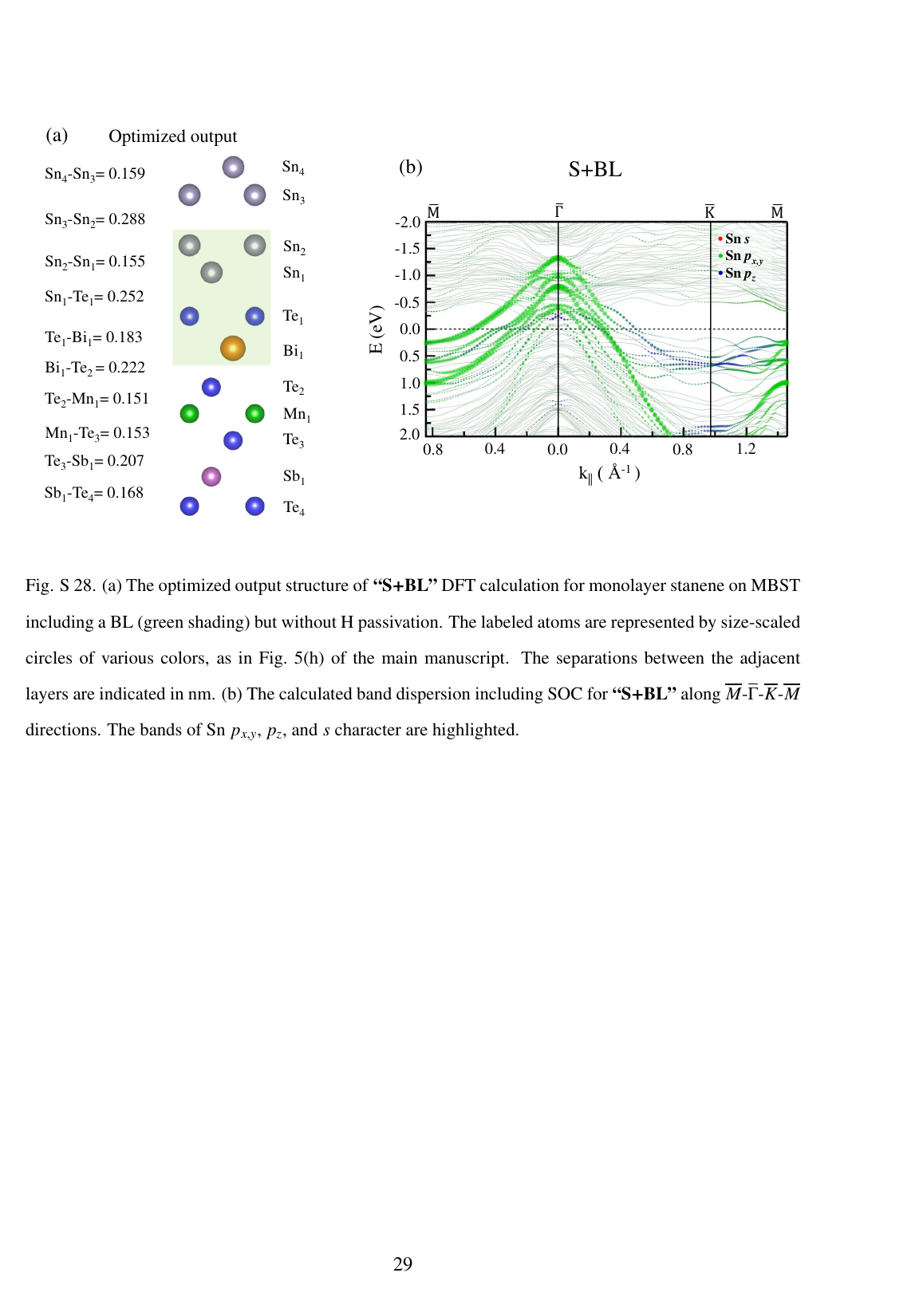} 
\end{figure}\newpage
\begin{figure}
	\includegraphics[page=30,width=1.2\linewidth,keepaspectratio,trim={2cm 0 -1cm 0},clip]{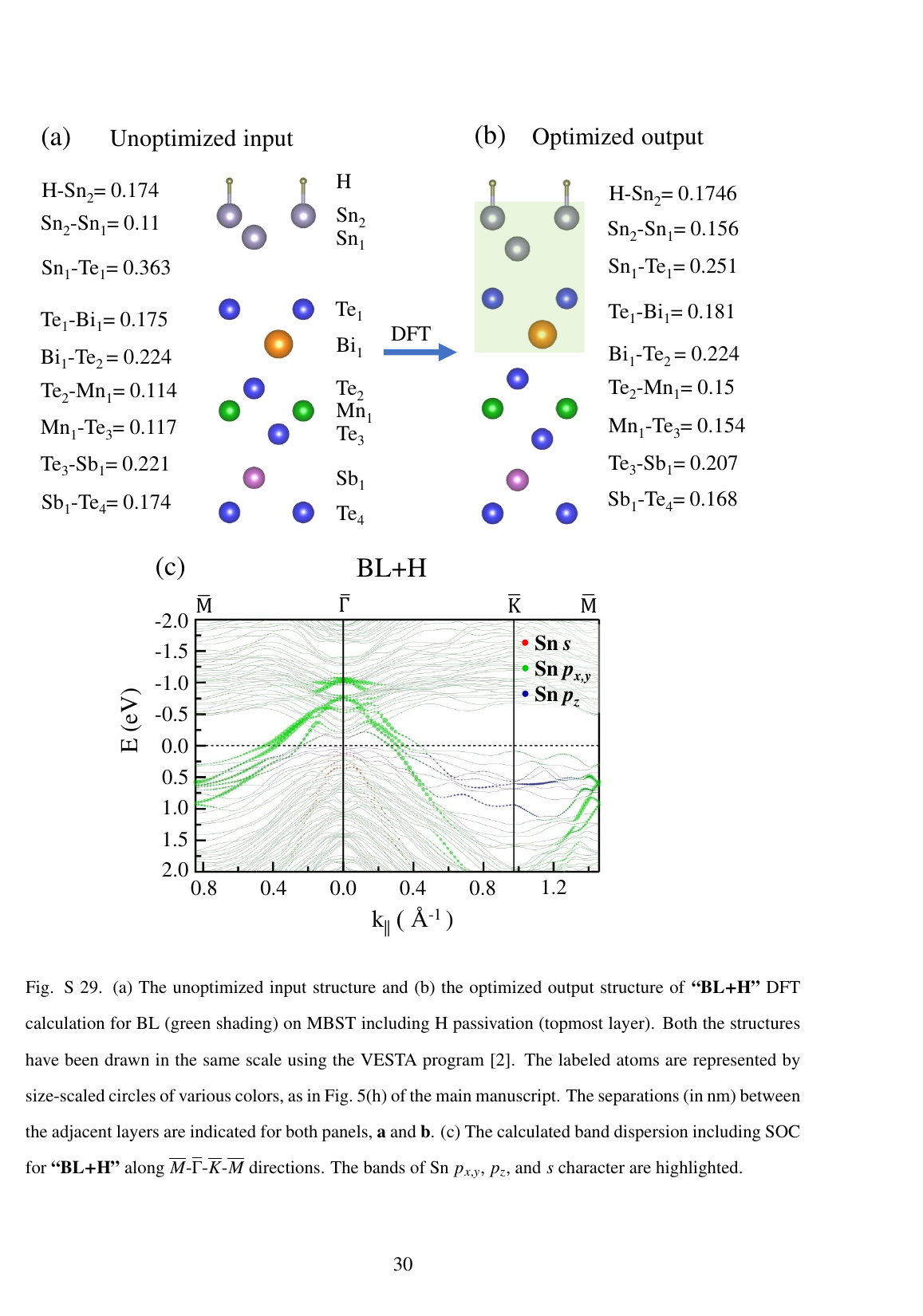} 
\end{figure}\newpage
\begin{figure}
	\includegraphics[page=31,width=1.3\linewidth,keepaspectratio,trim={2cm 0 -1cm 0},clip]{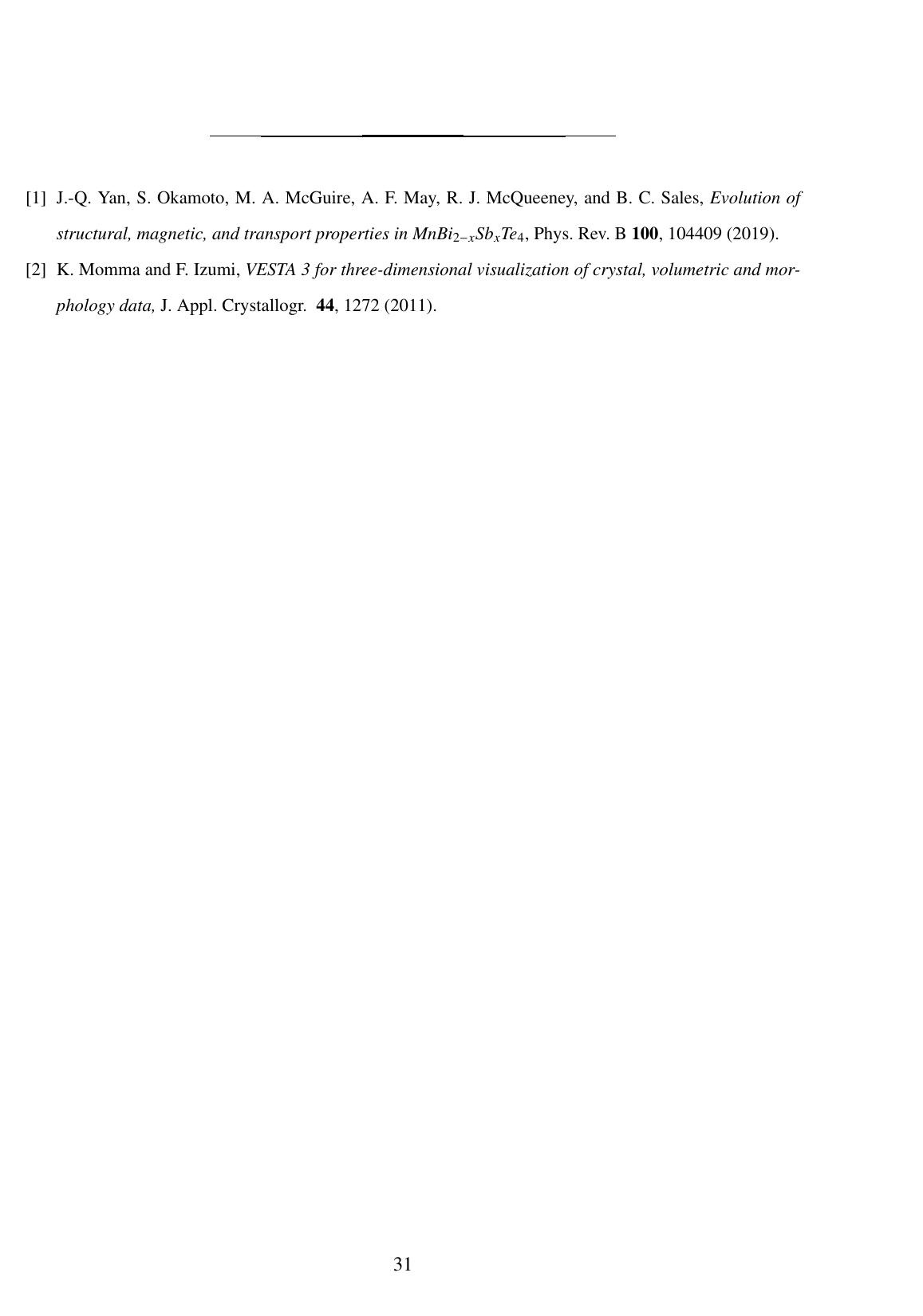} 
\end{figure}\newpage
\end{document}